\documentclass[preprint,12pt,authoryear]{elsarticle}

\usepackage{amssymb}
\usepackage{booktabs}
\usepackage{array}
\usepackage[shortcuts,acronym]{glossaries}
\usepackage[margin=2.5cm]{geometry}
\usepackage{rotating}
\usepackage{graphicx}
\usepackage{arydshln}
\usepackage{makecell}
\usepackage{multirow}
\usepackage{adjustbox}
\usepackage{amsmath}
\usepackage{natbib}
\usepackage{xcolor}
\newacronym{sc}{SC}{Silhouette score}
\newacronym{rbs}{RBS}{rule-based system}
\newacronym{fp}{FP}{false positives}
\newacronym{fn}{FN}{false negatives}
\newacronym{tp}{TP}{true positives}
\newacronym{tn}{TN}{true negatives}
\newacronym{ml}{ML}{machine learning}
\newacronym{nn}{NN}{neural network}
\newacronym{fmcw}{FMCW}{frequency-modulated continuous wave}
\newacronym{ai}{AI}{artificial intelligence}
\newacronym{id}{ID}{in-distribution}
\newacronym{ood}{OOD}{out-of-distribution}
\newacronym{ffnn}{FFNN}{feed-forward neural network}
\newacronym{iai}{IAI}{interpretable AI}
\newacronym{shap}{SHAP}{Shapley Additive Explanations}
\newacronym{xai}{XAI}{explainable AI}
\newacronym{er}{ER}{experience replay}
\newacronym{tl}{TL}{transfer learning}
\newacronym{gru}{GRU}{gated recurrent unit}
\newacronym{srv}{SRVs}{SHAP reference values}
\newacronym{euai}{EU AI Act}{European Union's Artificial Intelligence Act}
\newacronym{rnn}{RNN}{recurrent neural network}
\newacronym{hleg}{HLEG}{High-Level Expert Group}
\newacronym{fft}{FFT}{fast Fourier transform}
\newacronym{rf}{RF}{radio-frequency}
\newacronym{if}{IF}{intermediate frequency}
\newacronym{std}{STD}{standard deviation}
\newacronym{ae}{AE}{autoencoders}
\newacronym{vae}{VAE}{variational autoencoder}
\newacronym{hgr}{HGR}{hand gesture recognition}
\newacronym{mmwave}{mmWave}{millimeter-wave}
\newacronym{cnn}{CNN}{convolutional neural network}
\newacronym{lstm}{LSTM}{long short-term memory}
\newacronym{svm}{SVM}{Suport vector machines}
\newacronym{knn}{kNN}{k-nearest neighbors}
\newacronym{hmm}{HMM}{Hidden markov models}
\newacronym{pca}{PCA}{principle component analysis}
\newacronym{resnet}{ResNet}{Residual networks}
\newacronym{rmd}{RMD}{relative Mahalanobis distance}
\newacronym{gan}{GAN}{generative adversarial networks}

\newacronym[\glsshortpluralkey={NNs}]{NN}{NN}{neural network}

\journal{Machine Learning with Applications}

\begin{document}
\sloppy
\begin{frontmatter}



\title{Complying with the EU AI Act: Innovations in Explainable and User-Centric Hand Gesture Recognition}


\author[label1,label2]{Sarah Seifi}
\ead{sarah.seifi@tum.de}
\author[label2,label3]{Tobias Sukianto}
\ead{tobias.sukianto@infineon.com}
\author[label2]{Cecilia Carbonelli}
\ead{cecilia.carbonelli@infineon.com}
\author[label1]{Lorenzo Servadei}
\ead{lorenzo.servadei@tum.de}
\author[label1,label4]{Robert Wille}
\ead{robert.wille@tum.de}

\affiliation[label1]{organization={Chair for Design Automation, Technical University Munich},
            addressline={Arcisstr.21}, 
            city={Munich},
            postcode={80333}, 
            state={Bavaria},
            country={Germany}}

\affiliation[label2]{organization={Infineon Technologies AG},
                    addressline={Am Campeon 1-15}, 
                    city={Neubiberg},
                    postcode={80939}, 
                    state={Bavaria},
                    country={Germany}}
\affiliation[label3]{organization={Institute for Signal Processing, Johannes Kepler University Linz},
            addressline={Altenbergerstraße 69}, 
            city={Linz},
            postcode={4040},
            country={Austria}}
        
\affiliation[label4]{organization={Software Competence Center Hagenberg GmbH (SCCH)},
            addressline={Softwarepark 32a}, 
            city={Hagenberg},
            postcode={4232},
            country={Austria}}
\begin{abstract}
The EU AI Act underscores the importance of transparency, user-centricity, and robustness in AI systems, particularly for high-risk applications. In response, we present advancements in XentricAI, an explainable \ac{hgr} system designed to meet these regulatory requirements. XentricAI addresses fundamental challenges in HGR, such as the opacity of black-box models using explainable AI methods and the handling of distributional shifts in real-world data through transfer learning techniques.

We extend an existing radar-based HGR dataset by adding 28,000 new gestures, with contributions from multiple users across varied locations, including 24,000 out-of-distribution gestures. Leveraging this real-world dataset, we enhance XentricAI's capabilities by integrating a variational autoencoder module for improved gesture anomaly detection, incorporating user-specific dynamic thresholding. This integration enables the identification of $11.50\%$ more anomalous gestures.

Our extensive evaluations demonstrate a $97.5\%$ success rate in characterizing these anomalies, significantly improving system explainability. Furthermore, the implementation of transfer learning techniques has shown a substantial increase in user adaptability, with an average performance improvement of at least $15.17\%$.

This work contributes to the development of trustworthy AI systems by providing both technical advancements and regulatory compliance, offering a commercially viable solution that aligns with the EU AI Act requirements.
\end{abstract}


\end{frontmatter}


\section{Introduction}
\label{sec:intro}

The increasing prevalence of \ac{ai} tools in our daily lives, such as OpenAI's release of ChatGPT\footnote{https://chatgpt.com/} in November 2022 and the text-to-video generator Sora\footnote{https://openai.com/index/sora/} in February 2024, has prompted a regulatory response from numerous countries, aiming to manage AI's development and deployment effectively. One of these requirements is the \ac{euai}\footnote{https://eur-lex.europa.eu/eli/reg/2024/1689/oj}. It plays a significant role in the governance of \ac{ai}, requiring that AI systems be transparent, user-centric, and robust. The act categorizes AI systems based on their risk levels and sets forth stringent requirements for compliance, including significant fines for violations. This regulatory framework underscores the critical need for clear guidelines and robust mechanisms to ensure AI safety and reliability. The principles of transparency and user-centric design emphasized by the EU AI Act are central to our ongoing research and development of AI systems.
Current regulations primarily address high-risk categories and general-purpose AI-based applications. However, the EU AI Act also recommends a code of conduct for all other applications, such as the one discussed in this paper. We believe that adopting best practices offers a competitive advantage, creates opportunities, and fosters further innovation.
The EU AI Act outlines obligations according to risk categories, but these obligations are still too abstract to be directly mapped onto technical requirements. It is now the task of standardization bodies and relevant organizations to translate these obligations into tests and concrete metrics. Our framework aims to contribute to building this benchmarking landscape.
Aligned with the EU AI Act’s focus, this paper presents advancements in \ac{hgr} systems, a key area that benefits from such regulatory guidance.

\ac{hgr} systems are designed to interpret and respond to human gestures, enabling intuitive and efficient human-machine interaction \citep{sarkar2013hand}. These systems provide an interface between individuals and technology, with valuable applications in areas such as smart homes, medical devices, and security systems \citep{wan2014gesture,wang2022medical,kabisha2022face}. However, their performance is often hindered by two significant challenges. Firstly, traditional \ac{ai} models, such as neural networks, are so-called black-box models, making it difficult to understand their internal mechanisms and decision-making processes \citep{molnar2020interpretable_book}. This lack of transparency can lead to a lack of trust in these systems. Secondly, the disparity between the training data and real-world deployment scenarios usually results in a distributional shift, leading to reduced accuracy and unexpected misclassifications \citep{cui2022stable}. This can ultimately undermine the usability and reliability of these systems.

Previously, we introduced XentricAI, an innovative approach that combines \ac{xai} with \ac{er} to create a user-centric framework for \ac{hgr} \citep{seifi2024xentricai}. XentricAI was designed to address the limitations posed by the lack of explainability and distributional shifts, enhancing the accuracy and user engagement of AI models in real-world applications. Building on this foundation, the current work focuses on the extensive evaluation, improvement, and extension of XentricAI, ensuring it meets the increasingly stringent requirements of AI applications.

In this paper, we present several key contributions that extend and enhance XentricAI:

\begin{enumerate}
    \item Innovative anomaly detection with a tailored \ac{vae} architecture: Our work introduces a tailored \ac{vae}-based module to address limitations in the original system, which relied on hardcoded constraints for anomaly detection. By incorporating user-specific dynamic thresholding based on reconstruction error, this module achieves an 11.50\% improvement in identifying and characterizing anomalous gestures. This enhancement provides a more robust and reliable solution to the challenges of \ac{hgr}, ensuring better accuracy and coverage in detecting varied gesture patterns. To our knowledge, this is the first application of user-specific thresholding in \ac{fmcw} radar-based \ac{hgr}.
    \item Extension of an existing dataset with \ac{ood} data\footnote{https://ieee-dataport.org/documents/60-ghz-fmcw-radar-gesture-dataset}: To evaluate model robustness and adaptability, we extend an existing publicly available \ac{fmcw} radar-based gesture sensing dataset with \ac{ood} data and additional users \citep{radar_data}. This extension involves collecting 24,000 anomalous gestures from eight users, based on three types of anomalies. Additionally, we add 4,000 nominal gestures from four users to the nominal gesture dataset, resulting in a total of 31,000 nominal gestures.
    \item User-centric design evaluation: We evaluate the adaptability of our models to a larger and more diverse user base to emphasize the importance of designing AI systems that cater effectively to the varied needs and behaviors of different users. This assessment, termed user-centric calibration, is extended to nine users utilizing our dataset. The evaluation also includes the model's behavior on anomalous data, ensuring a comprehensive understanding of its performance across different user scenarios.
    \item Precise analysis of gesture characterization capabilities: We thoroughly assess our explainable system's ability to characterize anomalous gestures. Evaluating a larger dataset with more users, we find that $97.5\%$ of gestures flagged as anomalous are accurately characterized by the system.
    \item Practical application example: Using radar-based \ac{hgr} and a commercial sensor as a case study, we demonstrate the real-world applicability of our findings, bridging the gap between theoretical research and applied \ac{ml}.
\end{enumerate}

Our research aligns with the EU AI Act by emphasizing transparency and user-centric design. This paper provides a comprehensive study on the evaluation, improvement, and extension of XentricAI, demonstrating its potential as a robust and reliable user-centric approach to \ac{hgr} systems. By examining the user adaptability of the models and assessing gesture characterization capabilities across a diverse user base, our work contributes significantly to the ongoing discourse on trustworthy AI.

\section{Foundations of Trustworthy AI}
\label{sec:background}

In this section, we lay the groundwork for understanding the principles and challenges associated with developing trustworthy AI systems. We begin by examining the EU AI Act, which provides a regulatory framework for ensuring AI systems are understandable, user-centric, and robust. Next, we delve into the essential terminology used in the context of trustworthy AI, clarifying key concepts and definitions that underpin this field. Finally, we explore the challenges faced in implementing trustworthy AI, highlighting the technical and ethical obstacles that must be navigated.

It is important to note that the proposed XentricAI algorithm is constructed from various building blocks. Thus, each building block is discussed in a dedicated subsection in the methodology section \ref{sec:methodology}, including the related work pertinent to that block.

\subsection{The European AI Act}

\begin{figure}
\centering
       \includegraphics[width = 1\linewidth]{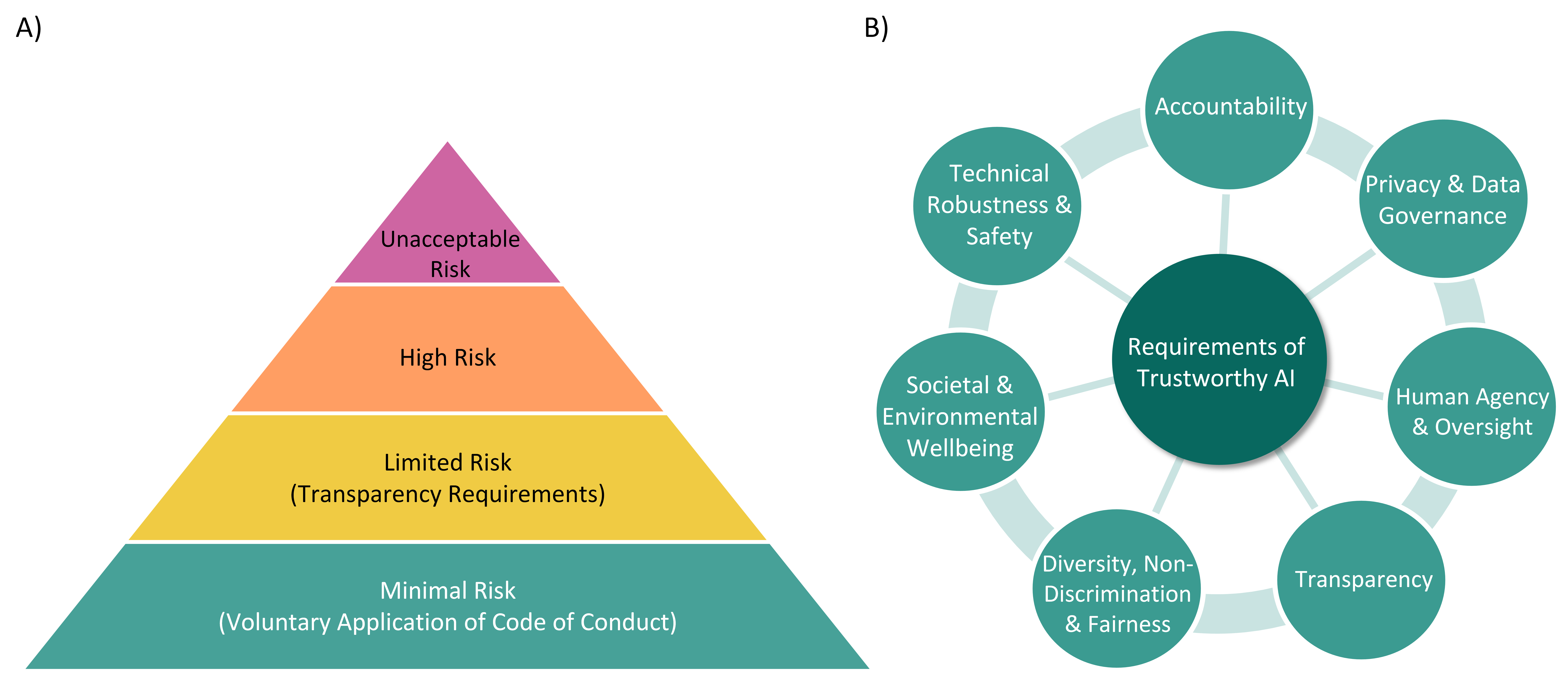}
      \vspace{-3mm}

    \caption{Overview of the EU AI Act's framework. A) Pyramid illustrating the risk-based approach of the EU AI Act. B) Ethical principles and requirements for trustworthy AI as outlined in the EU AI Act.}
    \vspace{-3mm}
\label{fig:trustworthy-ai}
 \end{figure}

The European Union has established regulatory frameworks for digital technologies, starting with the General Data Protection Regulation\footnote{https://eur-lex.europa.eu/eli/reg/2016/679/oj} (GDPR), which was focused on safeguarding personal data. Following the GDPR, the EU introduced the Ethics Guidelines for Trustworthy AI\footnote{https://ec.europa.eu/futurium/en/ai-alliance-consultation.1.html} in 2019, emphasizing transparency and the rights of individuals in the context of AI-driven decisions. Building on these foundations, the EU has now developed the AI Act, a comprehensive regulatory framework aimed at categorizing AI systems based on their risk levels, ranging from minimal to unacceptable risk, as pictured in Fig. \ref{fig:trustworthy-ai} panel A). Unacceptable risk AI systems are prohibited, as they pose a significant threat to individuals or society. Examples of such systems include social scoring systems and manipulative AI.
Moving up the risk scale, high-risk AI systems are subject to strict regulations and transparency obligations. These systems, which include those used in critical infrastructure, healthcare, and law enforcement, must be designed and developed with safety and transparency in mind. Providers of high-risk AI systems must ensure that their systems are safe, transparent, and accountable and that they can explain their decision-making processes.

Providers of high-risk AI systems have several obligations under the Act, including ensuring safety and transparency, providing technical documentation and instructions for use, conducting regular testing and validation, reporting incidents, and establishing procedures for human oversight and intervention. Users of high-risk AI systems also have obligations, including ensuring they have the necessary skills and domain expertise, following instructions for use, monitoring performance, and reporting incidents.

Limited-risk AI systems, such as chatbots and deepfakes, are subject to lighter transparency obligations, requiring providers to ensure that end-users are aware they are interacting with an AI system. Minimal-risk AI systems, which include the majority of AI applications currently available on the EU single market, are not regulated under the Act.

The Act also addresses general-purpose AI models, requiring providers to provide technical documentation and instructions for use, comply with copyright law, and publish a summary of the training data used. Providers of general-purpose AI models that present a systemic risk must comply with additional obligations, including conducting model evaluations and adversarial testing, tracking and reporting serious incidents, and ensuring cybersecurity protections.

Overall, the AI Act is designed to ensure AI safety and reliability by mandating robustness, accuracy, (cyber) security, transparency, interpretability, and user-centric design for AI systems, particularly those deemed high-risk. Companies that fail to comply with the Act face substantial fines, proportionate to the severity of their non-compliance and their size, demonstrating the EU's commitment to enforcing these regulations.

\subsection{Terminology in Trustworthy AI}
In the context of Trustworthy AI, a range of terms and concepts are used to describe the principles, methods, and practices that underpin the development and deployment of AI systems that are safe and transparent, and in this section, we provide definitions and explanations of key terminology to facilitate a common understanding of these concepts.
\subsubsection{Ethical Principles for AI} 
\label{subsubsec:ethical-principles-for-ai}
The EU AI Act \citep{EUAIACT} refers to the 2019 Ethics guidelines for Trustworthy AI, developed by the independent High-Level Expert Group on Artificial Intelligence (HLEG)  on AI appointed by the European Commission \citep{hleg}. These guidelines outline seven non-binding ethical principles to ensure AI systems are trustworthy and ethically sound.

The principles, highlighted in Fig. \ref{fig:trustworthy-ai} panel B), are as follows:
    \begin{itemize}
        \item Human agency and oversight:  Ensuring fundamental rights, human agency, and human oversight are preserved in AI systems.
        \item Technical robustness and safety: Guaranteeing resilience to attacks, security, fallback plans, and general safety, accuracy, reliability, and reproducibility. 
        \item Privacy and data governance: Including respect for privacy, quality and integrity of data, and access to data
        \item Transparency: Ensuring traceability, explainability, and effective communication of AI decision-making processes.
        \item Diversity, non-discrimination and fairness: Avoiding unfair bias, promoting accessibility and universal design, and encouraging stakeholder participation.
        \item Societal and environmental well-being: Fostering sustainability, environmental friendliness, social impact, and democratic values.
        \item Accountability: Ensuring auditability, minimizing and reporting negative impacts, and providing redress mechanisms.
    \end{itemize}
    
Different groups of stakeholders have distinct roles in meeting the requirements. Developers should integrate and apply these requirements throughout the design and development processes. Deployers should confirm that the systems they use and the products and services they offer comply with these standards. End-users and the broader society should be made aware of these requirements and have the ability to request their enforcement.

\subsubsection{Transparency}

Transparency in AI can be defined and categorized in various ways, reflecting the complexity and multifaceted nature of the concept. Below, we provide an overview of different definitions and levels of transparency as outlined in the literature.

One research group categorizes transparency into three distinct levels \citep{9613467}:
\begin{enumerate}
    \item \textbf{Simulatability}: A model is considered transparent at this level if a person, given the input data and model parameters, can produce a prediction by performing every calculation of the model in a reasonable amount of time.
    \item \textbf{Decomposability}: This level requires that each part of the model (input data, parameters, calculations) is interpretable, including the input features.
    \item \textbf{Algorithmic Transparency}: This level pertains to understanding how model parameters or predictions are generated. It involves knowledge of the algorithm's learning process and the types of relationships it can learn from the data. 
\end{enumerate}

Another research group divides transparency into multiple levels, emphasizing different aspects \citep{walmsley2021artificial}:
\begin{enumerate}
    \item \textbf{Transparency about values}: Clarity regarding the assumptions and cognitive attitudes of scientists towards theories and the role of values in their descriptions of scientific information needs to be provided.
    \item \textbf{Explicability}: This level requires open communication about AI systems' purpose and deployment reasons. Users and affected individuals must understand the values and motivations of AI system developers and deployers.
    \item \textbf{Outward descriptive transparency}: Clear statements about the capabilities and limitations of the AI system should be communicated.
    \item \textbf{User-facing outward transparency}: Informing users that they are interacting with an AI system ensures the system is identifiable.
\end{enumerate}

These definitions closely align with the concept of transparency as outlined in the EU AI Act. According to the EU AI Act, transparency means that AI systems are developed and used in a way that ensures appropriate traceability and explainability. This includes making humans aware that they are interacting with an AI system. Deployers should be made aware of the system's capabilities and limitations, while affected individuals should be informed about their rights.

To achieve traceability and transparency, the EU AI Act mandates that providers of high-risk AI systems document their assessment before placing the system on the market or putting it into service. This documentation must be made available to national competent authorities upon request.

\subsubsection{Black-Box Model} 
\label{subsubsec:black-box}
A black-box model is a system in which the internal workings are not transparent or understandable solely by examining its parameters. While the inputs and outputs of the black-box model are observable, the processes within the model remain hidden and opaque. Examples of black-box models include deep learning algorithms like \ac{nn} and machine learning algorithms such as random forests \citep{guidotti2018survey,molnar2020interpretable,blackbox}.

\subsubsection{Interpretable Model} An interpretable model, also known as a white box or transparent model, is a system whose internal workings are clear and understandable. The architecture of the model allows for a mathematical examination of the relationship between input and output. This enables users to examine how input variables influence output predictions. Examples of a transparent model are decision trees and linear regression \citep{frasca2024explainable,guidotti2018survey,minh2022explainable}.

\subsubsection{Explainability and Interpretability}
The terms "explainability" and "interpretability" lack standardized definitions within the AI community, resulting in diverse interpretations and minimal consensus regarding their precise meanings \citep{doshivelez2017rigorous}. The definitions of these terms, along with others like transparency, vary across different scientific disciplines and among stakeholders, including policymakers, applied AI practitioners, and legal scholars \citep{xairoleineuaiact}. Some sources use interpretability and explainability interchangeably, suggesting they convey the same concept \citep{adadi2018peeking,koh2017understanding,bojarski2017explaining}. Others provide a definition of interpretability but claim that a clear definition of explainability remains elusive \citep{doshivelez2017rigorous}. 

In this work, we define interpretability as the ability to present information in clear and comprehensible terms to a human and to explain the steps that led to a particular output decision, typically without requiring additional methods \citep{li2022interpretable}. It is about the extent to which cause-and-effect relationships within a system can be discerned and understood \citep{frasca2024explainable,linardatos2020explainable,miller2019explanation,minh2022explainable}. Interpretability can also be described as the extent to which a human is able to consistently predict the outcomes produced by the model \citep{kim2016examples}.

Conversely, explainability refers to the capability to provide explicit and comprehensible justifications or reasoning for specific model decisions. It involves identifying the parameters that influenced the model's decisions and elucidating their significance within the system \citep{frasca2024explainable,linardatos2020explainable}.

\begin{figure}
\centering
       \includegraphics[width = 1\linewidth]{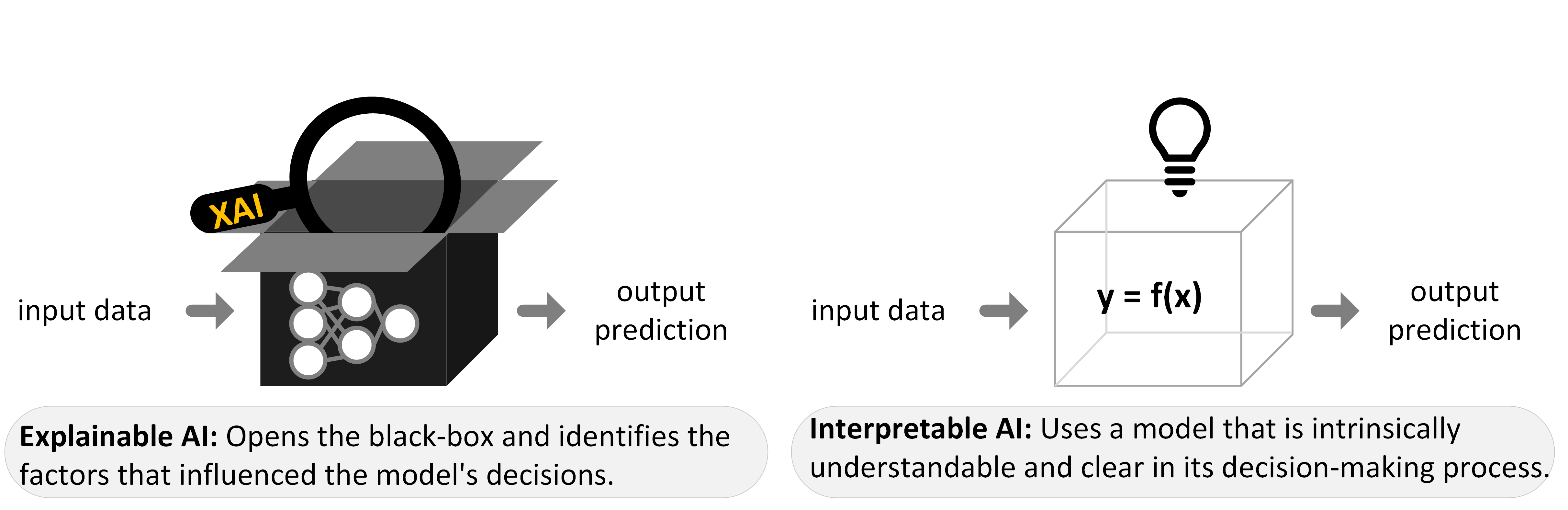}
      \vspace{-3mm}
    \caption{Comparison of AI models. Explainable AI involves opening the black-box to identify the factors that influenced the model's decisions, whereas interpretable AI uses a model that is inherently understandable and transparent in its decision-making process.}
    \vspace{-3mm}
\label{fig:xai_vs_iai}
 \end{figure}

Therefore, as highlighted in Fig. \ref{fig:xai_vs_iai}, interpretable AI is about working with interpretable models and explainable AI is the field of explaining black-box models.

According to the EU AI Act, explainability and interpretability are essential aspects of the transparency requirement for high-risk AI systems. These systems must be designed to help deployers understand their operation, evaluate their functionality, and identify their strengths and limitations. This requirement addresses concerns about the complexity and opacity of AI systems, ensuring they are transparent before being launched or used. In this context, opacity refers to the difficulty in understanding how AI systems function or make decisions.

XAI enhances transparency by clarifying AI decision-making processes, making models easier to use and more widely accepted. It also enables users to actively contribute to debugging and improving these systems \citep{minh2022explainable}. For example, studies have shown that incorporating XAI into virtual agents can significantly increase trust in AI systems, as users display greater confidence in the explanations provided and reduced skepticism towards the system \citep{10.1145/3308532.3329441}.

\subsubsection{Robustness}
Robustness refers to the ability of a model to consistently deliver reliable and accurate performance across a range of conditions, including unexpected or adversarial data, and to accurately quantify its confidence and uncertainties in its predictions \citep{pruksachatkun2023practicing}. This involves handling the complexities and variability of real-world data and performing well across diverse datasets and scenarios. A robust model is one that can maintain its performance even when faced with small changes or unusual data and is less likely to make unreliable decisions.
In the context of the EU AI Act, technical robustness and safety require that AI systems are designed and deployed to ensure they can withstand potential issues and resist attempts by third parties to manipulate their functionality or performance for illegal purposes, thereby minimizing the risk of unintended harm.

\subsubsection{Definition of Trustworthiness}
complying with all applicable laws and regulations;
2. it should be ethical, ensuring adherence to ethical principles and value

The European Commission's HLEG Expert Group on AI has emphasized the importance of trustworthy AI systems in their 2019 Ethics Guidelines for Trustworthy Artificial Intelligence. According to the guidelines \citep{hleg}, trustworthy AI systems must meet three key criteria: they must be lawful, complying with all applicable laws and regulations; ethical, adhering to ethical principles and values; and robust, ensuring both technical and social reliability to prevent unintentional harm. Furthermore, the guidelines outline the seven essential requirements for achieving trustworthy AI, as mentioned in Section \ref{subsubsec:ethical-principles-for-ai}. This underscores the idea that trustworthiness is a multifaceted concept that depends on meeting a range of practical and normative standards \citep{schmidt2020transparency}.

Various researchers and experts have shared their insights on trustworthiness in AI systems. For instance, some argue that monotonicity and usability are crucial factors influencing trust in AI models. A model that respects monotonicity constraints and provides usable explanations can increase user trust \citep{guidotti2018survey}.

Interestingly, some researchers suggest that trust is closely tied to uncertainty and vulnerability. They argue that increasing transparency, often seen as a key factor in building trust, can actually decrease the need for trust by reducing uncertainty \citep{reinhardt2023trust}. This raises an important question: Even with full transparency, can we truly trust an AI system? One possible answer is that while transparency can provide understanding, it does not necessarily eliminate doubt or uncertainty, and therefore, trust is still required.

The relationship between interpretability and trust is also a topic of discussion. Some researchers argue that the lack of interpretability raises concerns about the trustworthiness of deep models. They propose that trustworthy interpretation algorithms are those that reliably reveal the underlying rationale of a model's decisions, rather than providing irrelevant or desired results \citep{li2022interpretable}. However, the optimal metric for evaluating trustworthiness in interpretation algorithms remains an open question.

Overall, these perspectives highlight the complexity of trustworthiness in AI systems, and the need for ongoing research and discussion to develop a deeper understanding of this critical concept.

\subsection{Challenges to Implementing Trustworthy AI}
\label{subsec:obstacles}

Trustworthiness, encompassing transparency, explainability, and interpretability, is a crucial aspect of AI systems, driven by ethical, legal, and practical considerations \citep{von2021transparency}. However, several obstacles hinder their implementation, including \citep{frasca2024explainable}:
\begin{itemize}
    \item \textbf{Opacity of complex models:} Deep learning techniques, such as deep neural networks, create highly complex models with millions of parameters. Due to this intricacy, even the model’s developers may struggle to understand how the model arrives at its conclusions.
    \item \textbf{Trade-off between performance and interpretability:} Achieving higher performance in terms of accuracy and generalization often requires the use of more complex models, which are typically harder to interpret. Finding the right balance between performance and interpretability remains a significant challenge \citep{arrieta2020explainable}.
    \item \textbf{Bias in training data}: Models trained on biased data may propagate these biases, resulting in discriminatory or inaccurate decisions. Understanding how the model utilizes data and identifying any unintentional biases is crucial.
    \item \textbf{Interpretability of features:} Identifying which features are relevant to a model’s decisions is a key component of interpretability. This task becomes particularly challenging with complex models like deep neural networks, where it can be difficult to determine which factors influence the model.
    \item \textbf{Scalability:} Large-scale deep learning models pose additional challenges for achieving explainability and interpretability with their numerous parameters and high computational resource requirements. 
    \item \textbf{Changes in model behaviour:} The behavior of machine learning algorithms can change over time due to variations in input data or training processes. Maintaining interpretability in the face of such changes can be difficult.
    \item \textbf{Social acceptance:} Even if a model is theoretically explainable or even interpretable, convincing users to trust and understand the explanations provided can be challenging, especially if those explanations do not align with users’ intuitions.
    \item \textbf{Quantifying model trustworthiness, interpretability and explainability:} Until recently, there are not many metrics to quantify the trustworthiness of interpretation algorithms of a model \citep{li2022interpretable}. 
    Some researchers use the complexity of the model in terms of the model size as a component to measure its interpretability \citep{freitas2014comprehensible}. \citet{molnar2020interpretable} claims that a model is better interpretable than another model if its decisions are easier for a human to comprehend than decisions from the other model. Another research group proposes a three-level interpretability evaluation approach \citep{doshivelez2017rigorous}:
    \begin{itemize}
        \item Application-grounded evaluation: This approach involves real humans performing real tasks. The model is evaluated by conducting human experiments, and users are asked if the explanations provided by the model are helpful.
        \item Human-grounded metrics: This method uses real humans but with simplified tasks. It is ideal for situations where working directly with the target community is challenging. Non-expert participants can be used to assess the model's interpretability.
        \item Functionally-grounded evaluation: This level does not involve human participants. Instead, it uses proxy tasks with a formal definition of interpretability, such as demonstrating accuracy improvements with interpretable models or validating regularizers based on prior human experiments.
    \end{itemize}
\end{itemize}   

While trustworthiness is a crucial aspect of AI systems, significant obstacles remain in implementing transparency, explainability, and interpretability. Addressing these challenges will require ongoing research and innovation to ensure AI systems are developed and deployed responsibly and reliably.

The EU AI Act mandates the establishment of an advisory forum to ensure stakeholder involvement in the implementation and application of the regulation. This forum will provide technical expertise to the Board and the Commission. Additionally, the EU AI Act has tasked the European Committee for Standardization (CEN) and the European Committee for Electrotechnical Standardization (CENELEC) with translating its principles into technical requirements. 

CEN and CENELEC have accepted a standardization request on AI from the European Commission, and CEN-CLC/JTC 21 is currently developing European standards\footnote{https://www.cencenelec.eu/areas-of-work/cen-cenelec-topics/artificial-intelligence/}. These standards will aim to provide manufacturers with a presumption of conformity with the forthcoming AI regulations. Through our work, we seek to contribute to this effort by extending the set of available benchmarks in line with the initial technical recommendations of the EU AI regulations.

\section{Methodology}
\label{sec:methodology}

\begin{figure}
\centering
       \includegraphics[width = 1\linewidth]{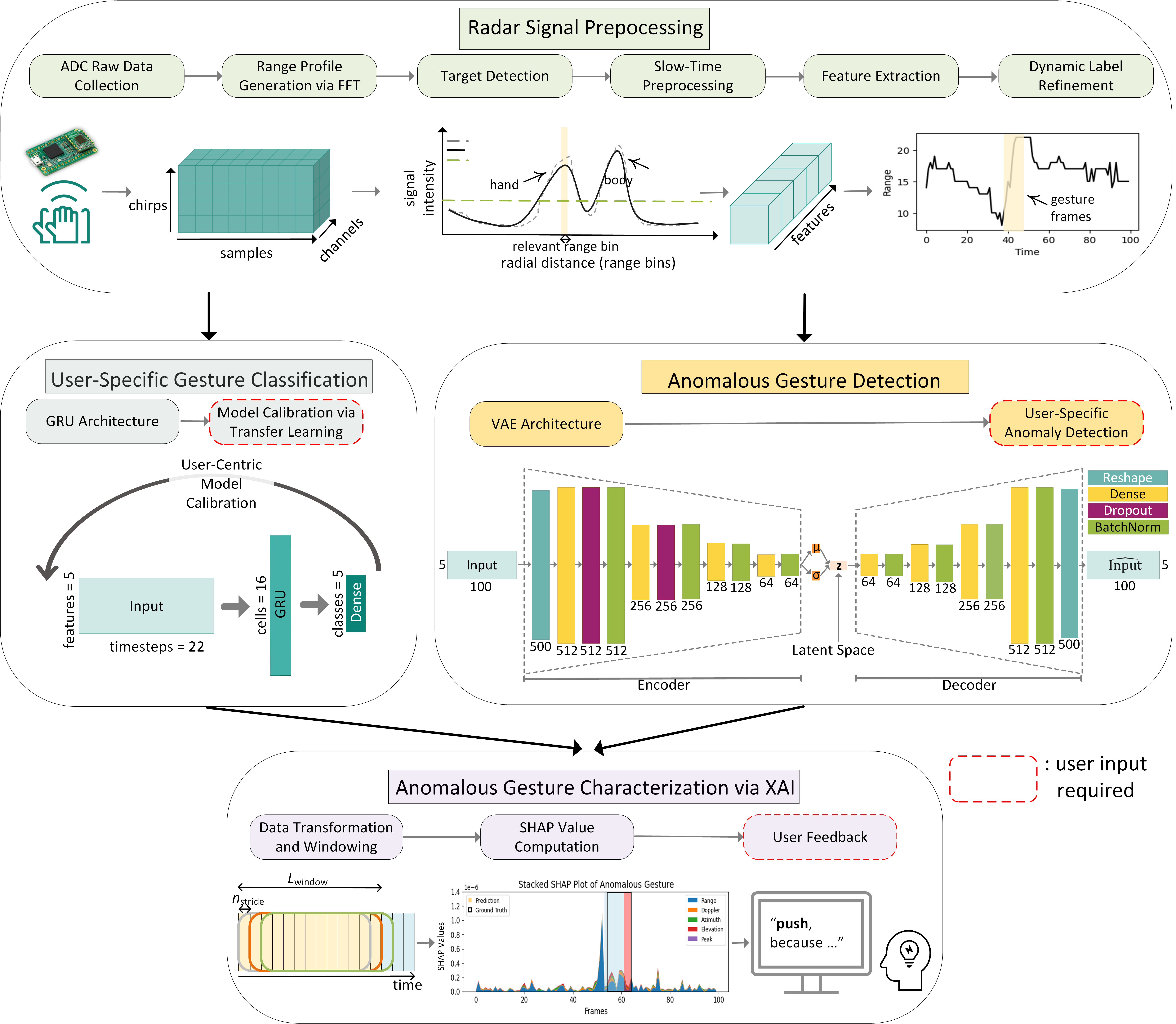}
      \vspace{-3mm}
    \caption{Overview of the XentricAI methodology. This comprehensive framework for HGR and analysis is composed of four interconnected building blocks}: the GRU-based model for gesture detection and classification (Section \ref{subsubsec:gru}), the transfer learning techniques for personalized \ac{hgr} (Section \ref{subsubsec:transfer-learning}), the VAE-powered anomaly detection module (Section \ref{subsubsec:vae}), and the XAI-based anomalous gesture characterization module (Section \ref{subsubsec:xai}). Each component is designed to enhance accuracy, personalization, and explainability of the HGR system. The radar signal preprocessing is detailed in Section \ref{subsec:radar-preprocessing}.
    \vspace{-3mm}
\label{fig:methodology}
 \end{figure}

In this work, we build upon our previously published e\underline{\textbf{x}}plainable and user-c\underline{\textbf{entric AI}} method, XentricAI, and extend it to create a comprehensive framework for \ac{hgr} and analysis compliant with the EU AI Act. XentricAI is an approach to enable personalized gesture calibration to combat distributional shifts between training data and real-world data discrepancies while allowing insights into the black-box model. Its methodology is highlighted in Fig. \ref{fig:methodology} and will be discussed throughout this section. The enhanced framework is designed to provide a more accurate and personalized \ac{hgr} system, while also reasoning about the underlying causes of mis- or unclassified gestures. To achieve this, we introduce a modular approach, consisting of four interconnected building blocks, each addressing a specific aspect of \ac{hgr} and interpretation.
At the core of the framework is a \ac{gru}-based model, responsible for detecting and classifying gestures. This model is then calibrated to individual users through the application of \ac{tl} techniques, ensuring a personalized \ac{hgr} experience. 
An optimized anomaly detection module powered by a \ac{vae} is employed to identify potential anomalies and misclassifications. Finally, an anomalous gesture characterization module, leveraging XAI, provides insights into the underlying causes of anomalies, enabling users to adapt their gestures and improve \ac{hgr} accuracy.
Through the integration of these building blocks, XentricAI provides a robust and explainable \ac{hgr} system, capable of adapting to individual users and providing actionable feedback. In the following sections, we will describe each building block. Additionally, the related work pertinent to each building block will be discussed within the corresponding subsection, providing a comprehensive understanding of the context and advancements underlying each component.

\subsection{Radar-Based Hand Gesture Recognition using a GRU-Based Neural Network}
\label{gru-model}
\subsubsection{Background and Related Work}

\ac{hgr} enables the interpretation of human body movements and gestures through computational algorithms. Their numerous applications have gained wide recognition and significant research interest, as highlighted in various review papers \citep{review1, review2, review3, review4}. Notable applications of \ac{hgr} include smart homes \citep{li2022trajectory}, smart vehicles \citep{geng2020using}, gaming and virtual reality \citep{taranta2015exploring, singh2018simulation}, and sign language communication \citep{lee2017smart}.

\ac{hgr} systems can be broadly categorized into two types: contact and non-contact devices. Wearable technologies, such as data gloves \citep{dg1,dg2,dg3} or smart bands \citep{smartband}, can capture and transmit gestures from an operator to a connected machine. However, they are inflexible to wear and can pose health risks, such as transmitting bacteria and viruses \citep{multimodal1}.
On the other hand, non-contact devices eliminate these drawbacks by using various sensing modalities to detect and interpret gestures without physical touch. These modalities include vision-based systems, which rely on cameras and image processing algorithms \citep{cam1,cam2}, multi-modal systems that combine various sensor inputs \citep{multimodal1, multimodal2}, WiFi \citep{wifi1,wifi2,wifi3} and radar-based systems \citep{li20224,yan2023mmgesture}. 
Vision-based \ac{hgr} systems, however, have several inherent drawbacks. They are highly sensitive to lighting and environmental conditions, e.g., weather, dust, and smoke \citep{review2}. Additionally, they typically have higher power consumption and raise privacy concerns \citep{li20224}. In contrast, WiFi and radar-based systems do not suffer from these constraints and offer better privacy preservation \citep{liu2022behavior}. 

However, WiFi-based \ac{hgr} systems have their own disadvantages, including shorter range resolutions and higher wavelengths, which makes it difficult to capture fine-grained motion information \citep{yan2023mmgesture}. On the other hand, the \ac{mmwave} radar can capture more subtle gesture information due to its higher frequency band and high range and angle resolution \citep{wang2023dcs}. As a result, \ac{mmwave} radar-based solutions have received more attention for their superior performance \citep{wang2024sensing}.
Moreover, the availability of commercialized, small, low-power radar-on-chip mmWave radars, such as Google Soli \citep{lien2016soli} and Infineon Technologies' XENSIV™ BGT60LTR13C $60\,$GHz \ac{fmcw} radar has facilitated the widespread adoption of this technology and was therefore used in this work \citep{jin2023interference,yan2023mmgesture}. 

Over the years, many techniques have been developed to improve the accuracy and robustness of \ac{hgr} systems. These techniques can be broadly categorized into classical \ac{ml} methods and deep learning approaches, each with distinct advantages and limitations.

Classical \ac{ml} methods leverage handcrafted features and simpler algorithms. \ac{svm}, for example, have been employed for feature-based \ac{hgr} with two-antenna Doppler radar systems, demonstrating their efficacy in high-dimensional feature spaces \citep{svm}. Similarly, \ac{knn} algorithms have been used for \ac{hgr}, particularly when combined with \ac{pca} for feature reduction \citep{knn}. Decision trees, such as the C4.5 algorithm, have been employed in \ac{hgr}, offering ease of interpretation and implementation despite their susceptibility to overfitting \citep{decisiontree}. \ac{hmm} have shown promise in modeling temporal sequences using micro-Doppler radar signatures \citep{hmm}. 
Rule-based systems like MIRA employ rule-based approaches for interpretable \ac{hgr}, though their complexity and adaptability are limited compared to deep learning methods \citep{seifi2024mira}.

Despite their advantages, classical \ac{ml} methods require extensive feature engineering and often struggle to handle the complexity and variability of real-world gestures.

Deep learning techniques, particularly \ac{cnn} and their variants, have emerged as a dominant approach due to their ability to automatically learn spatial features. Lightweight, multichannel \ac{cnn}s have been employed for robust \ac{hgr} on \ac{mmwave} \citep{cnn-multichannel}. They have also been utilized in automotive applications to distinguish between traffic gestures based on reconstructed Doppler spectrograms \citep{cnnautomotive}.

A \ac{rnn} based on \ac{lstm} units, for instance, using two mmWave \ac{fmcw} radar systems has been shown to achieve high accuracy \citep{lstm}.

Hybrid architectures that combine \ac{cnn}s for spatial extraction with \ac{lstm}s for temporal pattern learning have also been highly effective \citep{multimodal1,cnn_lstm,cnn_lstm_2,yan2023mmgesture,sluyters2023radarsense}. 
Transformer networks have recently been explored in \ac{hgr} in combination with \ac{cnn}s. For instance, a 2-D \ac{cnn}-transformer network has been developed for interference-robust \ac{mmwave} radar-based dynamic \ac{hgr} \citep{2d_transformer_cnn}. This system uses \ac{cnn}s to extract local features of gestures and transformer modules to capture deeper, more effective features from range-time and Doppler-time maps. Another approach, DCS-CTN, uses a time-distributed \ac{cnn}, position encoding, and a transformer to capture both local and global features for \ac{hgr} \citep{wang2023dcs}.

\ac{resnet} offer another deep learning architecture for \ac{hgr}, alleviating the vanishing gradient problem and enabling deep network training. In one study, a ResNet50-based system was employed for \ac{hgr} using interferometric MIMO radar \citep{resnet50_cnn}.

Another approach involves real-time spiking \ac{cnn} for \ac{hgr}, which bypasses complex \ac{fft} processing to streamline gesture recognition \citep{shaaban2024rt}. Using untrained resonate-and-fire neurons for efficient \ac{hgr}, RT-SCNNS have achieved competitive accuracy for various gestures.

While AI techniques have shown several advancements, they also present challenges. First, many models rely on input data in the form of images, particularly spectrum map videos such as range-Doppler maps, range-angle maps, and Doppler-angle maps. These representations are difficult to interpret and explain, making them less suitable for systems where user explainability is a priority. Second, achieving efficient deployment on edge devices requires smaller architectures with fewer parameters. Lastly, many existing methods rely on multiple radar sensors, whereas in this work, a robust solution using only one radar sensor is preferred for edge computing applications.

To address these challenges, \ac{rnn}s based on \ac{gru} cells have been applied in \ac{hgr}, offering efficient temporal modeling with fewer parameters compared to \ac{lstm}s. The \ac{gru}-approach described in \citet{strobel} enables robust \ac{hgr} with a lightweight architecture that utilizes features with physically intuitive meanings. This is especially ideal for applications requiring both user explainability and edge deployment with a single radar sensor.

Given the focus on user explainability in this work, we have adopted the \ac{gru}-based approach described in \citet{strobel}.

\subsubsection{GRU-Based Model Design and Training}
\label{subsubsec:gru}

Sensor data, being time-series in nature, requires a model proficient in time-series analysis. Therefore, we implemented a compact \ac{rnn} \citep{rnn} with \ac{gru} \citep{gru} cells, which are adept at retaining information across time points, crucial for time-series prediction. 
Our architecture is based on the design in \citet{strobel}, tailored for \ac{hgr} using \ac{fmcw} radar at the edge. It achieves high performance with a minimal number of parameters, making it suitable for the needed hardware embedding. The model consists of a single \ac{gru} layer with 16 units to learn patterns in the time-series data, followed by a densely connected layer with five neurons, corresponding to the output classes.
The softmax function \citep{softmax} activates the dense layer, which allows probabilistic predictions. Unlike rigid classification, this function assigns a probability to each potential class, thereby indicating the likelihood of any given sample belonging to each class. The sum of these probabilities across all nodes or classes is constrained to one. Ultimately, the sample is classified to the class with the highest Softmax probability.

Using this architecture, gesture classification of five classes is performed: \textit{SwipeUp}, \textit{SwipeDown}, \textit{SwipeLeft}, \textit{SwipeRight}, \textit{Push}, and a \textit{Background} class.
The input to the \ac{gru} architecture consists of data sequences characterized by five features. They serve to capture critical patterns discernible within radar data. These intuitively designed features are the radial distance (range), the radial velocity (Doppler), the vertical (elevation) and horizontal angle (azimuth), and the signal amplitude (peak).
Specifics regarding the preprocessing protocols employed are detailed in Section \ref{subsec:radar-preprocessing}.
In the following, we will refer to this specific \ac{gru} model as \textbf{baseline model}.

\subsection{User-Centric Model Calibration}
\label{subsec:model-calibration}
\subsubsection{Background and Related Work}

Calibrating machine learning models to individual users is favorable to enhancing user experience and aligning to regulatory standards with respect to user-centric designs such as those outlined in the \ac{euai}. Such calibration tailors models to users' unique behaviors and characteristics, improving the predictive accuracy and ensuring personalized interactions, thereby increasing user trust and engagement. It also addresses ethical concerns by reducing biases and respecting user autonomy, ensuring that AI systems do not discriminate and are transparent in their operations.

\ac{tl} is a \ac{ml} technique where a model trained on one task is retrained or fine-tuned for another related task, either by freezing certain layers to retain previously learned features or by allowing all layers to be retrained for better adaptation to the new task \citep{zhuang2020comprehensivesurveytransferlearning}. The idea is to leverage the knowledge and features learned from the initial task to improve performance on the new, unseen task.
To create a user-adapted \ac{ai} system, the initial pre-trained model undergoes a calibration process using \ac{tl} to tailor it to individual users. 

However, a significant obstacle encountered during \ac{tl} is catastrophic forgetting, a phenomenon where the model, after retraining, loses its ability to perform well on the original dataset it was trained on \citep{chen2019catastrophic}. This performance drop or forgetting rate is attributed to the overwriting of previously learned information with new data.

While there has been extensive research on \ac{tl} for \ac{hgr} in general \citep{yang2018gesture,demir2019surface,bird2021synthetic}, this section focuses specifically on user-centric \ac{tl}.

In earlier work from 2017, researchers demonstrated that certain early layers of a \ac{cnn} could be applied to unseen users when further training is performed on subsequent layers using new user data \citep{cote2019deep}. This method allowed tasks to be completed faster. However, while this approach uses deep \ac{cnn} architecture, which allows for performing \ac{tl} with freezing, it is not suitable for systems intended for deployment with computational constraints due to its large architecture.

A study on personalized view-invariant \ac{hgr} highlights the challenge of designing \ac{hgr} systems that work across different users \citep{costante2014exploiting}. They used low-cost RGB cameras and focused on creating a robust system for autonomous systems with limited power and load capacity. Essentially, users were asked to perform a set of gestures, feature augmentation was performed, and an \ac{svm} classifier was trained using both training and user data. Despite some improvements in \ac{hgr} accuracy through \ac{tl}, overall performance was not high. Additionally, it was unclear how much gesture data or how many frames were required to make a prediction, which is critical for model deployment for real-time inference.

Another study leveraged pre-trained models from one user and fine-tuned them for another user to reduce training overhead for 3D finger motion tracking using a \ac{mmwave} radar \citep{liu2022leveraging}. They explored different fine-tuning strategies, including updating only batch normalization layers or the entire network. While this approach reduced training data requirements, it involved a model with two million parameters, which is impractical for edge deployment.

The mmSign system used few-shot transfer learning for online handwritten signature verification with \ac{mmwave} radar \citep{han2024mmsign}. This method adapted a model trained on multiple users to a new user using a small amount of data and model-agnostic meta learning applicable to deep NNs. 
One research group developed an \ac{hgr} algorithm for electromyographic data covering five gestures using an ensemble of classical ML models \citep{dolopikos2021electromyography}. Model calibration was performed by retraining the model with user-specific data. However, similar to most previously discussed studies, they overlooked the model's performance on the original training dataset, thus neglecting the issue of catastrophic forgetting. In contrast to prior work assuming similar gesture execution patterns \citep{zhang2021unsupervised,liu2022mtranssee}, our approach does not assume uniform user gesture execution patterns. Instead, it accommodates variations in execution styles, such as differences in speed and proximity to the radar. 

In summary, while various methods for user-centric \ac{tl} in \ac{hgr} exist, they often face issues such as catastrophic forgetting, impracticality for edge deployment, or unclear data requirements. Our approach combines user-specific retraining with \ac{er}, merging a subset of the initial training data with user data during retraining. This strategy reduces forgetting rates and maintains the model's generalization, accommodating varied user execution styles for robust and personalized \ac{hgr}. Our contributions merge \ac{xai} and \ac{tl} for a human-centric approach, pioneering \ac{er} for model calibration in radar-based \ac{hgr} and addressing significant distributional shifts.

\subsubsection{Methodology of User-Centric Model Calibration}
\label{subsubsec:transfer-learning}

To create a user-adapted \ac{ai} system, the initial pre-trained model undergoes a calibration process using \ac{tl} to tailor it to individual users. 
In this work, the user is asked to perform a set of gestures which is then used to retrain the baseline model to personalize the \ac{ai} system's predictions.

To mitigate catastrophic forgetting, we employed a data mixing strategy known as \ac{er} \citep{rolnick2019experience}. Specifically, we retain a fraction of the original, diverse training dataset, and merge it with the user-specific dataset. This merged dataset is then used for retraining, ensuring the model's ability to generalize beyond the user-specific data. This strategy is particularly effective for performing \ac{tl} on tiny NNs with very few layers, where freezing the weights of layers is less effective.

We conducted a comparative analysis of ER with two other catastrophic forgetting mitigation techniques which are Elastic weight consolidation (EWC) \citep{ewc}, and synaptic intelligence (SI) \citep{synapticintelligence}. The evaluation was designed to measure their performance and ability to adapt to user-specific data while mitigating forgetting. Our findings indicate that ER is the most effective method for our use case, which involves learning the new task of adapting to user data while maintaining good performance on the original task of generalizing across the training dataset. Detailed experimental results and discussions of these methods are provided in \ref{app:catastrophic-forgetting}.

For ER, the proportion of the original dataset and the user dataset to be mixed is governed by a carefully chosen regularization coefficient, $n_\text{train}$-$n_{\text{user,}i}$, where $n_\text{train}$ represents the number of training gesture samples per class and $n_{\text{user,}i}$ denotes the number of user $i$ gesture samples per class. This coefficient balances the retention of original knowledge with adaptation to new user data. To determine its optimal value, an extensive study was conducted, with results presented in Section \ref{subsec:model-calibration-results}.

\begin{figure}
\centering
       \includegraphics[width = 1\linewidth]{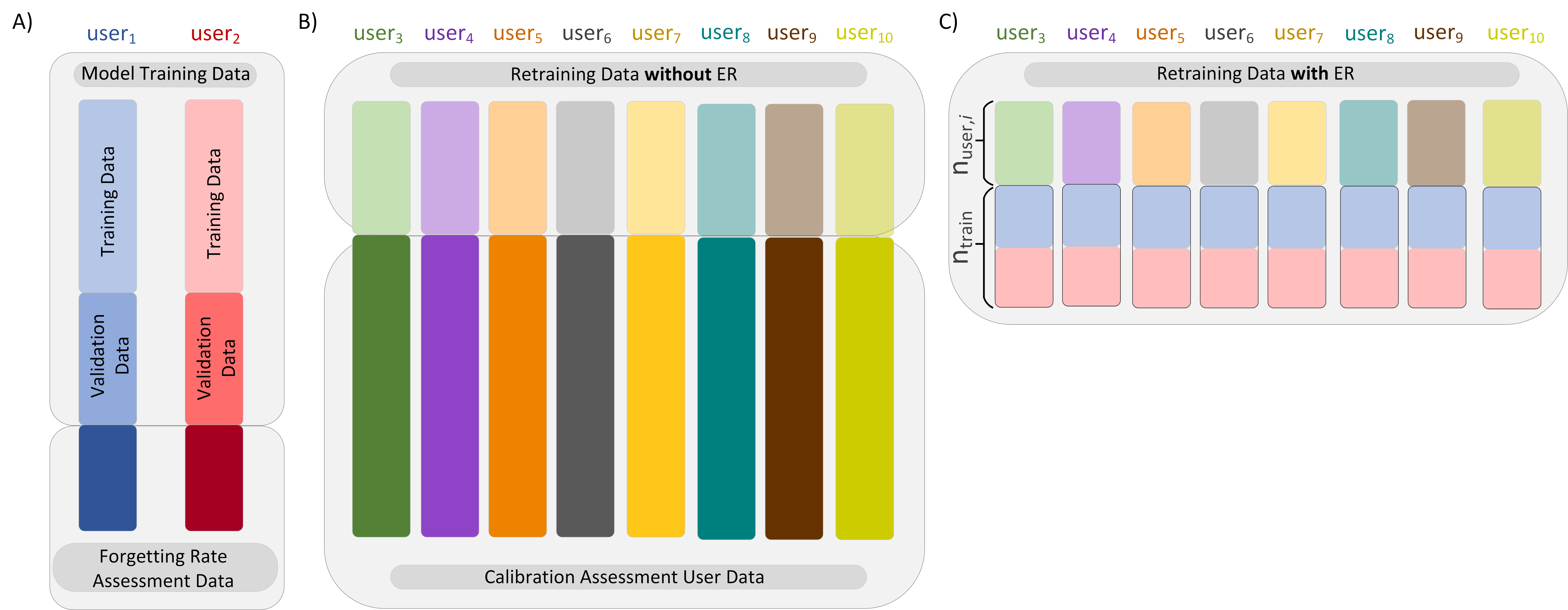}
      \vspace{-3mm}

    \caption{Exemplary illustration of dataset allocation for model training, calibration, and assessment. (A) Training and validation datasets are derived from most data from $user_1$ and $user_2$, with a small portion set aside for forgetting rate assessment. (B) Data from the remaining users is divided between model calibration without ER and calibration assessment. (C) For model calibration with ER, $n_\text{train}$-$n_{\text{user,}i}$ amount of user and training data is added to the retraining dataset, creating the ER-augmented retraining dataset.}
    \vspace{-3mm}
\label{fig:dataset-allocation}
 \end{figure}

The user dataset is then divided into two independent, non-overlapping parts. One part is used for model calibration without \ac{er} and is referred to as the retraining data without \ac{er} (Fig. \ref{fig:dataset-allocation} B). To perform model calibration with ER, we randomly select $n_{train}$ of the initial training dataset, which is based on the data of $n_\text{{users}}$ ($n_\text{{users}} = 2$ in Fig. \ref{fig:dataset-allocation} A). This is then added to the retraining data without ER, creating a novel dataset called retraining data with \ac{er} (Fig. \ref{fig:dataset-allocation} C). This combined dataset enables the model to adapt to the user's unique gesture patterns while preventing catastrophic forgetting.

To evaluate the calibrated model, we create two assessment datasets. The first dataset is called forgetting rate assessment data (Fig. \ref{fig:dataset-allocation} A), reserved exclusively for evaluating the forgetting rate. It consists of a withheld data segment of users whose data is used for model training. The second, called calibration assessment user data (Fig. \ref{fig:dataset-allocation} C), is based on the remaining part of the user dataset, comprising independent recordings from the model retraining dataset.

\subsection{Anomaly Detection via Variational Autoencoder}
\label{subsec:autoencoder}
\subsubsection{Background and Related Work}

Anomaly or \ac{ood} detection in radar-based \ac{hgr} is an emerging area of research with limited studies to date. Current \ac{ood} detection methods can be categorized into four main types, as extensively described in \citet{yang2024generalized}. Classification-based methods primarily rely on classifiers. Density-based methods detect \ac{ood} data by modeling data density. Distance-based methods identify \ac{ood} data using distance metrics. Lastly, reconstruction-based approaches focus on reconstructing data via reconstruction techniques. 
For radar-based \ac{hgr}, distance-based methods, such as the \ac{rmd} are used. They assess the difference between test and \ac{id} samples using distance metrics like the Mahalanobis distance, with \ac{ood} data typically showing a higher distance. For example, one work adopts the \ac{rmd} to enhance detection reliability by calculating confidence scores based on the distance between test samples and Gaussian class models \citep{choi2022fmcw}. Despite increasing reliability, \ac{rmd} faces limitations like overconfidence in high-dimensional spaces, and reliance on labeled data. Additionally, the assumption that each class follows a Gaussian distribution can be too simplistic for real-world data, particularly in applications involving dynamic gestures with significant variability in signal magnitude, range, velocity, and angles over time.

Reconstruction-based methods, such as \ac{ae} \citep{bank2023autoencoders} and \ac{gan} \citep{goodfellow2020generative}, operate by learning to recreate \ac{id} data and identifying \ac{ood} samples through high reconstruction error. 
These methods have shown promise in radar-based \ac{hgr}, as they learn the underlying structure of the input features in an unsupervised fashion making them suitable for cases with limited labeled ID samples. 
They are also adept at modeling complex, nonlinear relationships between radar features, capturing dynamic patterns more effectively than distance-based methods. For instance, GANomaly utilizes a GAN architecture to learn a representation of normal hand gestures for user authentication \citep{li2022new}. 

Similar to conventional \ac{ae}s, a \ac{vae} \citep{vae} learns a compact latent space representation of \ac{id} data and utilizes the reconstruction error to identify \ac{ood} data. 
However, VAEs have a distinct advantage by mapping the input data to a probability distribution rather than converting it into a fixed vector.

A VAE consists of three main components: the encoder, the latent space sampling, and the decoder. The encoder compresses the input data into a set of latent variables, specifically the mean and the logarithm of the variance, which describe a Gaussian distribution in the latent space. The sampling layer uses these parameters to generate latent variables, ensuring differentiability. Finally, the decoder reconstructs the input data from these latent variables.

This approach provides a probabilistic mapping to the lower-dimensional space, as opposed to the deterministic mapping found in traditional AEs. Consequently, VAEs offer a more robust probabilistic framework, which enhances generalization and improves the handling of data variability due to the regularization of their latent space.

In radar-based systems, where labeled data is limited and gestures exhibit high variability, VAEs present a promising solution for improving \ac{ood} detection and system reliability. Consequently, we have incorporated a VAE-based anomaly detection module into XentricAI to leverage these benefits. This integration allows us to significantly enhance anomalous gesture detection, offering greater transparency and robustness in real-world applications. 
While the \ac{vae} architecture is well-established in anomaly detection, its novelty in this work lies in its tailored adaptation for radar-based \ac{hgr}. Specifically, we address the limitations of prior systems \cite{seifi2024xentricai} by introducing user-specific dynamic thresholding based on reconstruction errors. Unlike previous methods such as mmFall \citep{jin2020mmfall}, which used variational recurrent autoencoders (VRAEs) for fall detection with centroid tracking, our approach focuses on gesture characterization and adaptability to individual users. This focus ensures robustness and personalized anomaly detection in FMCW radar-based \ac{hgr}, filling critical gaps in existing systems.

\subsubsection{Methodology of VAE-Enhanced Anomaly Detection}
\label{subsubsec:vae}
After calibrating the model to individual users, we aim to provide insights into the system and deliver user feedback. This is particularly important for gesture instances where the model fails to make a correct prediction. Various types of incorrect predictions can occur.
Firstly, there are cases of unpredicted gestures where the system fails to generate any prediction for a particular gesture instance. Secondly, we encounter mixed-class predictions when a single gesture is associated with predictions corresponding to multiple, different gesture classes. Then there are fully misclassified gestures. Lastly, our analysis includes instances of sparse predictions, where gesture classification is marked by intermittently occurring predictions at the level of individual frames.
Sparse and mixed-class predictions are so-called condition-flagged misclassifications. These are misclassified gestures that are flagged as anomalies based on predefined conditions.
While condition-flagged misclassifications were easily detected in our previous work, wrongly predicted and unpredicted gestures were analyzed but remained unflagged. 

To tackle this, we introduce a \ac{vae} module with user-specific thresholding into the XentricAI pipeline, enhancing anomaly detection capabilities beyond the condition-flagged misclassifications. 
So-called VAE-flagged misclassifications are misclassified gestures that are flagged as anomalies by the VAE.
Exclusive VAE-flagged gestures are flagged by the VAE as anomalies but are not part of the condition-flagged misclassifications. This means that they are fully misclassified or unpredicted gestures.
This approach allows us to increase the robustness of XentricAI and improve the quality of outlier detection.

To detect anomalies, we calculate the reconstruction error $e_{\text{rec}}$, which measures the discrepancy between the input data and its reconstruction:
\begin{equation}
    e_{\text{rec},m} =  \lVert \mathbf{\textbf{X}}_m - \mathbf{\hat{\textbf{X}}}_m \rVert_2 ^2
\end{equation}
where $\textbf{X} \in \mathbb{R}^{M\times T \times D} $ is the input data, with $M$ being the number samples in the dataset, $T$ being the number of timesteps, and $D$ being the number of features, $m$ denotes the $m$-th sample, $\mathbf{\textbf{X}}_m$ is the original input, and $\mathbf{\hat{\textbf{X}}}_m$ is the reconstructed output. Gestures with high reconstruction errors are flagged as anomalies, indicating deviations from the learned normal patterns.

Instead of using a validation dataset to determine a global threshold for anomaly detection, we implemented user-specific thresholding, similar to \citet{ko2022new,fatemifar2021client}. This personalized approach ensures the anomaly detection process is sensitive to individual variations in gesture execution. This process can be broken down into the following steps:
\begin{enumerate}
    \item \textbf{Model Calibration:} Users perform a set of calibration gestures, which are used to calibrate the model (Section \ref{subsec:model-calibration}). 
    \item \textbf{Threshold Determination:} Once the model is calibrated, we calculate the reconstruction errors for the calibration gestures. A user-specific threshold is then set based on a chosen percentile (e.g., 90th percentile) of these errors. 
    \item \textbf{Anomaly Detection:} During operation, the gestures performed by the user are compared against this user-specific threshold. If the reconstruction error of a gesture exceeds the threshold, the gesture is flagged as an anomaly. 
\end{enumerate}

To validate the effectiveness of the tailored VAE, we conducted a comparative analysis against four established anomaly detection methods: One-class \ac{svm} \citep{svm}, isolation forest \citep{liu2008isolation}, a density-based approach using local outlier factors \citep{breunig2000lof}, and generative adversarial networks (GANs) \citep{goodfellow2020generative}. Across all comparisons, the VAE demonstrated superior performance, consistently outperforming these alternative approaches. Detailed results of this analysis are provided in \ref{app:anomal-detection-methods}.

Beyond this comparison, we explored alternative VAE architectures, including VRAEs and standard AEs. These architectures, however, proved less effective at capturing the variability and complexity inherent to radar data. Consequently, they were excluded from further analysis due to their inferior performance.

A key innovation of this work is the personalized thresholding mechanism, introduced for the first time in the context of FMCW radar-based \ac{hgr}. This mechanism leverages the tailored VAE’s adaptability to align anomaly detection with individual user patterns. By accounting for the unique nuances of each user's gestures, it significantly enhances the robustness and reliability of the system. This personalized approach not only improves the system’s accuracy in detecting anomalies but also ensures its practical applicability in real-world scenarios, where variability and user-specific differences are critical considerations.

\subsection{Anomalous Gesture Feedback through Feature Analysis}
\label{subsec:anomalous-gesture-feedback}

\subsubsection{Background and Related Work}

NNs are frequently described as black-box models, as detailed in Section \ref{subsubsec:black-box}. The inner mechanisms of the \ac{nn} cannot be observed, hindering understanding for both \ac{ml} developers and specifically non-expert users. Especially in critical domains needing high safety measurements, like the medical \citep{tjoa2020survey} or financial sector \citep{vcernevivciene2024explainable}, providing understandable explanations to end-users is essential for maintaining safety, building trust, and ensuring compliance with the EU AI Act.

In the context of radar-based \ac{hgr}, studies providing explainability to end-users are particularly limited. Our previous work demonstrated a proof of concept and conducted a user study with XentricAI \citep{seifi2024xentricai}, yet, to our knowledge, no other studies have reported similar efforts. Recent advancements have introduced \ac{xai} techniques for \ac{hgr} in general, primarily focusing on surface electromyography (sEMG)-based systems. These techniques predominantly enhance training processes through feature selection or model pruning rather than providing end-user explainability.
For example, one study proposed using graph NNs for \ac{hgr} from raw EMG data, employing an \ac{xai} algorithm named GNN explainer \citep{ying2019gnnexplainer} to refine graph topology \citep{massa2023explainable}. This approach optimizes training by improving recognition rates and reducing computational costs but does not offer explanations to end-users. In another research, an \ac{xai} fusion-based feature selection framework for sEMG-based \ac{xai} was developed, which used \ac{shap}-based explainability \citep{lundberg2017unifiedapproachinterpretingmodel} to select relevant features, again focusing on training optimization rather than user-centric explanations \citep{gehlot2024surface}. 

Further, a feature ranking-based solution combined deep learning and game theory to achieve high \ac{hgr} accuracy and interpretability for sEMG-based similar gesture recognition \citep{wang2024explainable}. This method processed sEMG signals into color images and employed deep NNs for \ac{hgr}, using Shapley values \citep{shapley1953value} to quantify the contribution of each channel to recognizing similar gestures. However, the focus remained on enhancing model precision.

In contrast, a study on Indian sign language recognition using a 2D CNN employed the feature ranking \ac{xai} method called LIME \citep{lime} to explain predictions to users \citep{ghosh2024fedxai}. While the authors claimed that users could comment on model forecasts and that the \ac{xai} module identified key image features influencing predictions, they did not provide experimental results to substantiate these claims.

Our work diverges from these previous studies by proposing a user-centric AI solution that actively involves regular users. XentricAI utilizes SHAP to explain model misclassifications to users, encouraging more effective inputs and interactive feedback. By enhancing decision transparency through physical features, users are empowered to adapt their gestures, thereby improving model detection accuracy. This approach ensures that the end-users, including but not limited to domain or ML experts, benefit from the explainability features of the AI system.

\subsubsection{Methodology of Anomalous Gesture Feedback through Feature Analysis}
\label{subsubsec:xai}
The GRU-based model employed for \ac{hgr} is a black-box model. In order to gain insight into the relationships between the model's predictions and the input data sequences, we seek explanations for each prediction by attributing importance to the features. This is particularly crucial in \ac{hgr}, where anomalies can occur due to various factors, such as incorrect speed or distance from the radar.

Previous work has demonstrated the potential of \ac{xai} in reasoning about anomalies \citep{seifi_gas}. By analyzing the feature attributions of anomalous gestures, we can identify the underlying causes of the anomalies and provide feedback to the user. For instance, if a user performs a gesture at an incorrect distance from the radar, resulting in an anomalous model prediction, the feature importances can help the model reason about the cause of the anomaly and provide a clear explanation to the user, such as "the model could not detect your gesture properly because you were standing too far away from the radar".

Feature importances, or so-called attributions or explanations are defined as follows \citep{hedstrom2024sanity}:

Consider a machine learning model $f(x)$ that is a differentiable black-box function. This model maps input data $\textbf{X}_m$ from an input space $\mathbb{X}$ to an output space $\mathbb{Y}$. Formally, we can write:
$ f: \mathbb{X} \rightarrow \mathbb{Y}, \quad f \in \mathbb{F}$,
where $\mathbb{F}$ is the function space containing all possible functions $f(x)$. We assume that the input $\textbf{X}_m$ is a $D$-dimensional vector, where $D$ represents the number of features, and the output $y_m$ is a $C$-dimensional vector, representing $C$ different classes:
$ \textbf{X}_m \in \mathbb{R}^D, \quad y_m \in \mathbb{R}^C $.
We further assume that the model $f(x)$ is composed of $L$ layers of functions, denoted as $f_1(x), f_2(x), \ldots, f_L(x)$. The overall function $f(x)$ can thus be expressed as a composition of these layers:
$f(x) = f_L(x) \circ f_{L-1}(x) \circ \cdots \circ f_1(x)$.
To interpret the behavior of the model $f(x)$, local explanation methods are used. These methods assign importance scores to the features of the input $x$. The purpose is to understand the influence of each feature on the model's prediction $y$.
Given the model's prediction $y$, an explanation $\textbf{\textit{e}}$ is defined as a $D$-dimensional vector of importance scores: $\textbf{\textit{e}} \in \mathbb{R}^D$.
This explanation is derived through an explanation function $\phi_{\lambda}$. The function $\phi_{\lambda}(x)$ maps the input $x$, the model $f(x)$, and the prediction $y$ to the explanation $\textbf{\textit{e}}$: $\textbf{\textit{e}} = \phi(x, f(x), y; \lambda)$.
The space of all possible explanations with respect to the model $f(x)$ is denoted as $\mathbb{E}$ with $\phi_{\lambda} \in \mathbb{E}$.

The explanation function can be chosen using any feature attribution method, e.g., \citet{lime,integratedgradients,deeplift,lrp}.

To provide meaningful insights into the predictions of XentricAI, we adopt SHAP \citep{lundberg2017unifiedapproachinterpretingmodel} for feature attribution. SHAP is grounded in \textit{Shapley values} from game theory \citep{shapley1953value}, offering a robust framework based on established axioms: local accuracy, missingness, and consistency. These ensure fair and consistent feature importance scores.
 
\textit{Shapley values} are a method for fairly distributing a value among a group of individuals or features based on their contribution to the overall outcome \citep{shapley1953value}. The algorithm calculates the average marginal contribution of each feature value across all possible coalitions, which are combinations of present or absent features. To do this, it first generates predictions for different coalitions with and without the analyzed feature, then takes the difference between those predictions to calculate the marginal contribution of the feature. This process is repeated for all features, and the resulting values are the SHAP values, which are estimates of the \textit{Shapley values}, representing each feature's importance on the model’s prediction. This is defined as:
\begin{equation}
   g(z^ \prime )= \phi_0+ \sum_{j=1}^M  \phi_j z_j\prime
\end{equation}
with $g$ being the explanation model, $z\prime \in \{0, 1\}^M$ is the coalition vector, $\phi_j \in  \mathbb{R}$ is the SHAP value for a feature $j$, and $M$ is the maximum coalition size.
The SHAP algorithm provides explanations on a local (instance-specific) as well as a global (dataset-wide) level, which is achieved by averaging over all absolute local explanations:

\begin{equation}
    I_j= \frac{1}{M} \sum_{m=1}^M | {\phi_j}^m |
\end{equation}
with $I_j$ being the global explanation for feature $j$ and $M$ being the number of samples in the dataset.

Alternative feature ranking methods, such as LIME and Grad-CAM, are predominantly designed for generating local explanations, limiting their applicability to our objectives. Additionally, LIME is known to exhibit instability with random seeds and lower fidelity in high-dimensional spaces \citep{visani2020optilime}. In contrast, SHAP's theoretically sound framework and model-agnostic nature make it well-suited for diverse architectures, including the GRU-based model used in our study. Grad-CAM, however, is a model-specific method restricted to CNNs, rendering it incompatible with our approach.

While SHAP is widely recognized for its robust theoretical foundation, it is also computationally intensive. However, the specific characteristics of our study mitigate these challenges. The GRU model employed is compact and parameter-efficient, designed for computational simplicity. Additionally, the feature set consists of only five features, significantly reducing SHAP's computational demands. Furthermore, the use of the GradientExplainer, a sampling-based approximation of SHAP, enhances its efficiency without compromising interpretability. Although faster alternatives, such as FastSHAP \citep{jethani2021fastshap}, exist, these were not considered in this work. Our focus was on ensuring compliance with the EU AI Act and prioritizing model explainability and robustness.

Nevertheless, the current implementation of SHAP has a significant limitation: It is unable to handle three-dimensional data, which is a common characteristic of multi-variate time series data. To overcome this limitation, we employed a window-based approach to extract the SHAP values for each time step.

\begin{figure}[ht]
\centering
       \includegraphics{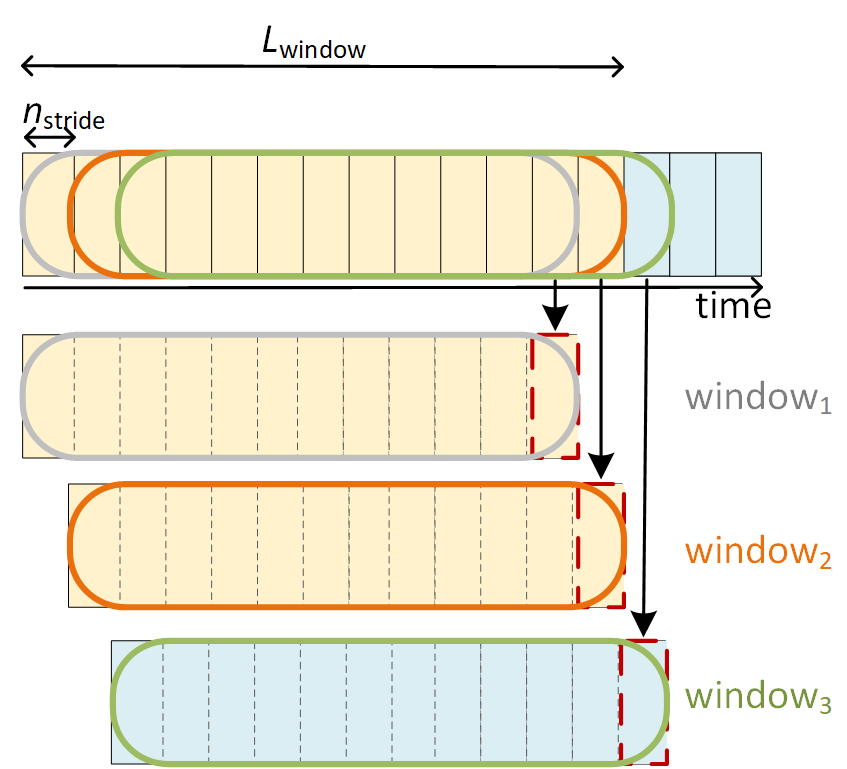}
    \caption{Windowing and preprocessing of gesture sequences for SHAP analysis.}
 \label{fig:windowing}
 \end{figure}

We transformed the dataset $\textbf{X}$ by concatenating each recording sequentially, thereby creating a contiguous sequence of time series data  $\widetilde{\textbf{X}} \in \mathbb{R}^{T \times D}$, where $T$ is the number of time steps, and $D$ is the number of features. The same transformation was applied to the label matrix $\textbf{Y} \in \mathbb{R}^{M \times C}$ leading to $\widetilde{\textbf{y}} \in \mathbb{R}^{C}$.  Subsequently, we divided $\widetilde{\textbf{X}}$ into overlapping windows, each comprising a short segment of the time series data (as illustrated in Fig. \ref{fig:windowing}). Each window has a length of $L_{\text{window}}$. The total number of windows, given a stride of $n_{\text{stride}} = 1$, is calculated as: $W = T - L_{\text{window}} + 1$.

For each window $w$, the window data $\widetilde{\textbf{X}}_w$ and the corresponding label $y_w$ are defined as:

\begin{equation}
    \widetilde{\textbf{X}}_w = \widetilde{\textbf{X}}[w:w+L_{\text{window}}, :]
\end{equation}

\begin{equation}
    y_w = \widetilde{\textbf{y}}[w + L_{\text{window}} - 1]
\end{equation}

where $w \in \{1, 2, \ldots, W\}$.

The SHAP values for the window $w$ are denoted as $\phi_w$, resulting in a window-specific SHAP value matrix $L_{\text{window}} \times D$:
$\phi_w = \text{SHAP}(\widetilde{\textbf{X}}_w, y_w)$, $\phi_w \in \mathbb{R}^{L_{\text{window}} \times D}$.

To ensure that the SHAP values for each time step are captured, we applied window striding with a step size of $n_{\text{stride}} = 1$. This approach involves moving the window one step at a time and repeating the aforementioned process for each stride: $\phi_{1}, \phi_{2}, \ldots, \phi_{W}$.
Further details about the preprocessing can be found in Section \ref{subsec:radar-preprocessing}.

In contrast to previous studies, such as \citet{seifi_gas}, where the SHAP values of all windows with the respective time step are averaged, we opted not to perform averaging in this paper. Our rationale is that the prediction of one window is based on the label of the last frame within that window, and we aim to gain a clear understanding of how each feature of each time step within this specific time window affected the model prediction of the last frame of that window. Averaging the SHAP values would obscure this clear assessment, which is essential for our frame-by-frame prediction approach.

Reasoning about the potential cause of the anomalous gesture is defined as "gesture characterization" in the following. The gesture characterization process is now two-split: 

\textbf{Initialization} This phase sets the foundation for gesture characterization. For each gesture class and for each feature $j$, the local SHAP values $\phi_j^m$ of $n$ nominal gestures are averaged to retrieve the global SHAP values $I_j$.

Using a thresholding mechanism, the range of acceptable global SHAP values, referred to as \ac{srv}, for each feature is deduced. Specifically, the upper $( I_j^{\text{max}} )$ and lower $(I_j^{\text{min}} )$  thresholds are determined by considering the minimum and maximum SHAP values across the nominal gestures: 
\begin{equation}
    I_j^{\text{max}} = \max_{m=1,\ldots,M} \phi_j^m
\end{equation}
\begin{equation}
    I_j^{\text{min}} = \min_{m=1,\ldots,M} \phi_j^m.
\end{equation}

To capture the relationships between feature importances, we gauge whether the ordering of feature importance changes between nominal and anomalous gestures. This involves calculating a median threshold $I_j^{\text{median}}$ using the median of the upper and lower threshold:
\begin{equation}
    I_j^{\text{median}} = \frac{I_j^{\text{min}} + I_j^{\text{max}}}{2}
\end{equation}

The slope between consecutive features of this median threshold aids in analyzing the alteration in feature importance ordering.

Specifically, the slope between consecutive features $j$ and $j+1$ is calculated as:

\begin{equation}
    \text{slope}_{j, j+1} = I_{j+1}^{\text{median}} - I_j^{\text{median}}.
\end{equation}

By monitoring these slopes, it is possible to detect changes in the ordering of feature importances, which may indicate a shift from nominal to anomalous gestures.

This initialization phase is performed once and prepares the model for subsequent utilization in characterizing future anomalous gestures.

\textbf{Explanation Phase:} In this phase, each anomalous gesture is characterized. Local and global SHAP values are computed for the anomalous gesture. The system asks for user input regarding the intended gesture class. Using the user input, the anomalous SHAP values and the \ac{srv} are leveraged to provide feedback to the user, aiding in understanding and refining the execution of the gesture.
The now-gained knowledge is then communicated with the end user.

\begin{figure}
\centering
       \includegraphics{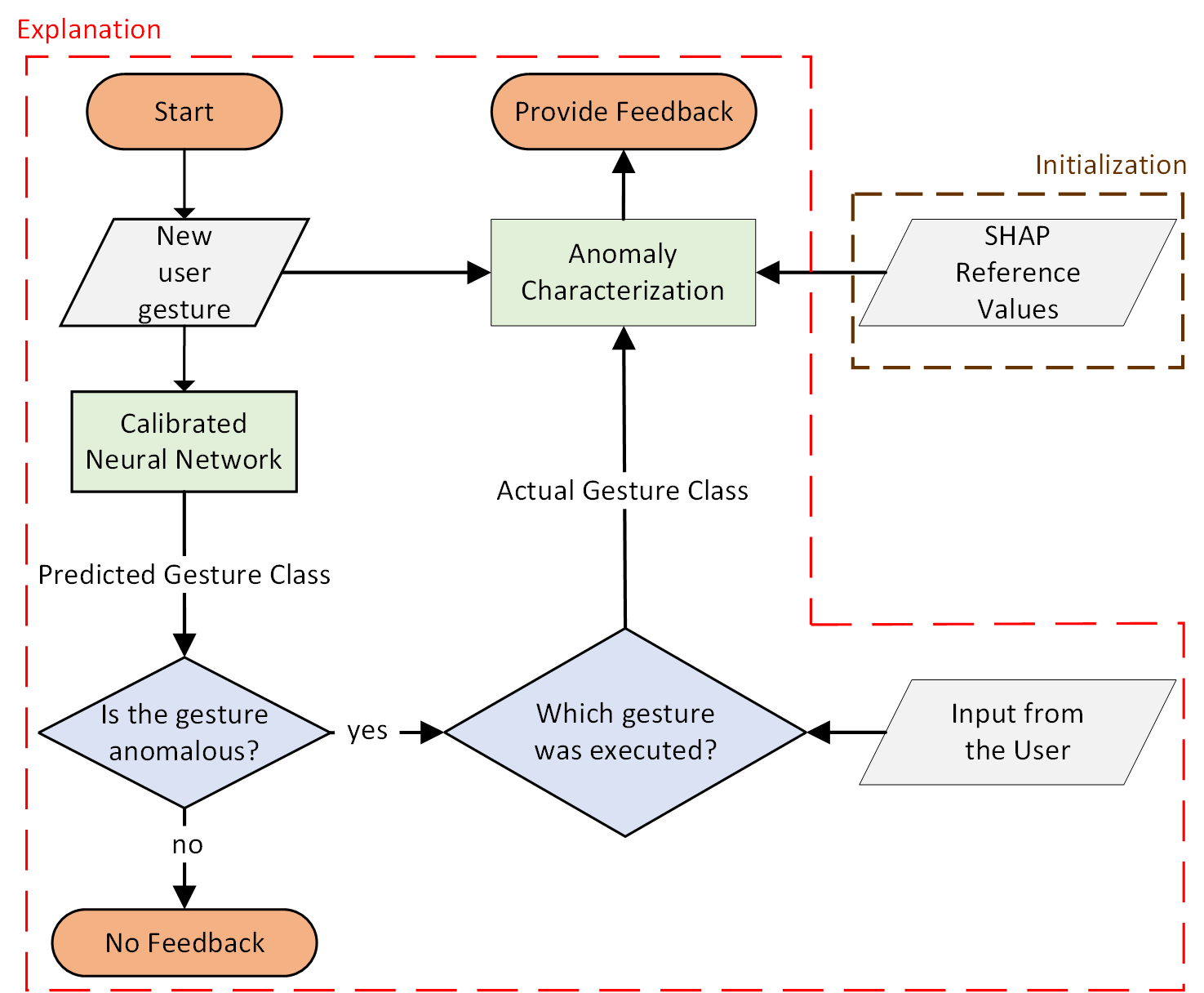}
    \caption{Flowchart illustrating the two-phase process for gesture characterization using SHAP values. The initialization phase sets the foundation by computing global SHAP values and determining the range of acceptable SRVs for each feature. The explanation phase characterizes each anomalous gesture by computing local and global SHAP values, and provides feedback to the user by comparing the anomalous SHAP values to the SRV, aiding in understanding and refining the gesture execution.}
 \label{fig:flowchart-gesture-characterization}
 \end{figure}

The complete process of gesture characterization, as illustrated in the flowchart (Fig. \ref{fig:flowchart-gesture-characterization}), provides a structured approach to analyzing and interpreting user input for feedback generation.

\section{Experimental Setup}
\label{sec:results}
This section outlines the experimental setup for evaluating XentricAI. We begin with the fundamentals and configurations of the FMCW radar system (Section \ref{subsec:fmcw-fundamentals}) and the collection of a gesture dataset comprising 24,000 anomalous and 4,000 nominal gestures (Section \ref{subsec:gesture-data-acq}). We then overview radar preprocessing (Section \ref{subsec:radar-preprocessing}) and the gesture sensing model architecture, its training, and evaluation, as well as the \ac{tl} setup (Section \ref{subsec:model-training}). As a last step, the VAE-based anomaly detection module is described (Section \ref{subsec:vae}).

\subsection{Fundamentals and Configuration of FMCW Radar System}
\label{subsec:fmcw-fundamentals}

\begin{figure}
\centering
       \includegraphics[width = 1\linewidth]{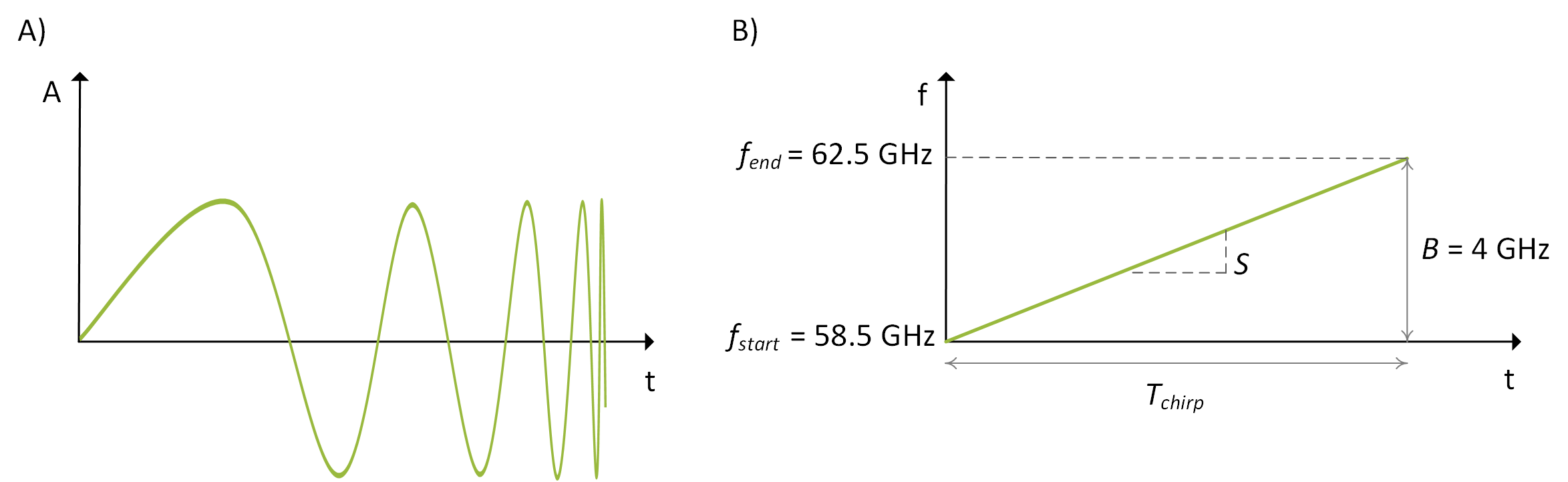}
    \caption{Illustration of a single chirp emitted by the radar device. (a) Time-domain representation of the chirp, showing the signal amplitude as a function of time. (b) Frequency-domain representation of the chirp, displaying the frequency sweep over time. It also highlights some of the radar configuration parameters used in this work, including the frequency range ($f_{\text{start}} = 58.5 \,$ GHz to $f_{\text{end}} = 62.5 \,$ GHz) and bandwidth ($B = 4 \,$ GHz).}
\label{fig:chirp}
 \end{figure}

\ac{fmcw} radar systems transmit an electromagnetic signal that objects reflect and then capture the reflected signals \citep{iovescu2017fundamentals}. The transmitted signal called a chirp has a linearly increasing frequency over time, characterized by its start and end frequency $f_{\text{start}}$ and $f_{\text{end}}$, bandwidth $B$, and duration $T_{\text{chirp}}$. The slope $S$ of the chirp captures the rate of change of frequency. This is also illustratively highlighted in Fig. \ref{fig:chirp}.

The main \ac{rf} components of an \ac{fmcw} radar system include a synthesizer, transmit antenna, receive antenna, and a frequency mixer. The synthesizer generates a chirp, while the mixer combines the received and transmitted signals to produce an \ac{if} signal.

The range resolution of a \ac{fmcw} radar system is its ability to distinguish between two or more objects. Increasing the chirp bandwidth increases the range resolution, with a bandwidth of a few GHz resulting in a range resolution of centimeters. Velocity measurement is achieved by transmitting two chirps separated by a time interval and then processing the reflected chirps using a \ac{fft} to determine the object's range. The resulting phase difference between the two range-\ac{fft}s corresponds to the object's velocity.

The two-chirp velocity measurement method has a significant limitation: it fails to accurately measure velocity when multiple objects with different velocities are at the same distance from the radar. To overcome this limitation, the radar system must transmit a set of $N$ equally spaced chirps, known as a burst, and use range-\ac{fft} and Doppler-\ac{fft} processing to resolve the objects.

Furthermore, \ac{fmcw} radar systems can estimate the angle of a reflected signal with respect to the horizontal plane by exploiting the fact that a small change in target distance results in a phase shift across two horizontally placed antennas. peak. This principle enables angular estimation using at least two virtual antennas.

For this work the XENSIV™ BGT60LTR13C $60\,$GHz FMCW radar from Infineon Technologies is utilized. The radar system was configured to operate within a frequency range of $58.5\,$GHz to $62.5\,$GHz, providing a range resolution of $37.5\,$mm and a maximum resolvable range of $1.2\,$m. The velocity resolution is approximately $0.26 \, m/s$ and the maximum velocity is around $4.17 \, m/s$. The radar employed a burst configuration consisting of $32$ chirps per burst, with a frame rate of $33\,$Hz and a chirp pulse repetition time of $300\, \mu$s. The 32-chirp bursts are arranged in a 3D array, denoted as $[R \times C \times S]$, where $R$ represents the three receive channels, $C$ corresponds to the slow time axis with $32$ elements, and $S$ represents the fast time axis with $64$ elements. In this context, the fast time axis corresponds to the time within a single chirp, capturing the rapid frequency modulation and allowing for range measurement. In contrast, the slow time axis refers to the time across multiple chirps, which is used for capturing the Doppler shifts related to object velocity. The three receive antennas are arranged in an L-shape, allowing for angle estimation of the horizontal (azimuth) and vertical (elevation) axis.
The slow time axis represents the time between consecutive radar pulses, while the fast time axis represents the time delay between the transmission of a radar pulse and the reception of its echo, with the slow time axis typically measured in milliseconds or seconds and the fast time axis in microseconds or nanoseconds.

\subsection{Gesture Dataset Acquisition}
\label{subsec:gesture-data-acq}

\begin{figure}[htbp]
\centering
       \includegraphics{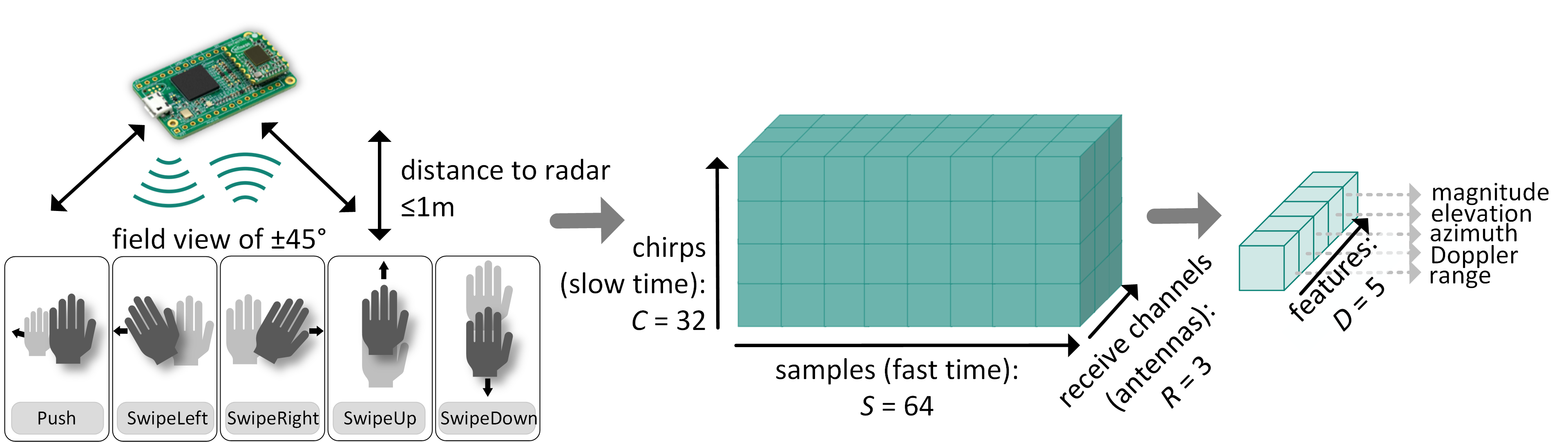}
    \caption{Data collection and processing pipeline. The 60 GHz FMCW radar captures gestures, producing a radar cube (32x64x3), which is then processed to features with a reduced dimension (1x5). This process is highlighted for a single frame.}
\label{fig:radar-preprocessing}
 \end{figure}

Using the $60 \, $ GHz \ac{fmcw} radar, we created a comprehensive gesture dataset by asking multiple individuals to perform five distinct gestures in various indoor locations. Twelve participants performed gestures in six different settings, including a gym, library, kitchen, bedroom, shared office room, and closed meeting room. The gestures were performed within a field of view of $\pm 45^{\circ}$ and at a distance of one meter or less from the radar, as highlighted in Fig. \ref{fig:radar-preprocessing}. The shortest arm length among the participants was $62\,cm$, and the longest was $73\,cm$. The dataset consists of six classes, comprising five distinct gesture types: \textit{SwipeLeft}, \textit{SwipeRight}, \textit{SwipeUp}, \textit{SwipeDown}, and \textit{Push}. Furthermore, a \textit{Background} class is included, which represents the absence of any gesture execution.

Each gesture recording lasted approximately $3\, s$ (100 frames), with a single gesture performed for an average duration of $0.3 \,s$ (ten frames). Participants were instructed to fully extend their arms during gesture execution.
Building upon a previously published dataset, we extend the dataset by including four additional users performing 1,000 gestures each, resulting in a total of 25,000 nominal gestures from twelve users \citep{radar_data}. The user labels and gender distribution are as follows: $\{\text{user}_i \mid i = 2, 3, 10\}$ were female, with the remaining users being male. To facilitate research in anomaly detection and reasoning, we introduce an additional extension to the dataset featuring anomalous gestures. Eight users performed three types of anomalous gestures: fast, slow, and wrist executions.
We selected anomalies related to execution speed and execution method based on our belief that these are the most common deviations in gesture execution observed across individuals. Variations in execution speed, such as fast or slow gestures, can be observed due to factors like user familiarity with the system, urgency, or personal physical abilities, all of which can influence how quickly a gesture is performed. These differences affect how the system interprets and responds to input.

Similarly, wrist executions are often a result of individuals using different parts of their body when performing gestures. Some users may rely more on wrist movements rather than broader arm gestures, leading to variations in how gestures are recognized. Others choose to perform gestures with minimal effort, using only their wrist rather than fully extending their arm. This can be due to a preference for a more relaxed or less physically demanding gesture execution. By focusing on these anomalies, we aim to capture some variations of human gesture execution.

Fast executions last approximately $0.1-, s$, and slow executions last $3\, s$, respectively. Wrist executions are performed with the wrist instead of a fully extended arm and the shoulder joint. Each of the eight users performed 1,000 gestures per anomaly, resulting in 24,000 anomalous gestures in total.

The dataset sequence of each gesture sample is saved as a NumPy array \citep{numpy} with four dimensions: $[100 \times 3\times32\times64]$. The first dimension represents the frame length of each gesture. The second, third, and fourth dimensions are, as described in Section \ref{subsec:fmcw-fundamentals}, the three receive channels, the slow time axis, and the fast time axis, respectively. 
In Fig. \ref{fig:radar-preprocessing}, the radar cube for a single frame is shown before and after running through the preprocessing pipeline. Details of the radar preprocessing are described in Section \ref{subsec:radar-preprocessing}.

\subsection{Radar Preprocessing}
\label{subsec:radar-preprocessing}

A lightweight radar preprocessing algorithm, as described in \citet{strobel}, is employed for frame-based labeling and feature extraction. This approach effectively isolates the hand from the body, thereby enabling efficient gesture labeling and \ac{fft} preprocessing to be performed solely within the relevant range bin. Since this radar processing algorithm requires significantly less computational resources than conventional approaches and provides intuitive features, it was adopted for this work.

\begin{figure}
\centering
    \includegraphics[width = 1\linewidth]{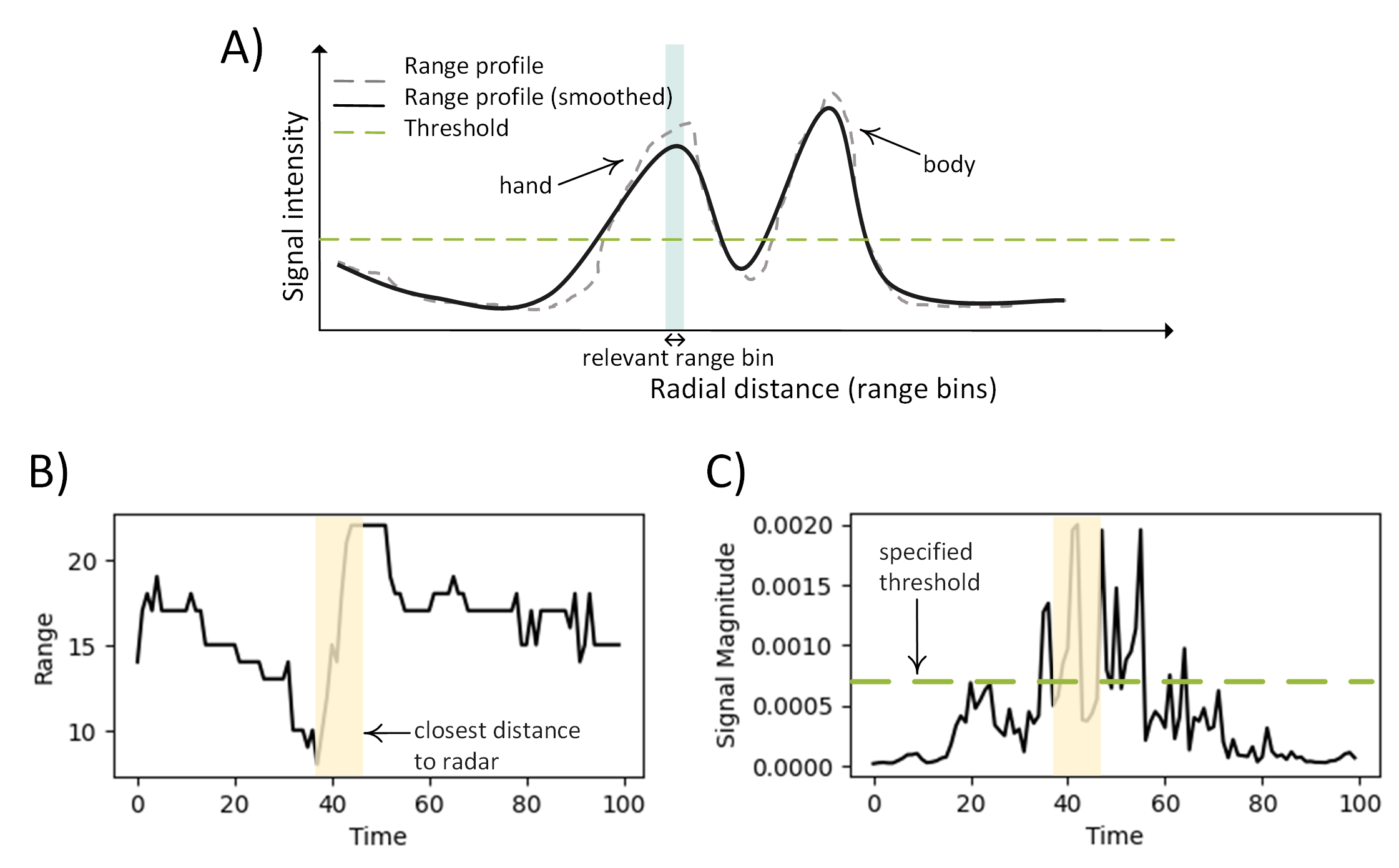}
    \caption{Lightweight preprocessing algorithm. A) Target detection for preprocessing: The range profile shows two targets above the threshold: the moving hand and the body of the person. The range bins of the closest target, the hand, are selected for further processing (highlighted in light blue). B) and C) Dynamic label refinement: Frames around the local maxima are labeled as the gestures (highlighted as light-orange area), and all remaining frames are labeled as background. B) The local peak search algorithm selects the closest target. C) The minimum radial distance is only searched for at frames with a signal amplitude above an empirically determined threshold.}
\label{fig:label-refinement}
\end{figure}

As a first step, the range profile is created by applying a \ac{fft}.
A range profile is a graphical representation of the reflected radar signal intensity with respect to the radar's operating distance, providing a one-dimensional representation of the radar scene. 
As depicted in Fig. \ref{fig:label-refinement} panel A), the x-axis represents the distance from the radar sensor and the y-axis represents the signal intensity.
Two maxima, which are so-called targets, can be seen: The moving hand and the body of the person. The signal strength of the hand is comparatively weaker than that of the body, primarily due to its smaller size, which results in lower reflectivity.
Instead of choosing the target with the highest signal strength, which is the body, the closest target, namely the hand, is chosen. 
This approach enables efficient and stable target detection, which is robust against random body movements of the person performing the gesture.

The features are then only extracted within the relevant range bin corresponding to the closest target to the radar (Fig. \ref{fig:label-refinement} light-blue area). 

More specifically, to detect the hand as a moving target, raw radar data is processed using fast-time processing, which involves removing DC (zero frequency) components, applying a \ac{fft} along the distinct chirps, and dropping the symmetric part of the spectrum. This yields complex range profiles. The complex mean over the slow time axis is then removed to eliminate (quasi-)static targets. To improve the signal-to-noise ratio, the magnitude of the complex data is integrated along the receive channels and chirps. The resulting one-dimensional vector represents the reflected energy of moving targets along the resolvable range.

Assuming the person performs the gesture towards the sensor, the hand can be assumed to be the closest moving target. Random body movements are more distant. Therefore, a local peak search is applied to select the closest target and retrieve its \textit{radial distance}. The one-dimensional vector is smoothed using a Gaussian filter and thresholded to suppress weak local maxima. If all elements are below the threshold, the global maximum is taken.

With the hand's range bin identified, the relevant scattering characteristics can be efficiently extracted. Instead of performing slow time \ac{fft}s across all range bins, \ac{fft}s are performed only on the hand's range bin across all receive channels. The resulting Doppler profile is integrated across the channels, and the maximum signal is searched for, which indicates the \textit{radial velocity} and \textit{signal magnitude}. Finally, the \textit{ horizontal} and \textit{vertical angles} are estimated using phase-comparison monopulse on the detected hand bin \citep{monopulse}.

In summary, the five gestures can be fully described by the following features:
\begin{itemize}
    \item \textit{Radial distance}: The distance from the radar to the hand, measured as the closest target to the radar.
    \item \textit{Radial velocity}: The speed at which the hand moves towards or away from the radar, derived from the Doppler shift in the detected range bin.
    \item \textit{Horizontal angle: }The azimuth angle of the hand relative to the radar, estimated using phase-comparison monopulse techniques.
    \item \textit{Vertical angle}: The elevation angle of the hand relative to the radar is also estimated using phase-comparison monopulse techniques.
    \item \textit{Signal magnitude}: The amplitude of the signal in the Doppler bin corresponding to the detected radial velocity, indicating the strength of the moving hand's reflection.
\end{itemize}

Following feature extraction, the dynamic label refinement approach proposed by \citet{strobel} is applied. 
A gesture recording lasts 100 frames, whereas the actual gesture execution is performed within a few hundred milliseconds averaging around ten frames. The remaining frames primarily consist of background noise. For a robust real-world application, it is essential to distinguish gesture frames from non-gesture frames to reduce false positives. 
Therefore, dynamically identifying the frame location of the actual gesture is needed. A common characteristic of the supported five gestures is that the hand reaches the closest distance to the radar during execution (Fig. \ref{fig:label-refinement} A)). This is used to anchor the beginning of the gesture label at the frame with the minimum radial distance.

To mitigate false alarms at the sequence boundaries, where noise dominates, the minimum radial distance is only searched for at frames with a signal amplitude above an empirically determined threshold (Fig. \ref{fig:label-refinement} B)). The label is then maintained for a fixed number of frames $L_{\text{gesture}}$ (in this work $L_{\text{gesture}}=10$, corresponding to the average gesture execution duration) before changing to the background class.
This leads to a dataset with the dimensions $[M \times F \times D]$, where $M$ is the total number of gesture recordings, $F$ is the number of frames per gesture recording and $D$ is the number of features.

As depicted in Fig. \ref{fig:windowing} and described in Section \ref{subsec:anomalous-gesture-feedback}, to generate SHAP values for smaller sequences, we apply windowing to the gesture sequences generating $W$ windows. In this work, the length of one window is set to $L_{\text{window}}=22$. This parameter was chosen to balance the trade-off between capturing sufficient historical context and avoiding the dominance of recent inputs. 
We set $n_{\text{stride}} = 1$ to enable frame-based feature attribution assessment. The resulting time-series windows are then fed into the \ac{nn} architecture.

The final dataset used for training has the shape $[W \times (L_{\text{window}} - W-1) \times D]$.

\subsection{GRU Model Architecture, Training and Evaluation}
\label{subsec:model-training}
In the implementation, we used TensorFlow v2.9.1\textsuperscript{TM}. We used the 6 Core Intel\textsuperscript \textregistered Core i7-9850H CPU as a processing unit. The results of each experiment are an average of six individual runs.

This study employs a tiny \ac{rnn}-based architecture to detect and classify gestures. The network consists of a \ac{gru} layer with 16 units to process time-series data and a dense layer with six neurons and Softmax activation to distinguish between five gestures (SwipeLeft, SwipeRight, SwipeUp, SwipeDown, and Push) and the Background class. 

For training, the Adam optimizer \citep{adam} is used with a learning rate of 0.001, sparse categorical cross-entropy loss function, and a batch size of 32 for 100 epochs. 

To further evaluate the robustness of the algorithm, we performed experiments using two distinct training setups. The first setup consisted of a simple training dataset with 2,000 gestures from two individuals ($\text{user}_4$ and $\text{user}_5$). The second setup featured a more diverse training dataset with 9,000 gestures from five individuals and four different environmental locations ($\{\text{user}_i \mid i = 1, 2, 4, 6 \}$). For each setup, $80\%$, $10\%$, and another $10\%$ of the dataset were divided into three independent, temporally non-overlapping parts. They were utilized as training, validation, and forgetting rate assessment datasets, respectively. For the first, simple train setup, $\{\text{user}_i \mid i = 1, 2, 3, 6, 7, 8, 9, 10, 11, 12 \}$ were used for model calibration. For the second, diverse train setup, the data of $\{\text{user}_i \mid i = 5, 7, 8, 9, 10, 11, 12 \}$ were used for calibration. 
We evaluated both setups based on two criteria: (1) the calibrated model's performance on user data with and without \ac{er}, and (2) the forgetting rate after recalibration with and without \ac{er}.

For the \ac{tl} experiments, retraining was performed without freezing any layers. Varying numbers of training recordings and user recordings per class were evaluated. The notation $n_\text{train} - n_{\text{user},i}$ indicates the number of training gesture samples per class and the number of user gesture samples per class, respectively. For experiments with ER, the configurations $\{(n_\text{train}, n_{\text{user},i}) \mid n_\text{train} \in \{5, 10, 25, 50, 100\}, n_{\text{user},i} \in \{5, 10, 25, 50, 100\}\}$ were assessed:.
For experiments without ER, the configurations $\{n_{\text{user},i} \mid n_{\text{user},i} \in \{5, 10, 20, 30, 35, 50, 55, 75, 100, 105, 110, 125\}\}$, corresponding to the sum of gestures used for experiments with ER, were assessed.


Three types of accuracy metrics were applied to provide a comprehensive model performance assessment.
First, the standard accuracy metric was used, which compares each frame of the prediction with each frame of the ground truth. 

Let $M$ be the number of gesture recordings, $F$ be the total number of frames in recording $m$, $\textbf{y}_{\text{true}, m}$ represent the true labels, and $\textbf{y}_{\text{pred}, m}$ represent the predicted labels. The accuracy for gesture sample $m$ is then given by:

\begin{equation}
    \text{accuracy}_m = \frac{1}{F} \sum_{j=1}^{F} \mathbb{I}(\textbf{y}_{\text{true}, m}[j] = \textbf{y}_{\text{pred}, m}[j])
\end{equation}

Where $\mathbb{I}(\cdot)$ is the indicator function, which is 1 if the condition inside is true, and 0 otherwise.

The overall accuracy $\text{acc}$ is then given by:

\begin{equation}
    \text{acc} = \frac{1}{M} \sum_{m=1}^{M} \text{accuracy}_m
\end{equation}

However, the accuracy metric can be misleading due to the class imbalance issue within individual recordings, where approximately $92\%$ of frames are labeled as background. For instance, even if the gesture is completely misclassified, a model that simply predicts the background class for all frames would still achieve an accuracy of $92\%$, which does not accurately reflect its performance.
To address this, two additional accuracy metrics were developed. 
The "gesture accuracy" metric only considers non-background frames when calculating accuracy, providing a more detailed evaluation of gesture frames and mitigating the impact of background frames.


Let $\mathbb{I}(\textbf{y}_{\text{true}, m}[j] = \textbf{y}_{\text{pred}, m}[j])$ be the indicator function, which returns 1 if the $j$-th frame in the $m$-th sequence is correctly predicted, and 0 otherwise. $B$ is the background class label and $G_m = \{j \mid \textbf{y}_{\text{true}, m}[j] \neq B\}$ represents the set of indices corresponding to gesture (non-background) frames in sequence $m$. The gesture window accuracy for sequence $m$ is given by:

\begin{equation}
    \text{gesture\_acc}_m = \frac{1}{|G_m|} \sum_{j \in G_m} \mathbb{I}(\textbf{y}_{\text{true}, m}[j] = \textbf{y}_{\text{pred}, m}[j])
\end{equation}

Where $|G_m|$ is the number of gesture frames in sequence $m$.

The overall gesture window accuracy $\text{gesture\_acc}$ is then given by:
\begin{equation}
\text{gesture\_acc} = \frac{1}{M} \sum_{m=1}^{M} \text{gesture\_acc}_m
\end{equation}

\begin{figure}[htbp]
\centering
    \includegraphics[width = 1\linewidth]{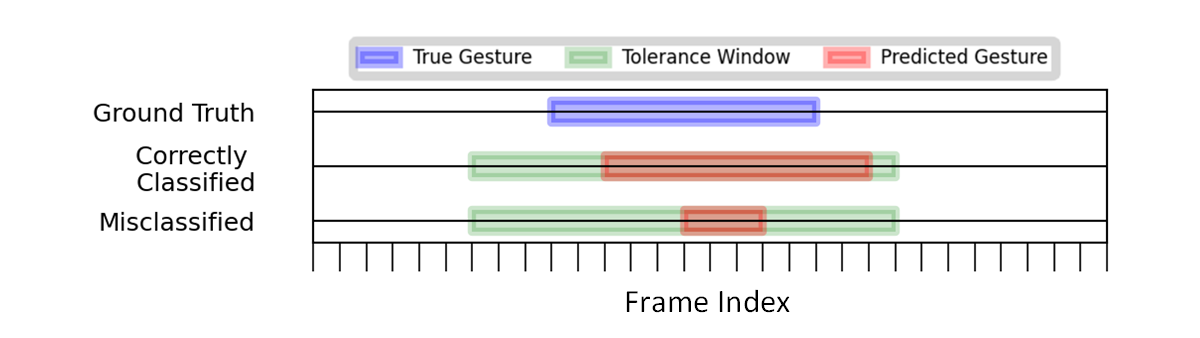}
    \caption{Visualization of the dynamic gesture accuracy metric dg\_acc for gesture classification. The blue box shows the true gesture, the green box represents the tolerance window, and the red box indicates the predicted gesture. The correctly classified gesture falls within the tolerance window and meets the duration requirement, while the misclassified gesture does not. This metric allows for slight offsets, reflecting a more practical assessment of model performance.}
\label{fig:dynamic_acc}
 \end{figure}

The "dynamic gesture accuracy" metric, based on the metric introduced in \citet{shaaban2024rt} and illustrated in Fig. \ref{fig:dynamic_acc}, assesses the practical value of predicted gestures. We argue that predicting a gesture with a slight offset (up to 3-4 frames) from the labeled window has negligible real-world implications. Therefore, if the predicted gesture is within this tolerance and longer than four frames, the entire recording is considered correctly classified. Otherwise, it is deemed misclassified. The formula for the dynamic gesture accuracy of each sequence $m$ is defined as follows:

\begin{equation}
    \text{dg\_acc}_m = 
\begin{cases} 
1 & \text{if } \text{unique}(\textbf{W}_{\text{pred}, m}) = \{\text{unique}(\textbf{W}_{\text{true}, m})\} \text{ and } |\text{non-zero}(\textbf{W}_{\text{pred}, m})| > 4, \\
0 & \text{otherwise}.
\end{cases}
\end{equation}

where $\mathbb{I}(\cdot)$ is the indicator function, which is 1 if the condition inside is true, and 0 otherwise.
$S_m$ is the set of indices where the true gesture occurs in sequence $m$. $s_m$ is the start of the extended window: $s_m = \min(S_m) - 3$, and $e_m$ is the end of the extended window: $e_m = \max(S_m) + 4$. $\textbf{W}_{\text{true}, m}$ represents the ground truth labels in the extended window: $\textbf{W}_{\text{true}, m} = \textbf{y}_{\text{true}, m}[s_m:e_m]$. $\textbf{W}_{\text{pred}, m}$ represents the predicted labels in the extended window: $\textbf{W}_{\text{pred}, m} = \textbf{y}_{\text{pred}, m}[s_m:e_m]$. $\text{unique}(Z)$ denotes the unique non-zero values in set $Z$ with $|Z|$ denoting the length of set $Z$.

The overall accuracy $\text{dg\_acc}$ is given by:

\begin{equation}
    \text{dg\_acc} = \frac{1}{M} \sum_{i=m}^{M} \text{dg\_acc}_m.
\end{equation}

\subsection{Variational Autoencoder-Based Anomaly Detection Setup and Implementation}
\label{subsec:vae}
The VAE architecture utilized in this study consists of an encoder and a decoder, both implemented with dense layers. The encoder compresses the input gesture data into a latent representation characterized by the mean $\mathbf{z_{\mu}}$ and the log variance $\mathbf{z_{\log\sigma^2}}$ of a Gaussian distribution. Specifically, the encoder comprises four dense layers with 512, 256, 128, and 64 units respectively, each followed by batch normalization and ReLU activation functions. A dropout rate of 0.3 is applied after the first and second dense layers to prevent overfitting.

The latent space sampling employs the reparameterization trick, using the mean and log variance to generate latent variables $\mathbf{z}$ from the Gaussian distribution. The decoder mirrors the encoder’s structure, with dense layers of 64, 128, 256, and 512 units, followed by Batch Normalization and ReLU activation functions. The output layer reconstructs the input gesture using a sigmoid activation function, reshaping it back to its original dimensions.

The VAE was trained with 6,000 gestures from $\{\text{user}_i \mid i = 3,4,5,8, 11, 12\}$ focusing on normal gesture patterns. The data preprocessing involved normalizing the gesture data to ensure consistency and splitting it into training and validation sets with a $65\%$ to $35\%$ ratio. The training configuration included 250 epochs, a batch size of 16, the Adam optimizer with a learning rate of 0.001 and early stopping with a patience of 20 epochs to prevent overfitting, with the restoration of the best weights based on validation loss.

The model was trained to minimize the combined and equally weighted reconstruction and KL divergence loss, using the validation data to monitor performance and adjust weights accordingly.

Post-training, the VAE detected anomalies in gesture data by calculating the reconstruction error for each gesture. For user-specific thresholding, calibration gestures were used to set user-specific thresholds based on the 90th percentile of their reconstruction errors.

\section{Results and Discussion}
In this section, we present and discuss the outcomes of our experiments. First, we evaluate the model calibration results using \ac{tl} with and without ER, assessed through both simple (Section \ref{subsubsec:simple}) and diverse training setups (Section \ref{subsubsec:diverse}). Next, we examine the performance of our VAE anomaly detection module (Section \ref{subsec:vae-results}). Additionally, we evaluate our gesture characterization algorithm to complete the analysis (Section \ref{subsec:xai-results}). As a last step, XentricAI's conformance with the EU AI Act is evaluated (Section \ref{subsec:xentric-and-aiact}) .

\subsection{Model Calibration Results}
\label{subsec:model-calibration-results}

The analysis will focus on comparing the performance of the model on user data with and without \ac{er}, allowing the evaluation of two key aspects: (1) the model's ability to adapt to different users through \ac{tl} and various learning strategies (see Table \ref{tab:er-calibrated-vs-uncalibrated}), and (2) the forgetting rate, which measures the decrease in performance on data with the same underlying distribution as the training dataset after retraining (see Table \ref{tab:forgetting-rate-calibrated-vs-uncalibrated}).

\subsubsection{Simple Training Setup for Gesture Recognition}
\label{subsubsec:simple}

The summary of the model calibration results evaluated on user data for calibration assessment is presented in Table \ref{tab:er-calibrated-vs-uncalibrated}. This table compares the baseline model's gesture accuracy to that of the calibrated models, which were retrained using varying numbers of training recordings and user recordings per class. The notation $n_\text{train} - n_{\text{user},i}$ indicates the number of training gesture samples per class and the number of user gesture samples per class, respectively. 

For experiments with ER, the following configurations were assessed: $\{(n_\text{train}, n_{\text{user},i}) \mid n_\text{train} \in \{5, 10, 25, 50, 100\}, n_{\text{user},i} \in \{5, 10, 25, 50, 100\}\}$.

For experiments without ER, the following configurations were assessed: $\{n_{\text{user},i} \mid n_{\text{user},i} \in \{5, 10, 20, 30, 35, 50, 55, 75, 100, 105, 110, 125\}\}$, corresponding to the sum of gestures used for experiments with ER. 
Table \ref{tab:er-calibrated-vs-uncalibrated} and Table \ref{tab:forgetting-rate-calibrated-vs-uncalibrated} present a comparison of calibration results using the same minimum and maximum numbers of user gesture samples, specifically $n_{\text{user},i} = 5$ and $100$. For a more detailed analysis, comprehensive tables are provided in the appendix, where the data is split by user and configuration (see Tables \ref{tab:full-results-with-er} - \ref{tab:full-std-traindata-without-er}). It should be noted that each value reported in the tables represents the average of six individual runs for each corresponding experiment, ensuring a robust representation of the results.

\begin{table}[h!]
\scriptsize
\centering
\begin{tabular}{ccccc}
\toprule
\textbf{User $i$} & \textbf{Baseline [\%]} & \textbf{\makecell{With ER [\%] \\ (5-100 samples)}} & \textbf{\makecell{Without ER [\%] \\ (5-100 samples)}}\\
\midrule
1   & 88.26 & 93.91 - 97.66 & 92.06 - 95.89\\
2   & 79.59 & 89.10 - 94.32 & 89.72 - 91.81\\
3   & 18.59 & 57.45 - 82.08 & 66.34 - 81.21\\
6   & 73.37 & 92.27 - 98.78 & 92.06 - 96.57\\
7   & 67.12 & 86.64 - 94.32 & 84.43 - 91.15\\
8   & 60.78 & 81.83 - 92.34 & 83.15 - 90.91\\
9   & 76.42 & 91.67 - 97.98 & 93.05 - 95.87\\
10  & 73.76 & 84.98 - 93.62 & 85.47 - 90.47\\
11  & 75.30& 89.62 - 94.89 & 89.83 - 93.94\\
12  & 82.40 & 93.43 - 98.10 & 94.35 - 96.89\\
\midrule
\textbf{Mean} & \textbf{69.56} & \textbf{84.73 - 93.89} & \textbf{87.13 - 92.53}\\
\bottomrule
\end{tabular}
\caption{Gesture accuracy on calibration assessment user data with and without ER.}
\label{tab:er-calibrated-vs-uncalibrated}
\end{table}

\begin{table}[h!]
\scriptsize
\centering
\begin{tabular}{cccc}
\toprule
\textbf{User $i$} & \textbf{\makecell{With ER [\%] \\ (5-100 samples)}} & \textbf{\makecell{Without ER [\%] \\ (5-100 samples)}}\\
\midrule
1   &  88.57 - 96.10 & 88.65 - 88.65 \\
2   &  89.47 - 96.10 & 92.54 - 92.54\\
3   &  83.24 - 91.33 & 58.78 - 58.78\\
6   &  91.44 - 98.78 & 89.81 - 89.81\\
7   &  88.67 - 94.32 & 86.01 - 86.01\\
8   &  88.02 - 96.33 & 78.15 - 78.15\\
9   &  91.67 - 97.98 & 88.15 - 88.15\\
10  &  87.06 - 93.62 & 82.01 - 82.01\\
11 & 88.43 - 93.59 & 61.95 - 89.87\\
12 & 89.74 - 96.36 & 65.43 - 92.58\\
\midrule
\textbf{Mean} &  \textbf{90.09 - 94.90} & \textbf{59.93 - 85.39}\\
\bottomrule
\end{tabular}
\caption{Forgetting rate assessment of calibrated models with and without ER.}
\label{tab:forgetting-rate-calibrated-vs-uncalibrated}
\end{table}

\textbf{Baseline vs. Calibrated Model}. As shown in Table \ref{tab:er-calibrated-vs-uncalibrated}, the baseline gesture accuracy is significantly lower for all users compared to any calibrated model. This demonstrates the effectiveness of \ac{tl} in improving performance through user-specific calibration. The average baseline gesture accuracy is $69.56\%$, while the average gesture accuracy across all calibrated models is much higher, starting at $84.73\%$ and reaching up to $93.89\%$. This shows that implementing transfer learning techniques using ER leads to improvements in user adaptability, with an average performance increase of at least $15.17\%$.
However, a potential reason for the low baseline gesture accuracy, mainly attributed to $user_3$, is the train dataset bias.
If the training data is biased towards the gestures of two men, it may not represent the gestures of women, including $user_3$. This is a classic example of a domain shift or dataset bias, where the training data does not match the characteristics of the testing data.
In this case, the model may have learned to recognize patterns in the gestures of men but not women. This also highlights the need for a diverse training dataset for good baseline performance.

\textbf{Performance on Calibration Assessment User Data.} 
The mean gesture accuracy across participants is consistently higher with \ac{er} across most gesture-per-class configurations compared to without \ac{er}. For example, the mean gesture accuracy for configurations with \ac{er} often exceeds $92\%$, while without \ac{er}, it stabilizes around $89-91\%$.
Users like $user_9$, $user_1$, and $user_6$ show particularly strong performance with \ac{er}, often achieving gesture accuracies above $95\%$ with the number of gestures per class increasing. This trend is somewhat mirrored without \ac{er} but with lower peak gesture accuracies.
$user_3$ and $user_8$, who initially had lower baseline gesture accuracies, benefit significantly from \ac{er}. Without \ac{er}, their gesture accuracy is more volatile and lower on average. $user_3$, for instance, improves from $18.59\%$ (baseline) to around $82\%$ with \ac{er} but stabilizes at a lower range ($76-79\%$) without \ac{er}.
The only exception where model calibration without ER achieves better results is when  $n_\text{user}$ is small, e.g., five recordings per gesture class. 
When retraining without ER, the model is exposed only to the new, limited set of user gesture samples. This can lead to overfitting because the model can become too specialized to these few samples, thus performing better on this specific, small dataset but failing to generalize. This effect is particularly evident when analyzing the forgetting rate.

\textbf{Assessment of Forgetting Rate.} 
The forgetting rate assessment evaluates how well the recalibrated models retain their performance on the initial training data after being fine-tuned on user-specific data. We will compare the performance with and without ER across different training configurations, as summarized in Table \ref{tab:forgetting-rate-calibrated-vs-uncalibrated}, and fully detailed in Tables \ref{tab:full-fr-results-with-er} and \ref{tab:full-fr-results-without-er}. 
In the following, the key statistics of both approaches evaluated on the forgetting rate assessment data are shown. The mean gesture accuracy is averaged over all users:

\begin{itemize}
    \item \textbf{With ER:}
    \begin{itemize}
        \item mean gesture accuracy: $90.09\% - 94.90\%$
        \item mean standard deviation: $0.69\% - 1.77\%$
    \end{itemize}
    \item \textbf{Without ER:}
    \begin{itemize}
        \item mean accuracy: $59.93\% - 85.39\%$
    	\item mean standard deviation: $2.03\% - 4.30\%$
    \end{itemize}
\end{itemize}

These results indicate that models calibrated with ER maintain a significantly higher accuracy on the original training data compared to those without ER. 
Without \ac{er}, there is a notable and sharp decline in gesture accuracy as higher $n_{\text{user,}i}$ becomes, which is indicative of catastrophic forgetting. For instance, for participants like $user_2$, the gesture accuracy drops from $92.54\%$ to around $64.96\%$ as more user gestures are added.
With \ac{er} on the other hand, the forgetting rate is much more controlled, and gesture accuracy remains relatively stable. For most participants, there is either a slight decline or even an increase in performance on the forgetting rate assessment data as the amount of user calibration gestures increases. This demonstrates that ER effectively mitigates forgetting. 

\begin{figure}
\centering
    \includegraphics[width = 1\linewidth]{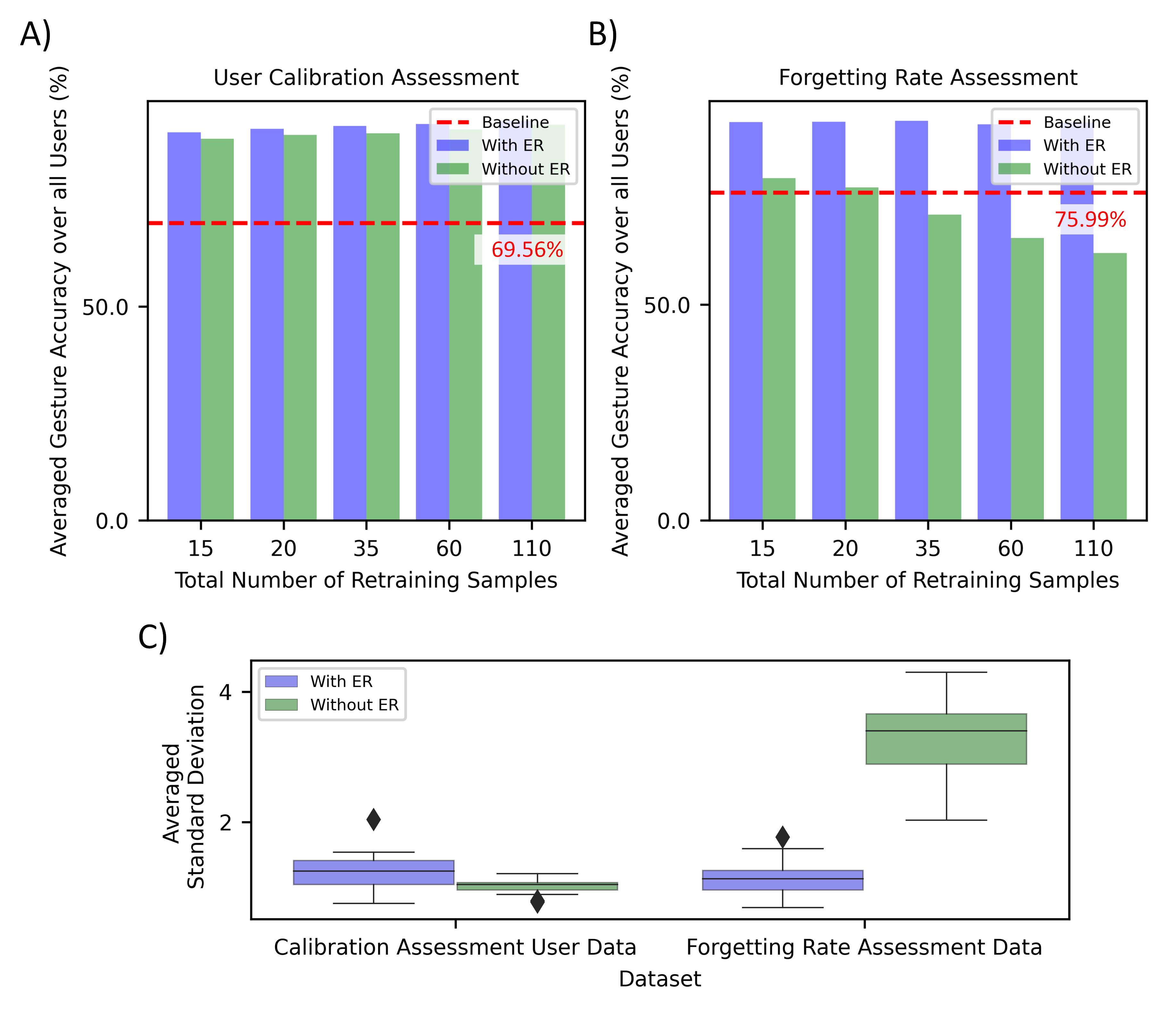}
    \caption{Comparison of gesture accuracy and model stability with (blue) and without ER (green) for user calibration and forgetting rate assessments. A) Comparison of gesture accuracy with and without ER for user calibration assessment. B) Comparison of gesture accuracy with and without ER for forgetting rate assessment. In both cases, the gesture accuracy with ER consistently outperforms the accuracy without ER across all retraining sample sizes. C)  Comparison of the STD of model performances with and without ER across calibration assessment user data and forgetting rate assessment data averaged over all configurations. These results show that incorporating ER into the model retraining can significantly improve gesture accuracy and model stability, particularly in scenarios with limited user data for calibration.}    
    
\label{fig:results-er}
 \end{figure}

Fig. \ref{fig:results-er} illustratively summarizes the performance of both approaches regarding user adaptation (panel A)) and forgetting rate panel B)).

\textbf{Robustness and Standard Deviation Analysis.} 
The robustness of the models is evaluated based on the \ac{std} of the gesture accuracy across different runs. The Tables \ref{tab:full-std-traindata-with-er}, \ref{tab:full-std-traindata-without-er}, \ref{tab:full-std-userdata-with-er} and \ref{tab:full-std-userdata-without-er}  as well as Fig. \ref{fig:results-er} panel C) allow for following key observations:

\begin{itemize}
    \item Standard Deviation on Calibration Assessment User Data:
    \begin{itemize}
        \item With ER: The \ac{std} is generally higher compared to the model without ER (averaged \ac{std} overall $n_\text{train}$-$n_{\text{user,}i}$: 1.24)
        \item Without ER: The \ac{std} is lower (averaged \ac{std} over all $n_{\text{user,}i}$: 1.00).
    \end{itemize}
    \item Standard Deviation on Forgetting Rate Assessment Data:
    \begin{itemize}
        \item With ER: The \ac{std} is lower compared to the model without ER (averaged \ac{std} overall $n_\text{train}$-$n_{\text{user,}i}$: 0.69)
        \item Without ER: The \ac{std} is higher (averaged \ac{std} over all $n_{\text{user,}i}$: 3.29).
    \end{itemize}
\end{itemize}

When using \ac{er}, the model is retrained on a combination of user data and samples from the original training data. This approach helps in adapting the model to the user while also retaining the knowledge from the original training data, reducing the forgetting rate. However, this process can introduce variability due to different reasons. The model must balance learning from the new user data while maintaining its original knowledge, which can lead to variability in how well the model adapts to different users across runs. The integration of user data and train data can create conflicts, leading to inconsistent performance in different runs, hence a higher \ac{std}.
Another reason is the impact of the gesture sample quality. When user recordings are limited, the quality of these samples becomes crucial. High-quality samples may lead to effective retraining, while poor-quality samples can result in less optimal model adaptation. This variability in the quality of user data can further contribute to the observed \ac{std} values, especially when combined with \ac{er}, where the model is also trying to retain previous knowledge.

Without \ac{er}, the model is retrained solely on the user data. This method focuses entirely on adapting the model to the user's gestures, without trying to retain knowledge from the original training data. This approach can lead to more consistent adaptation across different runs, resulting in a lower \ac{std}. 
However, the quality of the user data remains a significant factor. When user recordings are limited, poor-quality gesture samples can still lead to inconsistent retraining outcomes, but the effect is less pronounced compared to \ac{er}. Without the added complexity of balancing with training data, the model’s adaptation is more straightforward, leading to less variability in performance and, thus, a lower \ac{std}.

In the context of user data, the STD values reflect the differences in how the model adapts to user-specific gestures with and without ER. The main findings of the robustness analysis are that with ER, the model faces the challenge of balancing user adaptation with retaining prior knowledge, leading to higher variability in performance. Without ER, the model focuses entirely on the user data, resulting in more consistent performance and a lower \ac{std}. However, in both cases, the quality of the gesture samples plays a critical role, especially when the number of user recordings is low, contributing to the overall variability in model performance.

For the robustness analysis on the forgetting rate assessment data, the \ac{std} reflects the following factors.

When the model is retrained without \ac{er}, it focuses entirely on the new data. This means that any adaptation to new data comes at the cost of potentially forgetting some of the knowledge from the original training data, especially with longer training on new data. This forgetting can introduce variability in the calibrated model’s performance when being assessed with the forgetting rate assessment data.
Thus, the higher \ac{std} values of the calibrated model without ER reflect the variability introduced by the model’s potential forgetting and its complete reliance on the new user data for retraining, with the quality of this new user data being a key factor.

On the forgetting rate assessment data set, the \ac{std} values differ between using \ac{er} and not using it due to the differing strategies in retraining the model. With \ac{er}, the model tries to balance between retaining old knowledge and learning from new data, leading to moderate variability. Without \ac{er}, the model is more prone to forgetting the original data, resulting in higher variability, especially if the new data is inconsistent or of varying quality. In both cases, the quality of the new training data significantly impacts the model’s performance variability, especially when the data is limited.

In conclusion, models retrained with ER have higher \ac{std} values on the user dataset due to the re-exposure to varied and potentially noisy user data, increasing the model's sensitivity to this data.
On the other hand, \ac{er} shows reduced \ac{std} values on the forgetting rate assessment dataset. By mitigating the forgetting effect, a more consistent and robust performance on the seen training data can be observed.

In the following, the different retraining configurations  $n_\text{train}$-$n_{\text{user,}i}$ are analyzed in discussed.

\textbf{Effect of Increasing Training Recordings per Class}. Increasing the number of training recordings generally boosts performance, mainly when the number of user recordings is also high. This shows that a balance of general and user-specific data is beneficial. In Table \ref{tab:results-with-er}, some examples of retraining configurations are highlighted. The detailed table can be found in the appendix in Table \ref{tab:full-fr-results-with-er} For instance, $user_1$'s accuracy with ten training recordings and ten user recordings is $95.86\%$, which further improves to $97.05\%$ with 100 training recordings and 50 user recordings. As a more general example, when comparing the results of $n_\text{train}$-$n_{\text{user,}i} = $ 05-05 and 10-05, it is observable that increasing the number of training recordings per class from five to ten improves the accuracy for most users. Similarly, comparing the results of 10-05 and 25-05, we can see that further increasing the number of training recordings per class to 25 improves the accuracy for most users. 
However, the improvement in accuracy tends to plateau as the number of training recordings per class increases beyond a certain point. For example, the difference in accuracy between 25-05 and 50-05 is relatively minor for most users.

\begin{table}[h]
\scriptsize 
\label{tab:results-with-er}
\begin{tabular}{l|l|cccc:ccc:ccc:c:c}
\hline
\multirow{2}{*}{User $i$} & \multirow{2}{*}{Baseline} & \multicolumn{12}{c}{$n_\text{train}$-$n_{\text{user,}i}$}                                 \\ \cline{3-14} 
&                           & \multicolumn{1}{l}{05-05} & \multicolumn{1}{l}{05-10} & \multicolumn{1}{l}{05-25} & \multicolumn{1}{l}{05-50} & \multicolumn{1}{l}{10-05} & \multicolumn{1}{l}{10-10} & \multicolumn{1}{l}{10-25} & \multicolumn{1}{l}{25-05} & \multicolumn{1}{l}{25-50} & \multicolumn{1}{l}{25-100} & \multicolumn{1}{l}{50-05} & \multicolumn{1}{l}{100-50} \\ \hline
1 & 88.26 & 93.91 & 93.43 & 95.39 & 94.21 & 95.39 & 95.86 & 95.25 & 96.28 & 96.88 & 96.87 
  & 96.45 & 97.20 \\
2 & 79.59 & 89.10 & 91.02 & 91.59 & 92.77 & 91.00 & 92.21 & 93.44 & 92.19 & 93.17 & 94.63 
  & 91.71 & 93.77\\
3 & 18.59 & 57.45 & 68.31 & 76.49 & 78.98 & 76.91 & 77.43 & 78.94 & 72.82 & 79.48 & 82.38 
  & 69.43 & 79.99  \\
6 & 73.37 & 92.27 & 94.11 & 95.95 & 94.85 & 94.82 & 96.00 & 96.10 & 95.27& 96.65 & 97.28 & 94.85 & 98.23 \\
7 & 67.12 & 86.64 & 89.25 & 89.38 & 90.19 & 89.84 & 89.81 & 92.61 & 88.67& 92.15 & 93.68 & 91.72 & 92.86 \\
8 & 60.78 & 81.83 & 87.57 & 87.26 & 90.67 & 87.55 & 90.21 & 89.51 & 91.53 & 92.43 & 92.31 & 87.63 & 91.41 \\
9 & 76.42 & 91.67 & 92.54 & 93.91 & 94.63 & 94.69 & 94.73 & 95.77 & 94.76 & 96.52 & 97.36 & 94.95 & 97.12 \\
10 & 73.76 & 84.98 & 88.15 & 89.32 & 89.68 & 89.70 & 90.03 & 90.76 & 89.56 & 92.33 & 92.41 & 90.12 & 93.16 \\
11 & 75.30 & 89.62 & 91.23 & 91.94 & 92.94 & 91.54 & 92.88 & 93.04 & 91.97 & 94.34 & 94.89 & 91.94 & 94.46 \\
12 & 82.40 & 93.43 & 93.98 & 94.06 & 94.57 & 95.94 & 96.11 & 96.45 & 95.94 & 96.96 & 97.54 & 95.93 & 97.39 \\
\hline
\textbf{Mean} & 69.56 & 86.09 & 88.96& 90.53 & 91.35 & 90.74 & 91.53& 92.19 & 90.90 & 93.09 & 93.93 & 90.47 & 93.56 \\
\hline
\end{tabular}
\caption{Model performance on calibration assessment user data with ER (relevant parts)}
\end{table}

\begin{figure}
\centering
    \includegraphics[width = 1\linewidth]{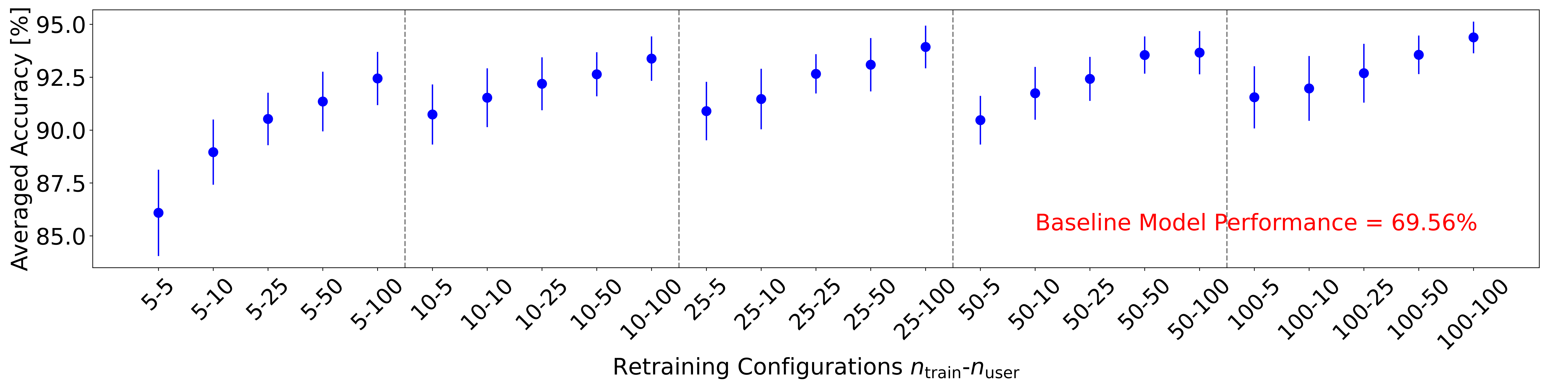}
    \caption{Error bar plot illustrating the mean performance of models across varying combinations of training data size $n_{\text{train}}$ and user data size $n_{\text{user}}$ (averaged over all users). Error bars represent the standard deviation of performance across multiple runs, indicating the robustness and variability of the model's gesture accuracy.}
    \label{fig:error-plot-userdata-er}
 \end{figure}

\textbf{Effect of Increasing User Recordings per Class}. The results show that increasing the number of user recordings per class generally improves the model performance. This is highlighted in Fig. \ref{fig:error-plot-userdata-er}.
For example, when comparing the results of 05-05 and 05-10, increasing the number of user recordings per class from 5 to 10 improves the accuracy for most users. 
Similarly, when comparing the results of 05-10 and 05-25, we can see that further increasing the number of user recordings per class to 25 improves the accuracy for most users.
Specifically, $user_2$'s accuracy increases from $89.10\%$ with five training recordings and five user recordings to $94.63\%$ with 25 and 100 user recordings.
However, the accuracy improvement tends to plateau again as the number of user recordings per class increases beyond a certain point. For example, the difference in accuracy between 05-25 and 05-50 is relatively minor for most users.

\textbf{User-Specific Variability.} Different users benefit differently from the calibration process, highlighting the importance of user-specific adaptation.
$user_3$, who had a low baseline accuracy of $18.59\%$, showed substantial improvement, reaching up to $82.38\%$ with 25 training recordings and 50 user recordings.
On the other hand, users like $user_1$ and $user_6$ already had high baseline accuracies ($88.26\%$ and $73.37\%$, respectively) and saw even more significant improvements, with $user_1$ reaching a maximum of $97.66\%$ and $user_6$ $98.78\%$.

\textbf{Optimal Combination of Training and User Recordings per Class.} The results suggest that the optimal combination of training and user recordings per class varies across users.
For example, for $user_2$, the best performance is achieved with ten training recordings per class and 25 user recordings per class (10-25).
For $user_9$, the best performance is achieved with 25 training recordings per class and 50 user recordings per class (25-50).
For $user_{10}$, the best performance is achieved with ten training recordings per class and ten user recordings per class (10-10).
This suggests that the optimal combination of training and user recordings per class depends on the individual user's characteristics and behavior.

Overall, the following conclusions can be drawn:
\begin{enumerate}
    \item Experience replay effectiveness: Models calibrated with ER consistently outperformed the baseline models across all users when a higher number of user gestures were used for calibration. The average baseline gesture accuracy of $69.56\%$ increased to a range of $84.73\%$ to $93.89\%$ with ER. For a low number of user gestures, the calibration without ER showed slightly better performance. However, this approach lacked generalizability and was prone to overfitting.
    \item Model robustness: Models retrained with \ac{er} demonstrate increased \ac{std} on user datasets, reflecting higher sensitivity to the varied and potentially noisy user data. However, ER significantly reduces the \ac{std} on the forgetting rate assessment dataset, indicating more consistent and robust performance on previously seen training data.
    \item Participant variability: While most participants benefit from ER, the degree of improvement varies. High initial performers (like $user_1$ and $user_6$) continue to excel, while low performers (like $user_3$) show substantial gains, though their performance still lags behind the high performers.
    \item Effect of training and user recordings on model performance: The results suggest that the model's performance can be improved by increasing the number of training recordings and user recordings per class. However, the improvement in accuracy tends to plateau as the number of recordings increases beyond a certain point.    
    \item Tailored calibration strategies for users: While ER helps, participants with initially low performance might benefit from additional strategies, such as targeted fine-tuning, to further boost their accuracy. Additionally, the optimal combination of training and user recordings per class could be explored further to develop a more personalized and effective calibration approach.
\end{enumerate}

This analysis indicates that \ac{er} is highly effective in improving model performance and mitigating the forgetting problem, especially when training involves new data being introduced incrementally.

\subsubsection{Diverse Training Data for Robust Gesture Recognition}
\label{subsubsec:diverse}
\begin{figure}
\centering
    \includegraphics[width = 1\linewidth]{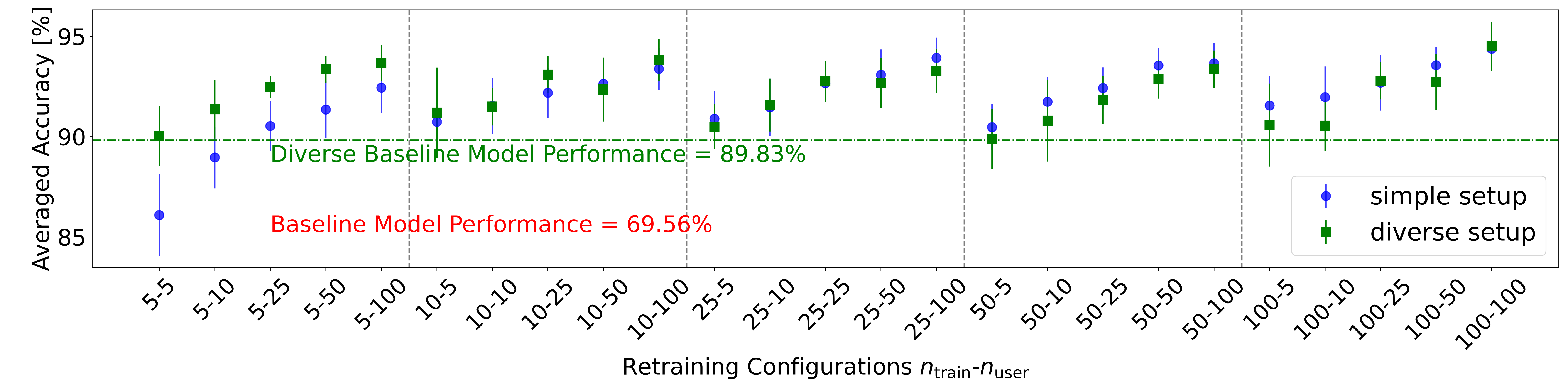}	
    \caption{Error bar plot illustrating the mean performance of two gesture recognition models (simple setup and diverse setup) across varying combinations of $n_\text{train}$-$n_{\text{user,}i}$ (averaged over all users user$_i$). The horizontal red dashed line marks the baseline gesture accuracy of the diverse uncalibrated model at $89.93\%$}. Error bars represent the standard deviation of performance across multiple runs, indicating the robustness and variability of each model's gesture accuracy.
    \label{fig:diverse-vs-simple-setup}
 \end{figure}

The performance of gesture recognition models can be significantly enhanced by using diverse training data.
As demonstrated in Fig. \ref{fig:diverse-vs-simple-setup}, the diverse baseline model achieves a substantial improvement in gesture accuracy, from $69.56\%$ to $89.83\%$, averaged over all $user_i$. 

The diverse baseline model proves highly effective when user recordings for calibration are limited, offering a strong starting point for model adaptation, as evidenced by lower \ac{std} values. However, as more user gesture samples are added to the retrain dataset, the performance of the simple setup converges with, and even surpasses, that of the diverse setup. This may be because the simple setup, unexposed to diverse data distributions during training, can better specialize to user gestures when sufficient data is available. These findings highlight that while diverse training data enhances robustness and accuracy for limited data scenarios, the trade-off between model complexity and data availability must be carefully considered, as overfitting can reduce the advantage of the diverse setup in data-rich conditions.

\subsection{Anomaly Detection Results}
\label{subsec:vae-results}
The previous section discussed the results of the GRU-based \ac{rnn}, while in this section, experiments for anomaly detection using a \ac{vae}-based architecture are evaluated. Following this, the anomaly reasoning capabilities of the proposed pipeline in XentricAI are assessed.

As mentioned before, anomalous gestures from multiple users were collected. By knowing the actual types of anomalies, the results of the anomaly reasoning provided by XentricAI were compared against the ground truth.

For further analysis, we adhered to the configuration of $n_\text{train}$-$n_{\text{user,}i}=$ 50-10. This choice is based on two considerations. First, the AI Act emphasizes limiting the amount of data collected to what is necessary, reducing the potential for misuse or accidental exposure of personal information. Second, the calibration effort required from users should be minimal. Asking a user to perform 100 gestures per class can be tedious, whereas 10 gestures are more manageable. We chose 50 training gestures because the performance improvement plateaus beyond 50 gestures, with no significant increase observed when moving from 50 to 100 gestures, compared to the jump from 25 to 50 gestures.
The primary task of the VAE module is to increase the number of flagged anomalies, particularly targeting misclassified or unpredicted gestures that were characterized but not automatically identified in previous work \citep{seifi2024xentricai}. 

Table \ref{tab:performance-anomalous-gestures} presents the performance of the baseline model and the calibrated model with ER on the anomalous gesture dataset, averaged over all users. After calibrating the model to each user, the performance on that user's anomalous data was evaluated.

While the dynamic gesture accuracy slightly improves from the baseline model to the calibrated model, there is a notable $6.45\%$ increase in gesture accuracy, demonstrating that model calibration enhances the model's performance for anomalous gestures.
\begin{table}[h]
\scriptsize
\centering
\label{tab:performance-anomalous-gestures}
\begin{tabular}{l|ccc}
\Xhline{2\arrayrulewidth}
 & \begin{tabular}[c]{@{}c@{}}Accuracy{[}\%{]}\end{tabular} & \begin{tabular}[c]{@{}c@{}}Gesture Accuracy {[}\%{]}\end{tabular} & \begin{tabular}[c]{@{}c@{}}Dynamic Gesture Accuracy {[}\%{]}\end{tabular} \\ \hline
Baseline Model  & 93.20  & 42.29  & 45.83   \\
Calibrated Model with ER & 93.75 & 48.74 & 47.24 \\ \hline
Improvement  & 0.55 & 6.45 & 1.41\\ 
\Xhline{2\arrayrulewidth}
\end{tabular}
\caption{Performance comparison on anomalous gesture dataset.}
\end{table}

In Table \ref{tab:vae-improvement}, we present the averaged improvement in the number of characterizable gestures using the VAE. Each user's results were averaged over six individual experimental runs, and these results were then further averaged across all users to obtain the final outcomes.

To validate the effectiveness of the tailored VAE, we conducted a comparative analysis against four established anomaly detection methods, which are one-class \ac{svm}, isolation forest, local outlier factors, and GANs. Across all comparisons, the VAE demonstrated superior performance, consistently outperforming these alternative approaches. Detailed results of this comparative analysis are provided in \ref{app:anomal-detection-methods}.

We evaluated the performance of the VAE by measuring the increase in the detection of misclassified or unpredicted gestures that were actually mispredicted by the GRU-based \ac{hgr} model but were not flagged by predefined criteria (sparse predictions or mixed-class predictions). The first column of Table \ref{tab:vae-improvement} shows the percentage of correctly condition-flagged gestures, representing the percentage of correctly flagged gestures that were anomalies to the GRU system before using the VAE, which amounted to $57.60\%$. The second column indicates the percentage of correctly flagged gestures that were anomalies to the GRU system with the VAE, including both condition-flagged and exclusive VAE-flagged anomalies. With the help of the VAE and the user-specific thresholding technique, $11.50\%$ more unpredicted and misclassified gestures were detected and subsequently characterized on average.

\begin{table}[h]
\label{tab:vae-improvement}
\centering
\scriptsize
\begin{tabular}{cc|c}
\Xhline{2\arrayrulewidth}
\begin{tabular}[c]{@{}c@{}}Correctly Flagged Gestures {[}\%{]}\\ (Condition-Flagged)\end{tabular} & \begin{tabular}[c]{@{}c@{}}Correctly Flagged Gestures {[}\%{]}\\ (Condition-Flagged+VAE)\end{tabular} & Percentage Increase \\ \hline
57.60 & 69.10 & 11.50   \\ \Xhline{2\arrayrulewidth}
\end{tabular}
\caption{Improvement in detecting characterizable gestures using the tailored VAE-based anomaly detection method}.
\end{table}

The results demonstrate that the VAE module enhances the detection and characterization of anomalous gestures, especially those that were misclassified or unpredicted by the GRU-based model. The $6.45\%$ increase in gesture accuracy and the $11.50\%$ increase in correctly flagged gestures highlight the effectiveness of incorporating \ac{tl} and the VAE into the XentricAI framework. This improvement ensures that more anomalies are accurately identified, providing better feedback to users and enhancing the overall reliability of the \ac{hgr} system.

\subsection{Gesture Characterization Results}
\label{subsec:xai-results}
After calibrating the model, we also aim to provide feedback in cases of anomalous gestures. Whenever the model fails to properly predict, the user is asked for input regarding the actual gesture performed. Based on the real gesture, the SHAP values of the anomalous gesture are analyzed. Using these SHAP values, XentricAI identifies the potential cause of the anomaly and provides feedback to the user accordingly

After flagging a gesture as anomalous, gesture characterization was performed. For this purpose, a set of nominal gestures used for model calibration was utilized to retrieve their global feature attributions. 

Based on these feature rankings, the \ac{srv} for each individual user were determined. Minimum and maximum SHAP values for each feature and gesture class within the nominal feature rankings were obtained and used as lower and upper bounds for gesture characterization.

Two different quantities of nominal feature rankings (i.e., 10 and 100) were used to derive SRVs, but no significant difference in gesture characterization was observed. Consequently, a lower number of gestures, consistent with those used for model calibration, was adopted for further analysis.

\begin{figure}[htpb]
    \centering
    \includegraphics[width = 1\linewidth]{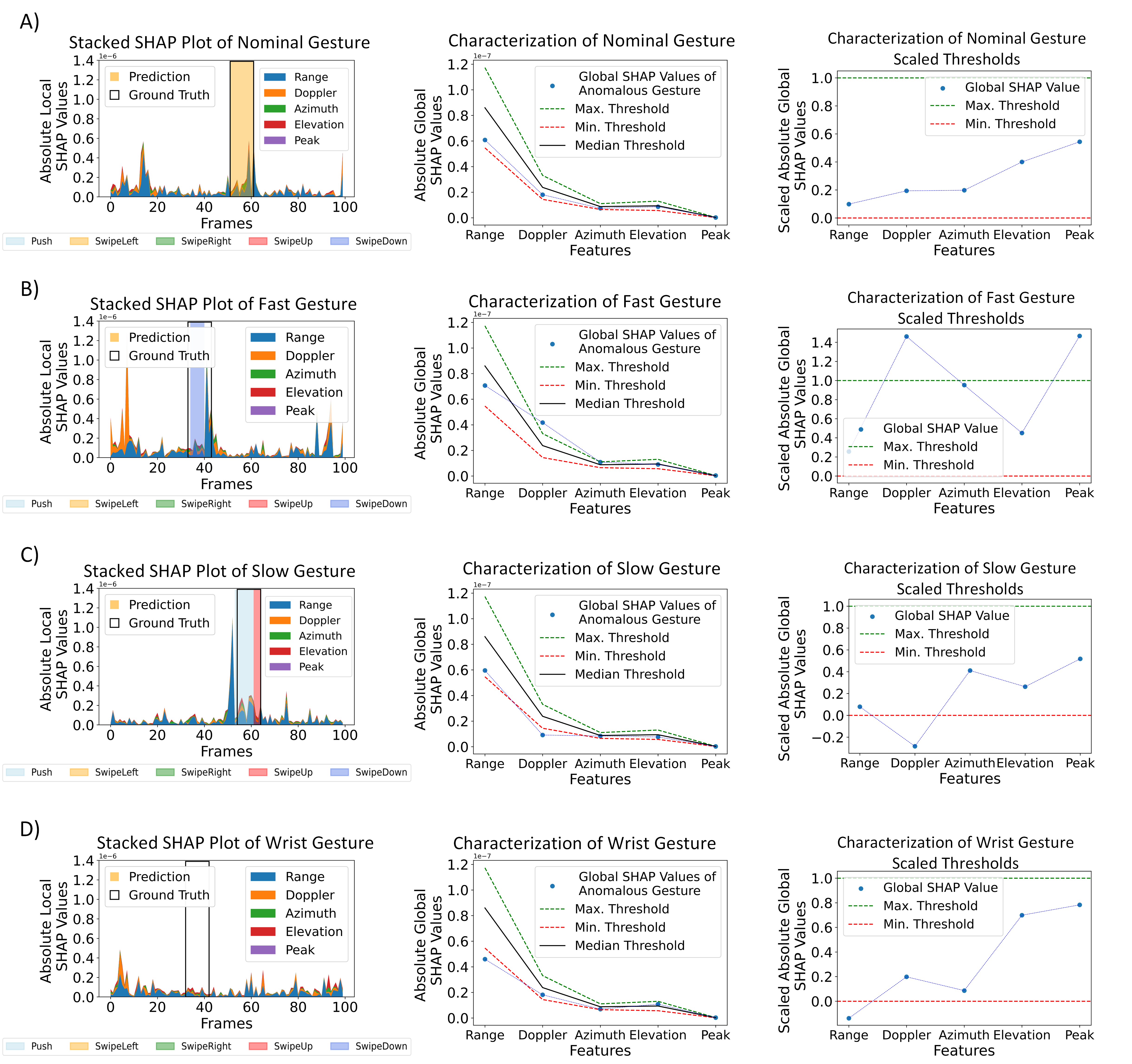}	
    \caption{Gesture characterization of a SwipeLeft gesture. A) The gesture characterization of a nominal SwipeLeft gesture is shown. B) Gesture characterization of a fast-paced gesture misclassified as a SwipeUp gesture. C) Gesture characterization of a slow-paced gesture misclassified as mixed-class predictions. D) Gesture characterization of an undetected wrist execution.}
    \label{fig:gesture-characterization}
 \end{figure}

For each user, the method’s effectiveness in characterizing gestures performed at slow and fast speeds, as well as wrist executions, was assessed. The analysis focused on how many flagged anomalous gestures exhibited deviations in feature rankings that enabled characterization, and how well these deviations contributed to understanding the anomalies. Additionally, XentricAI’s ability to handle mixed-class predictions, fully incorrect gestures, and undetected gestures across all three anomaly types and gesture classes was evaluated.
Fig. \ref{fig:gesture-characterization} illustrates the gesture characterization of a SwipeLeft gesture performed at different speeds (fast (panel B), slow (panel C)), and execution styles (wrist instead of extended arm (panel D)) by $user_2$.

The left column of panel A-D presents the absolute local SHAP values for each feature over time. The gesture period is marked with a black box labeled 'Ground Truth,' and the model’s predictions are color-coded by gesture class. Panel A depicts a correctly classified gesture, while Panel B and C display misclassifications, and Panel D shows an undetected gesture.

The middle and right columns of the remaining panels display the minimum, maximum, and median thresholds derived from the user’s input on the actual gesture class. The middle column shows actual SHAP values and the right column presents standardized SHAP values based on nominal SRVs.

In panel A) a correctly classified gesture is shown. This row indicates that the range feature has the highest influence, followed by Doppler. Depending on the performed gesture, azimuth or elevation ranks next. The median threshold reveals that the range-Doppler relationship should exhibit a downward slope.

Panel B) shows a fast-paced gesture misclassified as SwipeDown instead of SwipeLeft. Since this gesture was flagged as anomalous, user confirmation of the gesture was prompted. The gesture characterization revealed that the SHAP value of the Doppler feature exceeded the nominal upper limit, and the expected downward slope between range and Doppler was absent. This pattern, including elevated Peak feature values, is characteristic of fast-paced gestures. Feedback to the user suggested performing gestures more slowly for improved recognition.

Panel C) displays a slow SwipeLeft gesture misclassified as multiple gestures, so-called mixed-class predictions. Gesture characterization indicated that slow gestures consistently show lower Doppler SHAP values compared to nominal gestures.

Finally, Panel D) presents a SwipeLeft gesture performed with the wrist rather than the full arm, resulting in the gesture being undetected. Characterization indicated that the range feature values were lower than those of nominal SwipeLeft gestures, suggesting that performing gestures closer to the radar device enhances detection.

It is important to note that deviations in global SHAP values from nominal thresholds were consistently observed across all gesture classes and users when anomalies occurred. There were no significant differences in results between different gesture types, and the execution location had no impact on XentricAI's performance. Feedback was provided based on the specific type of deviation identified.

While the examples demonstrate the effectiveness of XentricAI's explainable AI mechanism in characterizing gestures, some issues persist. The thresholding mechanism successfully identified deviations in 97.5\% of anomalously flagged gestures, indicating a well-functioning system. However, the characterization of certain gestures, particularly wrist executions and mixed-class predictions, is not always as clear. In these cases, features like azimuth, elevation, or their combinations deviated from the norm in ways dependent on the incorrectly predicted gesture class. Recent research has shown that unclear explanations of AI models can lead to a lack of trust in AI systems \citep{schmidt2020transparency}. Therefore, it is important to further enhance the robustness and reliability of these explanations. One possibility would be to consider not just the absolute SHAP values but also the direction (positive or negative) of these values and the predicted gesture class. A positive SHAP value indicates that a feature contributed correctly to the prediction, while a negative SHAP value suggests that the feature negatively impacted the model, pushing it toward the wrong prediction. Incorporating this analysis could significantly improve the reliability and robustness of the explanations. 

Like all feature attribution methods, SHAP explanations are true to the model rather than true to the data \citep{chen2020true}. This means that these explanations reflect the behavior of the model, not the actual ground truth. Therefore, if the model itself is not robust and accurate, the explanations provided will not be reliable in understanding real-world scenarios. This issue became apparent when using both simple and diverse training datasets. Although the model performed better overall with the diverse dataset, the characterization of anomalous gestures became less distinct, as the model struggled to identify clear deviations in feature rankings. The diverse training data, incorporating a wide range of user behaviors, made it more challenging to reason clearly about flagged anomalies. This highlights a trade-off between model performance or data complexity and the transparency or explainability of the system, a typical obstacle as mentioned in Section \ref{subsec:obstacles}.

In conclusion, XentricAI's explainable AI mechanism proves highly effective in characterizing anomalous gestures, successfully identifying deviations in the vast majority of cases. This capability enhances transparency and aligns with the EU AI Act's focus on making AI systems understandable and trustworthy. While there is room for further refinement, especially in handling complex cases, XentricAI is well-positioned to meet the Act's standards, offering a robust and user-centric approach to AI transparency.

To conclude the experimental section, Table \ref{tab:experiments-overview} summarizes all the experiments conducted in this work and references the corresponding sections.

\begin{table}[h]
\scriptsize
\centering
\label{tab:experiments-overview}
\begin{tabular}{l|l|l}
\Xhline{2\arrayrulewidth}
\multicolumn{1}{c|}{Experiment}  & \multicolumn{1}{c|}{Section}                                                 & \multicolumn{1}{c}{Performance Indicators}                                     \\ \Xhline{2\arrayrulewidth}\hline
Model Calibration                & \ref{subsec:model-calibration-results} & effect of retrain configurations, accuracy, forgetting rate, robustness        \\ \hline
Anomaly Detection                & \ref{subsec:vae-results}                                    & accuracy metrics, detection success rate                                       \\ \hline
Gesture Characterization         & \ref{subsec:xai-results}                                    & nominal feature ranking configurations, detection success rate, effectiveness \\ \hline 
\end{tabular}
\caption{Summary of experiments and corresponding sections.}
\end{table}

\subsection{Evaluation of XentricAI's Conformance with the European AI Act: Transparency, Human Oversight, Privacy, and Robustness}
\label{subsec:xentric-and-aiact}

XentricAI aligns with the principles outlined in the EU AI Act, addressing key areas such as transparency, human oversight, privacy, and data governance. 

The EU AI Act defines obligations based on risk categories, but these obligations remain abstract and do not include specific benchmarks, metrics, or tests for compliance. For instance, the Act mandates "appropriate traceability and explainability" but does not clarify what qualifies as "appropriate" in technical terms. This lack of specifity creates challenges for practitioners aiming to meet these requirements \citep{guldimann2024compl}.

\begin{figure}[htbp]
    \centering
    \includegraphics[width = 1\linewidth]{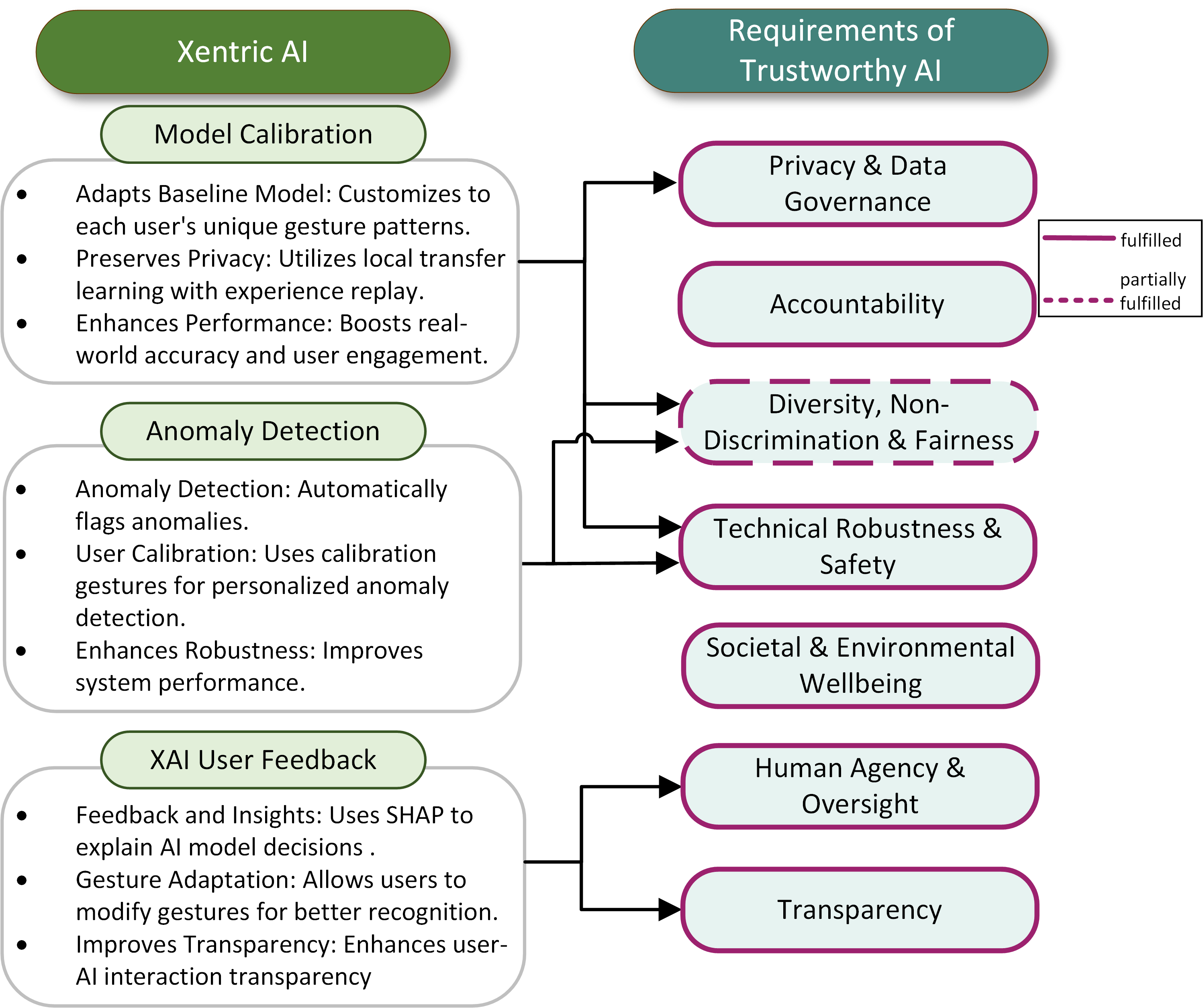}	
    \caption{Mapping of XentricAI's conformance with the European AI Act's principles, including transparency, human oversight, privacy, and robustness. The figure illustrates how XentricAI's design and features align with the AI Act's requirements, highlighting areas of conformance and opportunities for future development.}
    \label{fig:eu_ai_act_mapping}
 \end{figure}

To address this gap, the European Commission has tasked standardization bodies like CEN and CENELEC with developing standards for trustworthy AI to translate these obligations into actionable tests\footnote{https://www.cencenelec.eu/areas-of-work/cen-cenelec-topics/artificial-intelligence/}. Although formal benchmarks have yet to be established, XentricAI proactively incorporates emerging best practices from these standardization bodies to align with the current guidelines. Table \ref{tab:regulatory_compliance} systematically maps specific articles of the EU AI Act to the corresponding requirements for high-risk AI systems and lists the potential compliance measures that can be taken to comply with the requirement. Measures taken by XentricAI are highlighted in bold in Table \ref{tab:regulatory_compliance} and illustrated in Fig. \ref{fig:eu_ai_act_mapping} and discussed throughout this section.

\begin{table}[h]
\centering
\scriptsize 
\begin{tabular}{|>{\raggedright}p{4cm}|p{3.5cm}|p{7cm}|}
\hline
\textbf{Regulatory Reference} & \textbf{Regulatory Requirement} & \textbf{Compliance Measure} \\ \hline
Risk Management System (Art. 9) & Risk Management & - Risk management across the ML lifecycle.\newline - Test, monitor, and mitigate risks continuously.\newline - Maintain risks within acceptable levels. \\ \hline
Data and Governance (Art. 10) & Data Quality & - \textbf{Ensure data completeness, accuracy, and relevance.}\newline - Validate statistical properties for fairness and diversity.\newline \textbf{- Use automated tools for quality checks. } \\ \hline
Data and Governance (Art. 10) & Bias Mitigation & - Audit for bias (e.g., social, temporal, representation).\newline - Apply mitigation techniques (e.g., re-sampling, re-weighting). \\ \hline
Technical Documentation (Art. 11) & Technical Documentation & \textbf{- Document system design, components, and development process.}\newline \textbf{- Include testing methods and intended use cases.} \\ \hline
Record-keeping (Art. 12) & Record-keeping & - Maintain logs for operations, risks, and user interactions.\newline  - Ensure traceability for monitoring and verification. \\ \hline
Transparency and Provision of Information to Deployers (Art. 13) & Provision of Information & \textbf{- Provide clear user manuals covering capabilities and limitations.}\newline \textbf{- Tailor instructions to users’ expertise levels.} \\ \hline
Transparency and Provision of Information to Deployers (Art. 13) & Explainability and Interpretability (Model) & - Use feature importance ranking tools (e.g. \textbf{SHAP}, LIME, saliency maps).\newline - Use interpretable pre-hoc or post-hoc methods.\newline - Provide \textbf{global/local}/contrastive explanations.\\ \hline
Human Oversight (Art. 14) & Human Oversight & \textbf{- Include human-in-the-loop mechanisms.}\newline - Provide manual override and monitoring capabilities.\newline - Adjust oversight level to system risks. \\ \hline
Accuracy, Robustness and Cybersecurity (Art. 15) & Model Accuracy (Fairness) & - Use metrics like disparate impact ratio, statistical parity difference, and average odd difference.\newline - Conduct regular fairness audits and implement mitigation strategies. \\ \hline
Accuracy, Robustness and Cybersecurity (Art. 15) & Model Accuracy (Quality metrics) & \textbf{- Use metrics like accuracy, precision, recall, ROC curves, and confusion matrices.}\newline \textbf{- Regularly validate performance with real-world data.} \\ \hline
Accuracy, Robustness and Cybersecurity (Art. 15) & Model Robustness & \textbf{- Handle distribution shifts} and concept drift.\newline - Apply uncertainty estimation and adversarial testing. \newline \textbf{- Include feedback loops} \\ \hline
Accuracy, Robustness and Cybersecurity (Art. 15) & Model Cybersecurity & - Protect against data and model poisoning. \\ \hline
\end{tabular}
\caption{Potential compliance measures for regulatory requirements in the EU AI Act}
\label{tab:regulatory_compliance}
\end{table}

XentricAI is designed to prioritize transparency in the form of explainability by providing users with feedback and insights into its decision-making process. Achieving full transparency is typically associated with white-box models, as discussed in previous work \citep{seifi2024mira}, where rule-based systems classify gestures in an interpretable manner. Nevertheless, XentricAI, which uses a black-box \ac{hgr} model, makes a significant effort to be as transparent as possible. By explaining its predictions in the form of feature importance rankings using SHAP and maintaining a user-centric approach, the system aligns with the AI Act's emphasis on ensuring that AI systems are understandable and accessible to end-users.

XentricAI effectively incorporates human agency and oversight into its design. The system is designed to actively involve users in the decision-making loop, asking them to record and confirm gestures, and offering feedback. This approach ensures that users can monitor, intervene, and control AI decisions when necessary, fulfilling the AI Act's criteria for human oversight.

Regarding privacy and data governance, XentricAI adheres closely to the AI Act's principles. The system implements data minimization. We systematically evaluated the ideal balance between training samples ($n_\text{train}$) and user-specific recordings ($n_{\text{user},i}$) to reduce the calibration requirements, ensuring data efficiency. By collecting only the necessary data for model calibration, it ensures that this data is processed locally at the end-user site, without being shared externally. 
This design minimizes the risk of data misuse or accidental exposure, aligning with the AI Act’s requirements for protecting personal data and giving individuals control over their information.

While XentricAI strives to promote non-discrimination by tailoring its model to individual users' distinct gesture patterns, a crucial aspect that warrants further development is the evaluation of model fairness and bias. This oversight could be addressed to better align XentricAI with the AI Act's principles. Additionally, the study did not explore the critical dimension of trust and trustworthiness, specifically how end-users perceive and trust the explainable mechanisms of XentricAI. This gap highlights a potential area for future investigation.

To evaluate model accuracy, we implemented three metrics: the common accuracy metric, gesture accuracy, and dynamic gesture accuracy. Each metric reflects different aspects of performance, such as overall model performance, sensitivity to specific gestures, and real-world application effectiveness. For a detailed description, refer to Section \ref{subsec:model-training}.

In terms of robustness and reliability, XentricAI customizes the baseline model to each user's gesture execution patterns to ensure accurate predictions. By extending the existing FMCW radar-based gesture dataset with OOD data, XentricAI evaluates the algorithm's ability to handle diverse and unexpected inputs. While some anomalous gestures are correctly classified, others that are misclassified benefit from insights into the model's decision-making process. To improve model robustness and account for data distribution shits, TL and catastrophic mitigation techniques such as ER were applied. A user-specific VAE module was introduced, enabling the system to flag undetected or misclassified gestures as anomalous. This enhances automated and reliable anomaly detection.
The EU AI Act defines robustness as not just performing well and reliably across diverse datasets but also accurately assessing the model's confidence and uncertainties in its predictions, whether these arise from data variability (aleatoric) or from the model’s limitations (epistemic).
XentricAI currently lacks uncertainty quantification, which is crucial for fully meeting the AI Act's requirements in this area. Incorporating uncertainty measures in future iterations could ensure that the model handles real-world data variability more effectively and provides accurate confidence levels in its predictions.

Regarding the principle of societal and environmental well-being, XentricAI promotes small and efficient neural networks for HGR. The system enhances social skills, such as communication and interaction, by providing a user-centric HGR system.
By reporting on the algorithm, and conducting extensive experiments, evaluations, and academic research, XentricAI ensures accountability. We report on how the AI model makes decisions and discuss the inherent trade-offs between model performance and explainability. By providing transparency about the data used, the data preprocessing steps, and the algorithm design, and by enabling user feedback and thorough assessment of the algorithm, we comply with accountability principles.
These two principles, accountability, and societal and environmental well-being cannot be mapped to a specific building block of XentricAI but are rather addressed by the algorithm as a whole. Therefore, in Figure \ref{fig:eu_ai_act_mapping}, these principles are indicated as being met without any specific arrows connecting them to individual building blocks of the algorithm.

To extend XentricAI's approach beyond \ac{hgr}, the system's compliance with the EU AI Act's principles provides a foundation for its application across various AI domains. Key principles such as user-centered oversight, transparency, and data governance align with the EU AI Act’s mission to promote safe and accountable AI. For example, XentricAI's use of local SHAP values for anomaly characterization ensures the model is explainable, which is crucial in fields like healthcare diagnostics, where understanding model decisions is essential.

XentricAI also prioritizes human oversight, enabling end-users to verify and adjust model decisions. This feature is especially valuable in sectors like finance or criminal justice, where AI outputs directly impact individual rights. Additionally, its data minimization approach aligns with the EU AI Act's data governance standards, making the system suitable for privacy-sensitive fields like personalized healthcare or legal AI applications. 

The system’s robustness is enhanced through tailored calibration to user-specific inputs, making it adaptable for contexts that require accuracy and responsiveness, such as predictive maintenance. By following these best practices, XentricAI not only adheres to the principles of the EU AI Act but is also positioned for future compliance as formal standards are developed.

 This alignment exemplifies how expanding these principles to other domains will ensure that AI systems perform effectively. It also ensures that these systems uphold the EU AI Act’s principles, promoting technical robustness and transparency in applications that impact society.

\section{Conclusion}
\label{sec:conclusion}

In this study, we presented significant advancements in XentricAI, an explainable and user-centric hand gesture recognition (HGR) system, designed to comply with the EU AI Act. Our work addresses critical challenges in HGR, including the opacity of black-box models and the handling of distributional shifts in real-world data. By integrating a variational autoencoder module with user-specific dynamic thresholding into the XentricAI pipeline, we enhanced the system's anomaly detection capabilities. This led to the identification of $11.50\%$ more anomalous gestures.

Additionally, we expanded an existing radar-based HGR dataset, adding 28,000 new gestures from multiple users across varied locations, including 24,000 anomalous gestures. This comprehensive dataset enabled robust training and evaluation of our model, enhancing the system's overall diversity and robustness. We demonstrated a high success rate of $97.50\%$ in characterizing anomalies, significantly improving the system's explainability and reliability.

Furthermore, implementing transfer learning techniques using experience replay led to improvements in user adaptability, with an average performance increase of at least $15.17\%$. This adaptability ensures that XentricAI can provide personalized and accurate gesture recognition across different users and environments.

The advancements presented in this work contribute to the development of trustworthy AI systems by combining technical innovation with regulatory compliance. XentricAI not only meets the obligations of the EU AI Act but also offers a commercially viable solution for industrial applications. 
Our findings underscore the importance of explainable AI in machine learning applications and demonstrate the potential of advanced anomaly detection techniques and user-specific calibration in improving the performance and reliability of HGR systems.

\section*{Declaration of Interests}

The authors declare that they have no known competing financial interests or personal relationships that could have appeared to influence the work reported in this paper.

\section*{Data Availability}

The data has been made publicly available \citep{radar_data}.

\section*{Declaration of Generative AI and AI-Assisted Technologies in the Writing Process}

During the preparation of this work the authors used ChatGPT and Grammarly in order to improve writing clarity and coherence. After using this tool/service, the authors reviewed and edited the content as needed and take full responsibility for the content of the published article.


\bibliographystyle{elsarticle-harv}
\bibliography{references.bib}

\appendix
\section{Quantitative Evaluation of Catastrophic Forgetting Mitigation Approaches}
\label{app:catastrophic-forgetting}

To further validate the effectiveness of Experience Replay (ER) in mitigating catastrophic forgetting, we conducted a comparative evaluation of ER, Elastic Weight Consolidation (EWC), and Synaptic Intelligence (SI) on their effectiveness in mitigating catastrophic forgetting in gesture recognition tasks. 
These methods were assessed under varying configurations, measuring gesture accuracy and their ability to adapt to user-specific data, with the evaluation conducted on two fronts: (1) calibration assessment user data to measure transfer learning capabilities and (2) forgetting rate assessment data to assess the forgetting rate.
The proportion of the original train dataset and the user dataset to be mixed is governed by a carefully chosen regularization coefficient, $n_\text{train}$-$n_{\text{user,}i}$, where $n_\text{train}$ represents the number of training gesture samples per class and $n_{\text{user,}i}$ denotes the number of user $i$ gesture samples per class.
Tables \ref{tab:appendix-cma-caud} and \ref{tab:appendix-cma-frad} summarize the results, presenting the averaged gesture accuracy across all users and six experimental runs.

ER combats forgetting by storing a subset of past experiences and replaying them during training. For this study, ER was configured with a batch size of 32 and trained for 10 epochs.

EWC mitigates forgetting by adding a regularization term that penalizes changes to parameters crucial for previous tasks. Its effectiveness depends on the importance of weights measured by the Fisher information matrix. The model was trained with \( \lambda_{\text{ewc}} = 10 \), a batch size of 32, and for 10 epochs.

SI operates similarly to EWC but estimates parameter importance based on accumulated gradients during training. This study configured SI with \( \lambda_{\text{si}} = 0.01 \), a learning rate of \( 1 \times 10^{-4} \), a batch size of 32, and 10 epochs.

Table \ref{tab:appendix-cma-caud} presents the results of the calibration assessment task. ER demonstrated superior performance across all configurations, achieving the highest accuracy of 94.38\% for the largest dataset ($n_\text{train}$-$n_{\text{user,}i}$=100-100). EWC showed competitive performance, particularly with small datasets, peaking at 74.80\%. SI performed well for intermediate configurations but exhibited a significant drop in gesture accuracy for larger datasets, suggesting its limited scalability in this context.

\begin{table}[h]
\centering
\scriptsize
\begin{tabular}{l|c|c:c:c:c:c}
\hline
\multirow{2}{*}{Method} & \multirow{2}{*}{Baseline}& \multicolumn{5}{c}{Configuration ($n_\text{train}$-$n_{\text{user,}i}$)} \\ \cline{3-7} 
& & \multicolumn{1}{c}{05-05} & \multicolumn{1}{c}{10-10} & \multicolumn{1}{c}{25-25} & \multicolumn{1}{c}{50-50} & \multicolumn{1}{c}{100-100} \\ \hline
ER & \multirow{3}{*}{69.56} & \textbf{86.09} & \textbf{91.53} & \textbf{92.66} & \textbf{93.55} & \textbf{94.38} \\ 
EWC & & 70.42 & 69.80& 71.73 & 69.55 & 74.80 \\ 
SI & & 79.09 & 80.70 & 83.16 & 75.86 & 57.80 \\ 
\hline
\end{tabular}
\caption{Gesture accuracy comparison of different catastrophic forgetting mitigation methods on calibration assessment user data under varying configurations and averaged over all users.}
\label{tab:appendix-cma-caud}
\end{table}

Table \ref{tab:appendix-cma-frad} reports results on forgetting rate assessment data. ER again outperformed both EWC and SI across most configurations, with accuracies of 86.09\% to 94.38\%. Notably, EWC surpassed ER in smaller dataset configurations (05-05 and 10-10), achieving accuracies of 88.37\% and 92.76\%, respectively. However, ER maintained a consistent advantage as dataset sizes increased, reflecting its robustness in handling larger-scale data. SI struggled significantly, with accuracy declining sharply for larger datasets, from 73.55\% in the smallest configuration to 35.93\% in the largest.

\begin{table}[h]
\centering
\scriptsize

\begin{tabular}{l|c|c:c:c:c:c}
\hline
\multirow{2}{*}{Method} & \multirow{2}{*}{Baseline} & \multicolumn{5}{c}{Configuration ($n_\text{train}$-$n_{\text{user,}i}$)} \\ \cline{3-7} 
& & \multicolumn{1}{c}{05-05} & \multicolumn{1}{c}{10-10} & \multicolumn{1}{c}{25-25} & \multicolumn{1}{c}{50-50} & \multicolumn{1}{c}{100-100} \\ \hline
ER &  & 86.09 & 91.53 & \textbf{92.66} & \textbf{93.55} & \textbf{94.38} \\ 
EWC & 76.01 & \textbf{88.37} & \textbf{92.76} & 92.04 & 93.04 & 93.37 \\ 
SI &  & 73.55 & 71.55 & 64.87 & 56.08 & 35.93 \\ 
\hline
\end{tabular}
\caption{Gesture accuracy comparison of different catastrophic forgetting mitigation methods on forgetting rate assessment data under varying configurations and averaged over all users.}
\label{tab:appendix-cma-frad}
\end{table}

The results validate the effectiveness of ER in mitigating catastrophic forgetting and adapting to user-specific gesture data, particularly for larger datasets. While EWC displayed competitive performance for smaller datasets, its scalability is limited compared to ER. SI, though effective under certain conditions, proved less robust overall, highlighting the importance of tailoring mitigation strategies to dataset size and task complexity. This comprehensive analysis underscores ER as the most reliable method for catastrophic forgetting mitigation for XentricAI.

\section{Evaluation of Anomaly Detection Methods}
\label{app:anomal-detection-methods}

To validate the efficacy of the VAE-based approach for anomaly detection, we conducted a comparative evaluation against four alternative methods: One-class \ac{svm}, isolation forest (IF), a density-based approach based on local outlier factors (LOF), and generative adversarial networks (GAN). Each method was assessed for its ability to improve the detection of anomalous gestures missed by the GRU-based HGR model under predefined criteria (e.g., sparse predictions or mixed-class predictions).

In our experiments, we used a condition requiring a minimum of 95\% accuracy on normal gestures to ensure that they were not mistakenly flagged as anomalies. Table \ref{tab:vae-improvement-appendix} summarizes the results, presenting the averaged improvement in correctly flagged gestures across all users and six experimental runs. Each model was trained on 6,000 normalized gestures with a $65\%-35\%$ training-validation-split.  

As described in Section \ref{subsec:vae}, the VAE was trained for 250 epochs, using the Adam optimizer (learning rate = 0.001) and a batch size of 16. Early stopping (patience = 20) restored the best weights based on validation loss. The loss function combined reconstruction and KL divergence terms. Anomalies were detected using reconstruction errors, with user-specific thresholds set from the 90th percentile of calibration gesture errors.

The SVM approach employs a radial basis function kernel to learn the boundary between normal and anomalous gestures in feature space. The kernel's flexibility was controlled by setting the kernel coefficient \(\gamma\) to 0.0025, while the proportion of training errors allowed was determined by setting the error margin parameter \(\nu\) to 0.095. 

The IF method detects anomalies by isolating instances using randomly generated decision trees. The number of trees in the forest was set to 250, and the expected proportion of anomalies in the dataset was defined as 5.5\%.

The LOF technique identifies outliers by comparing the local density of a point to the densities of its neighbors. The number of neighbors considered for local density estimation was set to 35.

The GAN is trained to capture the distribution of normal gestures using two components: a generator that synthesizes data and a discriminator that distinguishes between real and generated samples. The generator consists of five dense layers with 512, 256, 128, 64, and output-dimension units, each followed by batch normalization, ReLU activation, and a dropout rate of 30\%. 
The final dense layer maps to the input dimensionality with a tanh activation function. The discriminator consists of a mirrored architecture with dense layers of sizes 512, 256, 128, and 64 units, each followed by batch normalization, LeakyReLU activation ($\alpha = 0.2$), and dropout. The output layer contains a single unit with a sigmoid activation to classify the input as real or generated. 
The GAN's latent dimension was set to 16, with a batch size of 64 and a learning rate of 0.0002. The model was trained for 100 epochs using the Adam optimizer. Anomalies were detected by calculating the reconstruction error between input data and generated samples, using a threshold based on the 90th percentile of normal data scores.

\begin{table}[h]
\centering
\scriptsize
\begin{tabular}{c|cc|c}
\Xhline{2\arrayrulewidth}
Method & 
\begin{tabular}[c]{@{}c@{}}Correctly Flagged Gestures {[}\%{]}\\ (Condition-Flagged)\end{tabular} & 
\begin{tabular}[c]{@{}c@{}}Correctly Flagged Gestures {[}\%{]}\\ (Condition-Flagged + Method)\end{tabular} & 
Percentage Increase \\ 
\hline
VAE & 57.60 & 69.10 & 11.50 \\ 
SVM & 57.60 & 64.42 & 6.82 \\ 
IF & 57.60 & 64.22 & 6.62 \\ 
LOF & 57.60& 66.37 & 8.77 \\ 
GAN & 57.60 & 67.70 & 10.10 \\ 
\Xhline{2\arrayrulewidth}
\end{tabular}
\label{tab:vae-improvement-appendix}
\caption{Improvement in detecting characterizable gestures using various anomaly detection methods.}
\end{table}

The results in Table B.10 demonstrate the VAE’s superior performance in improving the detection of correctly flagged gestures. The percentage increase in gesture characterization with the VAE (11.50\%) outperformed other methods, including GAN (10.1\%), LOF (8.77\%), SVM (6.82\%), and IF (6.62\%). The VAE's ability to leverage probabilistic mapping and user-specific calibration enabled it to adapt effectively to the variability of radar data. Furthermore, while GANs exhibited strong performance, they were more sensitive to mode collapse compared to the VAE.

These findings highlight the performance of the VAE with user-specific thresholding and transfer learning in FMCW radar-based gesture recognition systems, validating its suitability for real-world anomaly detection scenarios.

\section{Detailed Tables of the Experimental Results}

\begin{sidewaystable}
\centering
\label{tab:full-results-with-er}
\scriptsize 
\begin{adjustbox}{max width=\textwidth, max height=\textheight, keepaspectratio}
\begin{tabular}{l|c|ccccc:ccccc:ccccc:ccccc:ccccc}
\hline
\multirow{2}{*}{User $i$} & \multirow{2}{*}{Baseline} & \multicolumn{16}{c}{$n_\text{train}$-$n_{\text{user,}i}$}                                 \\ \cline{3-27} 
& & \multicolumn{1}{l}{05-05} & \multicolumn{1}{l}{05-10} & \multicolumn{1}{l}{05-25} & \multicolumn{1}{l}{05-50} & \multicolumn{1}{l}{05-100} & \multicolumn{1}{l}{10-05} & \multicolumn{1}{l}{10-10} & \multicolumn{1}{l}{10-25} & \multicolumn{1}{l}{10-50} & \multicolumn{1}{l}{10-100} &\multicolumn{1}{l}{25-05} &\multicolumn{1}{l}{25-10} & \multicolumn{1}{l}{25-25}& \multicolumn{1}{l}{25-50} & \multicolumn{1}{l}{25-100} & \multicolumn{1}{l}{50-05} & \multicolumn{1}{l}{50-10} & \multicolumn{1}{l}{50-25} & \multicolumn{1}{l}{50-50} & \multicolumn{1}{l}{50-100} & \multicolumn{1}{l}{100-05}& \multicolumn{1}{l}{100-10}& \multicolumn{1}{l}{100-25}& \multicolumn{1}{l}{100-50}& \multicolumn{1}{l}{100-100} \\ \hline

    1 & 88.26 & 93.91 & 93.43 & 95.39 & 94.21 & 96.16 & 95.39 & 95.86 & 95.25 & 95.87 & 96.95 & 96.28 & 96.77 & 96.59 & 96.88 & 96.87 & 96.45 & 95.63 & 96.69 & 97.33 & 97.27 & 96.55 & 96.86 & 97.05 & 97.20 & 97.66 \\
    2 & 79.59 & 89.10 & 91.02 & 91.59 & 92.77 & 94.00 & 91.00 & 92.21 & 93.44 & 93.21 & 93.42 & 92.19 & 91.47 & 93.90 & 93.17 & 94.63 & 91.71 & 92.81 & 93.31 & 94.17 & 94.38 & 92.20 & 92.68 & 93.69 & 93.77 & 94.32 \\
    3 & 18.59 & 57.45 & 68.31 & 76.49 & 78.98 & 78.21 & 76.91 & 77.43 & 78.94 & 81.21 & 81.16 & 72.82 & 75.96 & 79.60 & 79.48 & 82.38 & 69.43 & 77.56 & 76.56 & 78.77 & 78.36 & 76.53 & 76.61 & 78.47 & 79.99 & 82.08 \\
    6 & 73.37 & 92.27 & 94.11 & 95.95 & 94.85 & 95.32 & 94.82 & 96.00 & 96.10 & 96.07 & 97.59 & 95.27 & 96.22 & 96.83 & 96.65 & 97.28 & 94.85 & 96.45 & 97.77 & 97.39 & 97.67 & 97.39 & 97.24 & 97.92 & 98.23 & 98.78 \\
    7 & 67.12 & 86.64 & 89.25 & 89.38 & 90.19 & 92.22 & 89.84 & 89.81 & 92.61 & 91.25 & 93.46 & 88.67 & 90.31 & 91.28 & 92.15 & 93.68 & 91.72 & 91.04 & 91.66 & 92.61 & 94.02 & 91.62 & 91.60 & 91.45 & 92.86 & 94.32 \\
    8 & 60.78 & 81.83 & 87.57 & 87.26 & 90.67 & 91.12 & 87.55 & 90.21 & 89.51 & 91.31 & 91.72 & 91.53 & 89.44 & 90.18 & 92.43 & 92.31 & 87.63 & 88.71 & 91.48 & 93.09 & 92.45 & 86.63 & 88.14 & 90.35 & 91.41 & 92.34 \\
    9 & 76.42 & 91.67 & 92.54 & 93.91 & 94.63 & 96.20 & 94.69 & 94.73 & 95.77 & 96.44 & 96.90 & 94.76 & 95.01 & 96.20 & 96.52 & 97.36 & 94.95 & 95.81 & 95.07 & 97.26 & 97.51 & 95.13 & 95.61 & 96.13 & 97.12 & 97.98 \\
    10 & 73.76 & 84.98 & 88.15 & 89.32 & 89.68 & 90.37 & 89.70 & 90.03 & 90.76 & 90.97 & 91.89 & 89.56 & 90.49 & 91.12 & 92.33 & 92.41 & 90.12 & 90.77 & 91.46 & 92.86 & 92.83 & 91.80 & 90.99 & 91.73 & 93.16 & 93.62 \\
    11 & 75.30 & 89.62 & 91.23& 91.94 & 92.94 & 94.25 & 91.54 & 92.88 & 93.04 & 93.27 & 93.53 & 91.97 & 92.67 & 94.00 & 94.34 & 94.89 & 91.94 & 92.85 & 93.35 & 94.21 & 94.28 & 92.28 & 93.06 & 93.48 & 94.46 & 94.56 \\
    12 & 82.40 & 93.43 & 93.98 & 94.06 & 94.57 & 96.56 & 95.94 & 96.11 & 96.45 & 96.82 & 97.23 & 95.94 & 96.42 & 96.90 & 96.96 & 97.54 & 95.93 & 95.78 & 96.89 & 97.80 & 97.79 & 95.36 & 96.92 & 96.62 & 97.39 & 98.10 \\ \hline
    \textbf{Mean} & 69.56 & 86.09 & 88.96 & 90.53 & 91.35 & 92.44 & 90.74 & 91.53 & 92.19 & 92.64 & 93.38 & 90.90 & 91.47 & 92.66 & 93.09 & 93.93 & 90.47 & 91.74 & 92.42 & 93.55 & 93.66 & 91.55 & 91.97 & 92.69 & 93.56 & 94.38 \\

\hline
\end{tabular}
\end{adjustbox}
\caption{Model performance on calibration assessment user data with ER.}

\centering
\scriptsize 
\label{tab:full-results-wo-er}
\begin{adjustbox}{max width=\textwidth, max height=\textheight, keepaspectratio}

\begin{tabular}{l|c|cccccccccccccc}
\hline
\multirow{2}{*}{User $i$} & \multirow{2}{*}{Baseline} & \multicolumn{13}{c}{$n_{\text{user,}i}$}                                 \\ \cline{3-15} 
& & \multicolumn{1}{l}{05} & \multicolumn{1}{l}{10} & \multicolumn{1}{l}{15} & \multicolumn{1}{l}{20} & \multicolumn{1}{l}{30} & \multicolumn{1}{l}{35} & \multicolumn{1}{l}{50} & \multicolumn{1}{l}{55} & \multicolumn{1}{l}{75} & \multicolumn{1}{l}{100} &\multicolumn{1}{l}{105} &\multicolumn{1}{l}{110} & \multicolumn{1}{l}{125} \\ 
\hline
1 & 88.26 & 92.06 & 92.46 & 92.91 & 93.71 & 93.66 & 93.91 & 95.68 & 94.61 & 94.94 & 95.89 & 95.37 & 96.40 & 96.47 \\
2 & 79.59 & 89.72 & 90.49 & 91.51 & 91.94 & 92.26 & 92.05 & 92.81 & 92.32 & 93.32 & 91.81 & 91.83 & 93.28 & 92.47 \\
3 & 18.59 & 66.34 & 68.09 & 72.86 & 72.09 & 76.61 & 75.05 & 77.79 & 78.78 & 77.41 & 81.21 & 79.38 & 79.01 & 79.17 \\
6 & 73.37 & 92.06 & 93.73 & 94.56 & 95.38 & 95.02 & 95.97 & 95.59 & 96.44 & 96.79 & 96.57 & 96.61 & 97.47 & 97.26 \\
7 & 67.12 & 84.43 & 86.45 & 89.19 & 88.87 & 87.58 & 89.11 & 89.39 & 89.61 & 89.44 & 91.15 & 90.83 & 90.53 & 89.89 \\
8 & 60.78 & 83.15 & 84.12 & 84.08 & 88.55 & 87.43 & 86.81 & 89.79 & 89.41 & 89.89 & 90.91 & 91.34 & 91.21 & 90.89 \\
9 & 76.42 & 93.05 & 94.25 & 93.40 & 95.20 & 95.17 & 95.88 & 95.21 & 95.24 & 95.50 & 95.87 & 97.10 & 97.09 & 96.86 \\
10 & 73.76 & 85.47 & 88.56 & 88.61 & 89.71 & 90.03 & 89.03 & 89.44 & 90.51 & 90.15 & 90.47 & 90.78 & 89.52 & 89.84 \\
11 & 75.30 & 89.83 & 90.12 & 90.47 & 90.85 & 91.18 & 91.56 & 91.94 & 92.30 & 92.60 & 92.95 & 93.25 & 93.94 & 92.73 \\
12 & 82.40 & 94.35 & 94.52 & 94.68 & 94.85 & 95.12 & 95.38 & 95.64 & 95.85 & 96.08 & 96.37 & 96.62 & 96.89 & 96.35 \\
\hline
\textbf{Mean} & 69.56 & 87.13 & 88.28 & 89.23 & 90.11 & 90.41 & 90.47 & 91.33 & 91.51 & 91.61 & 92.32 & 92.31 & 92.53 & 92.19 \\
\hline
\end{tabular}
\end{adjustbox}
\caption{Model performance on calibration assessment user data without ER.}
\end{sidewaystable}

\begin{sidewaystable}
\centering
\label{tab:full-fr-results-with-er}
\scriptsize 
\begin{adjustbox}{max width=\textwidth, max height=\textheight, keepaspectratio}

\begin{tabular}{l|c|ccccc:ccccc:ccccc:ccccc:ccccc}
\hline
\multirow{2}{*}{User $i$} & \multirow{2}{*}{Baseline} & \multicolumn{16}{c}{$n_\text{train}$-$n_{\text{user,}i}$}                                 \\ \cline{3-27} 
& & \multicolumn{1}{l}{05-05} & \multicolumn{1}{l}{05-10} & \multicolumn{1}{l}{05-25} & \multicolumn{1}{l}{05-50} & \multicolumn{1}{l}{05-100} & \multicolumn{1}{l}{10-05} & \multicolumn{1}{l}{10-10} & \multicolumn{1}{l}{10-25} & \multicolumn{1}{l}{10-50} & \multicolumn{1}{l}{10-100} &\multicolumn{1}{l}{25-05} &\multicolumn{1}{l}{25-10} & \multicolumn{1}{l}{25-25}& \multicolumn{1}{l}{25-50} & \multicolumn{1}{l}{25-100} & \multicolumn{1}{l}{50-05} & \multicolumn{1}{l}{50-10} & \multicolumn{1}{l}{50-25} & \multicolumn{1}{l}{50-50} & \multicolumn{1}{l}{50-100} & \multicolumn{1}{l}{100-05}& \multicolumn{1}{l}{100-10}& \multicolumn{1}{l}{100-25}& \multicolumn{1}{l}{100-50}& \multicolumn{1}{l}{100-100} \\ 
\hline

1 &  & 93.83 & 92.08 & 92.28 & 88.57 & 90.12 & 93.24 & 93.41 & 91.86 & 92.16 & 92.03 & 94.73 & 94.27 & 94.48 & 94.41 & 92.81 & 94.99 & 94.12 & 94.73 & 95.30 & 93.98 & 96.25 & 96.22 & 96.48 & 96.10 & 95.84 \\
2 &  & 92.74 & 93.24 & 92.52 & 89.47 & 89.31 & 92.95 & 92.77 & 93.33 & 91.95 & 91.02 & 94.47 & 93.51 & 94.47 & 93.07 & 94.18 & 94.45 & 95.26 & 94.87 & 95.22 & 94.34 & 96.12 & 96.23 & 96.17 & 96.10 & 95.96 \\
3 &  & 89.78 & 88.52 & 87.58 & 85.72 & 83.24 & 89.24 & 88.62 & 89.03 & 89.50 & 88.34 & 91.69 & 92.63 & 92.24 & 91.44 & 91.33 & 93.06 & 94.27 & 93.15 & 92.90 & 92.91 & 76.53 & 76.61 & 78.47 & 79.99 & 82.08 \\
6 &  & 92.68 & 92.25 & 91.44 & 94.85 & 95.32 & 94.82 & 96.00 & 96.10 & 96.07 & 97.59 & 95.27 & 96.22 & 96.83 & 96.65 & 97.28 & 94.85 & 96.45 & 97.77 & 98.15 & 97.67 & 97.39 & 97.24 & 97.92 & 98.23 & 98.78 \\
7 &  & 86.64 & 89.25 & 89.38 & 90.19 & 92.22 & 89.84 & 89.81 & 92.61 & 91.25 & 93.46 & 88.67 & 90.31 & 91.28 & 92.15 & 93.68 & 91.72 & 91.04 & 91.66 & 92.61 & 94.02 & 91.62 & 92.40 & 92.74 & 92.86 & 94.32 \\
8 &  & 90.98 & 91.91 & 88.20 & 90.02 & 88.02 & 91.31 & 92.27 & 90.62 & 91.03 & 88.70 & 93.82 & 94.14 & 93.74 & 93.96 & 93.25 & 94.40 & 94.71 & 94.86 & 95.53 & 94.63 & 95.08 & 96.59 & 95.34 & 95.70 & 96.33 \\
9 &  & 91.67 & 92.54 & 93.40 & 94.63 & 96.20 & 94.69 & 94.73 & 95.77 & 96.44 & 96.90 & 94.76 & 95.01 & 96.20 & 96.52 & 97.36 & 94.95 & 95.81 & 95.07 & 97.26 & 97.51 & 95.13 & 95.61 & 96.13 & 97.12 & 97.98 \\
10 &  & 91.20 & 91.93 & 89.58 & 90.71 & 87.84 & 92.64 & 92.20 & 92.80 & 87.06 & 92.02 & 88.73 & 90.49 & 91.12 & 92.33 & 92.41 & 90.12 & 90.77 & 91.46 & 92.86 & 92.83 & 91.80 & 90.99 & 91.73 & 93.16 & 93.62 \\
11 &  & 90.23 & 92.10 & 91.32 & 88.58 & 88.43 & 91.27 & 90.58 & 91.48 & 90.32 & 90.01 & 92.57 & 92.94 & 92.71 & 92.84 & 92.63 & 91.46 & 92.17 & 92.54 & 93.40 & 93.25 & 90.47 & 90.74 & 91.63 & 93.17 & 93.59 \\
12 &  & 93.51 & 91.87 & 92.06 & 89.74 & 90.24 & 92.83 & 93.11 & 91.84 & 92.06 & 91.75 & 92.45 & 94.47 & 94.26 & 94.32 & 93.82 & 94.35 & 94.53 & 95.53 & 95.79 & 94.84 & 95.68 & 96.03 & 96.36 & 96.05 & 95.47 \\
\hline
\textbf{Mean} & 75.99 & 91.32 & 91.57 & 90.78 & 90.25 & 90.09 & 92.28 & 92.35 & 92.54 & 91.78 & 92.18 & 92.71 & 93.40 & 93.73 & 93.77 & 93.88 & 93.43 & 93.91 & 94.16 & 94.90 & 94.60 & 92.61 & 92.87 & 93.30 & 93.85 & 94.40 \\
\hline
\end{tabular}
 \end{adjustbox}
\caption{Forgetting rate assessment of calibrated model with ER.}

\centering
\label{tab:full-fr-results-without-er}
\scriptsize 
\begin{adjustbox}{max width=\textwidth, max height=\textheight, keepaspectratio}
\begin{tabular}{l|c|cccccccccccccc}
\hline
\multirow{2}{*}{User $i$} & \multirow{2}{*}{Baseline} & \multicolumn{13}{c}{$n_{\text{user,}i}$}                                 \\ \cline{3-15} 
& & \multicolumn{1}{l}{05} & \multicolumn{1}{l}{10} & \multicolumn{1}{l}{15} & \multicolumn{1}{l}{20} & \multicolumn{1}{l}{30} & \multicolumn{1}{l}{35} & \multicolumn{1}{l}{50} & \multicolumn{1}{l}{55} & \multicolumn{1}{l}{75} & \multicolumn{1}{l}{100} &\multicolumn{1}{l}{105} &\multicolumn{1}{l}{110} & \multicolumn{1}{l}{125} \\ 
\hline
1 &  & 91.26 & 88.65 & 83.03 & 85.38 & 80.10 & 77.38 & 80.24 & 77.02 & 76.89 & 78.91 & 77.69 & 77.14 & 72.66 \\
2 &  & 92.79 & 92.54 & 86.93 & 86.47 & 84.94 & 81.68 & 76.04 & 75.17 & 73.37 & 67.22 & 63.78 & 68.24 & 64.96 \\
3 &  & 51.18 & 58.78 & 55.42 & 52.91 & 45.82 & 46.18 & 42.49 & 41.83 & 37.14 & 32.99 & 38.27 & 38.80 & 35.36 \\
6 &  & 93.39 & 89.81 & 82.34 & 76.73 & 74.40 & 69.93 & 65.73 & 64.03 & 63.00 & 62.65 & 61.32 & 61.50 & 58.86 \\
7 &  & 88.18 & 86.01 & 85.75 & 82.76 & 79.22 & 77.98 & 76.34 & 72.46 & 70.19 & 67.88 & 68.25 & 64.71 & 64.50 \\
8 &  & 79.75 & 78.15 & 71.61 & 68.83 & 63.87 & 61.54 & 60.86 & 59.64 & 57.03 & 54.51 & 55.26 & 55.44 & 53.66 \\
9 &  & 92.29 & 88.15 & 80.21 & 77.36 & 72.39 & 71.79 & 65.51 & 60.44 & 58.96 & 54.76 & 57.93 & 57.58 & 56.31 \\
10 &  & 82.61 & 82.01 & 76.54 & 74.44 & 69.80 & 64.66 & 65.46 & 64.94 & 62.77 & 63.09 & 65.83 & 64.50 & 65.62 \\
11 &  & 89.87 & 87.56 & 85.77 & 83.62 & 81.91 & 79.63 & 77.21 & 74.91 & 72.85 & 70.45 & 67.61 & 64.73 & 61.95 \\
12 &  & 92.58 & 89.72 & 85.38 & 82.79 & 79.34 & 77.39 & 75.78 & 71.80 & 70.28 & 65.97 & 67.93 & 66.83 & 65.43 \\
\hline
\textbf{Mean}  & 75.99 & 85.39 & 84.14 & 79.30 & 77.13 & 73.18 & 70.82 & 68.57 & 66.22 & 64.25 & 61.84 & 62.39 & 61.95& 59.93 \\
\hline
\end{tabular}
\end{adjustbox}
\caption{Forgetting rate assessment of calibrated model without ER.}
\end{sidewaystable}

\begin{sidewaystable}
\scriptsize 
\centering
\label{tab:full-std-userdata-with-er}
\begin{adjustbox}{max width=\textwidth, max height=\textheight, keepaspectratio}
\begin{tabular}{l|ccccc:ccccc:ccccc:ccccc:cccccc}
\hline
\multirow{2}{*}{User $i$} & \multicolumn{16}{c}{$n_\text{train}$-$n_{\text{user,}i}$}                                 \\ \cline{2-27} 
& \multicolumn{1}{l}{05-05} & \multicolumn{1}{l}{05-10} & \multicolumn{1}{l}{05-25} & \multicolumn{1}{l}{05-50} & \multicolumn{1}{l}{05-100} & \multicolumn{1}{l}{10-05} & \multicolumn{1}{l}{10-10} & \multicolumn{1}{l}{10-25} & \multicolumn{1}{l}{10-50} & \multicolumn{1}{l}{10-100} &\multicolumn{1}{l}{25-05} &\multicolumn{1}{l}{25-10} & \multicolumn{1}{l}{25-25}& \multicolumn{1}{l}{25-50} & \multicolumn{1}{l}{25-100} & \multicolumn{1}{l}{50-05} & \multicolumn{1}{l}{50-10} & \multicolumn{1}{l}{50-25} & \multicolumn{1}{l}{50-50} & \multicolumn{1}{l}{50-100} & \multicolumn{1}{l}{100-05}& \multicolumn{1}{l}{100-10}& \multicolumn{1}{l}{100-25}& \multicolumn{1}{l}{100-50}& \multicolumn{1}{l}{100-100} \\ \hline
    1 & 0.57 & 1.20 & 0.63 & 0.77 & 0.74 & 1.23 & 0.56 & 0.77 & 1.33 & 0.79 & 0.69 & 0.65 & 0.72 & 0.74 & 0.66 & 0.62 & 0.87 & 0.53 & 0.53 & 0.57 & 0.36 & 0.43 & 0.53 & 0.68 & 0.52 \\
    2 & 0.99 & 1.24 & 0.73 & 1.51 & 1.03 & 1.76 & 1.71 & 0.63 & 1.17 & 0.28 & 0.51 & 1.61 & 0.56 & 1.39 & 0.92 & 0.73 & 0.92 & 0.79 & 0.63 & 0.43 & 1.10 & 1.11 & 0.78 & 1.15 & 0.70 \\
    3 & 7.97 & 3.60 & 1.86 & 3.41 & 3.93 & 2.30 & 1.70 & 2.43 & 2.06 & 2.23 & 4.77 & 4.93 & 0.80 & 2.95 & 2.50 & 2.60 & 2.66 & 2.09 & 1.98 & 3.12 & 4.37 & 2.94 & 5.49 & 1.70 & 1.40 \\
    6 & 1.37 & 0.62 & 0.60 & 0.91 & 0.76 & 1.18 & 0.57 & 0.98 & 0.63 & 0.92 & 1.26 & 0.63 & 0.52 & 0.60 & 0.96 & 1.73 & 1.78 & 0.41 & 0.61 & 0.70 & 0.61 & 0.78 & 0.72 & 0.47 & 0.36 \\
    7 & 1.15 & 2.30 & 2.26 & 1.08 & 1.67 & 2.05 & 1.99 & 2.19 & 1.42 & 1.59 & 1.64 & 1.04 & 1.77 & 1.91 & 1.60 & 0.55 & 1.75 & 1.45 & 1.78 & 1.40 & 2.40 & 2.35 & 1.00 & 0.91 & 1.49 \\
    8 & 2.07 & 0.91 & 1.65 & 1.85 & 1.40 & 1.56 & 2.29 & 1.49 & 0.60 & 1.33 & 2.33 & 2.20 & 1.54 & 1.23 & 0.56 & 1.12 & 1.13 & 1.47 & 0.96 & 1.51 & 1.28 & 2.60 & 1.85 & 1.30 & 1.18 \\
    9 & 1.53 & 1.22 & 1.32 & 0.98 & 0.25 & 0.87 & 1.18 & 0.64 & 0.45 & 0.82 & 0.53 & 0.65 & 0.98 & 0.79 & 0.50 & 1.12 & 0.71 & 0.54 & 0.49 & 0.54 & 1.34 & 1.21 & 0.98 & 0.48 & 0.20 \\
    10 & 2.01 & 1.81 & 1.17 & 1.23 & 1.12 & 0.82 & 1.10 & 1.97 & 1.21 & 1.09 & 0.70 & 0.48 & 0.84 & 0.92 & 0.75 & 1.14 & 1.02 & 1.67 & 0.60 & 0.95 & 1.04 & 1.57 & 0.73 & 0.72 & 0.47 \\
    11 & 1.12 & 1.38 & 0.90 & 1.29 & 1.06 & 1.45 & 1.67 & 0.75 & 1.04 & 0.52 & 0.71 & 1.36 & 0.60 & 1.25 & 1.01 & 0.88 & 0.91 & 0.81 & 0.56 & 0.47 & 1.04 & 1.19 & 0.93 & 1.08 & 0.65 \\
    12 & 1.65 & 1.12 & 1.25 & 1.07 & 0.64 & 1.03 & 1.12 & 0.62 & 0.49 & 0.93 & 0.60 & 0.75 & 0.95 & 0.84 & 0.62 & 1.03 & 0.79 & 0.59 & 0.66 & 0.50 & 1.17 & 1.08 & 0.88 & 0.58 & 0.49 \\ \hline
    \textbf{Mean} & 2.04 & 1.54 & 1.24 & 1.41 & 1.26 & 1.42 & 1.39 & 1.25 & 1.04 &
    1.05 & 1.38 & 1.43 & 0.93 & 1.26 &
    1.01 & 1.15 & 1.25 & 1.04 & 0.88 &
    1.02 & 1.47 & 1.53 & 1.39 & 0.91 &
    0.75 \\
\hline

\hline
\end{tabular}
\end{adjustbox}
\caption{Standard deviation of model performance on calibration assessment user data with ER.}

\scriptsize 
\centering
\label{tab:full-std-userdata-without-er}
\begin{adjustbox}{max width=\textwidth, max height=\textheight, keepaspectratio}
\begin{tabular}{l|ccccccccccccccc}
\hline
\multirow{2}{*}{User $i$}  & \multicolumn{13}{c}{$n_{\text{user,}i}$}                                 \\ \cline{2-15} 
& \multicolumn{1}{l}{05} & \multicolumn{1}{l}{10} & \multicolumn{1}{l}{15} & \multicolumn{1}{l}{20} & \multicolumn{1}{l}{30} & \multicolumn{1}{l}{35} & \multicolumn{1}{l}{50} & \multicolumn{1}{l}{55} & \multicolumn{1}{l}{75} & \multicolumn{1}{l}{100} &\multicolumn{1}{l}{105} &\multicolumn{1}{l}{110} & \multicolumn{1}{l}{125} \\ 
\hline
     1 & 0.71 & 0.68 & 0.69 & 0.73 & 0.55 & 0.99 & 0.75 & 0.75 & 0.59 & 0.34 & 1.26 & 0.44 & 0.29 \\
    2 & 0.67 &  0.96 & 1.17 & 0.91 & 0.91 & 0.66 & 0.95 & 1.21 & 0.82 & 0.70 & 1.11 & 0.66 & 0.64 \\
    3 & 1.07 & 2.58 & 1.53 & 1.23 & 2.53 & 1.37 & 1.40 & 1.55 & 1.64 & 1.44 & 1.23 & 2.47 & 2.28 \\
    6 & 0.74 & 1.06 & 0.93 & 0.84 & 1.28 & 0.55 & 1.11 & 0.73 & 0.63 & 1.20 & 1.12 & 0.67 & 1.00 \\
    7 & 0.51 & 2.38 & 2.23 & 1.79 & 1.02 & 1.12 & 2.38 & 0.92 & 1.52 & 1.34 & 0.65 & 1.04 & 1.26 \\
    8 & 1.39 & 0.80 & 1.54 & 1.37 & 0.84 & 1.42 & 0.63 & 1.46 & 0.83 & 2.20 & 0.47 & 0.88 \\
    9 & 0.61 & 0.68 & 0.38 & 0.77 & 0.67 & 0.66 & 0.61 & 0.91 & 1.04 & 0.89 & 0.57 & 0.60 & 0.70 \\
    10 & 1.26 & 0.73 & 1.12 & 1.18 & 0.97 & 0.19 & 0.91 & 1.75 & 0.71 & 0.93 & 0.95 & 1.32 & 1.44 \\
 11 & 0.72 & 0.95 & 1.10 & 0.93 & 0.85 & 0.69 & 0.90 & 1.17 & 0.88 & 1.04 & 0.75 & 0.63 & 0.84 \\
 12 & 0.58 & 0.70 & 0.44 & 0.80 & 0.65 & 0.63 & 0.60 & 0.89 & 1.00 & 0.86 & 0.56 & 0.61 & 0.68 \\ \hline
 \textbf{Mean} & 0.79 & 1.21 & 1.04 & 1.07 & 1.08 & 0.77 & 1.10 & 1.05 & 1.03 & 0.96 & 1.04 & 0.89 & 1.00 \\
\hline
\end{tabular}
\end{adjustbox}
\caption{Standard deviation of model performance on calibration assessment user data without ER.}
\end{sidewaystable}

\begin{sidewaystable}
\scriptsize 
\centering
\label{tab:full-std-traindata-with-er}
\begin{adjustbox}{max width=\textwidth, max height=\textheight, keepaspectratio}
\begin{tabular}{l|ccccc:ccccc:ccccc:ccccc:cccccc}
\hline
\multirow{2}{*}{User $i$} & \multicolumn{16}{c}{$n_\text{train}$-$n_{\text{user,}i}$}                                 \\ \cline{2-27} 
& \multicolumn{1}{l}{05-05} & \multicolumn{1}{l}{05-10} & \multicolumn{1}{l}{05-25} & \multicolumn{1}{l}{05-50} & \multicolumn{1}{l}{05-100} & \multicolumn{1}{l}{10-05} & \multicolumn{1}{l}{10-10} & \multicolumn{1}{l}{10-25} & \multicolumn{1}{l}{10-50} & \multicolumn{1}{l}{10-100} &\multicolumn{1}{l}{25-05} &\multicolumn{1}{l}{25-10} & \multicolumn{1}{l}{25-25}& \multicolumn{1}{l}{25-50} & \multicolumn{1}{l}{25-100} & \multicolumn{1}{l}{50-05} & \multicolumn{1}{l}{50-10} & \multicolumn{1}{l}{50-25} & \multicolumn{1}{l}{50-50} & \multicolumn{1}{l}{50-100} & \multicolumn{1}{l}{100-05}& \multicolumn{1}{l}{100-10}& \multicolumn{1}{l}{100-25}& \multicolumn{1}{l}{100-50}& \multicolumn{1}{l}{100-100} \\ \hline

1 & 0.60 & 1.56 & 0.92 & 2.01 & 1.68 & 1.65 & 1.09 & 0.60 & 2.56 & 1.57 & 1.03 & 1.34 & 1.29 & 1.12 & 1.28 & 0.55 & 0.88 & 0.70 & 0.85 & 1.11 & 0.48 & 0.51 & 0.73 & 0.87 & 0.84 \\
2 & 0.87 & 1.06 & 1.91 & 1.12 & 2.38 & 1.55 & 2.01 & 0.59 & 1.36 & 0.73 & 1.04 & 1.26 & 0.84 & 1.49 & 1.08 & 0.45 & 0.50 & 0.89 & 0.76 & 1.13 & 0.64 & 0.65 & 1.04 & 0.77 & 0.86 \\
3 & 1.37 & 2.03 & 1.65 & 0.96 & 2.79 & 1.14 & 0.70 & 1.11 & 1.46 & 1.24 & 0.83 & 1.23 & 1.14 & 0.59 & 0.99 & 1.42 & 1.03 & 0.84 & 0.85 & 1.20 & 4.37 & 2.94 & 5.49 & 1.70 & 1.40 \\
6 & 0.48 & 1.35 & 1.70 & 0.91 & 0.76 & 1.18 & 0.57 & 0.98 & 0.63 & 0.92 & 1.26 & 0.63 & 0.52 & 0.60 & 0.96 & 1.73 & 1.78 & 0.41 & 0.29 & 0.70 & 0.61 & 0.78 & 0.72 & 0.47 & 0.36 \\
7 & 1.97 & 1.62 & 1.67 & 0.95 & 1.61 & 1.23 & 0.65 & 1.00 & 1.26 & 1.64 & 0.74 & 0.67 & 1.55 & 0.96 & 1.42 & 0.43 & 0.58 & 0.91 & 0.70 & 1.32 & 0.96 & 0.39 & 0.97 & 0.68 & 0.42 \\
8 & 2.07 & 0.91 & 1.65 & 1.85 & 1.40 & 1.56 & 2.29 & 1.49 & 0.60 & 1.33 & 2.33 & 2.20 & 1.54 & 1.23 & 0.56 & 1.12 & 1.13 & 1.47 & 0.96 & 1.51 & 1.28 & 1.85 & 1.10 & 1.30 & 1.18 \\
9 & 1.53 & 1.22 & 1.63 & 0.98 & 0.25 & 0.87 & 1.18 & 0.64 & 0.45 & 0.82 & 0.53 & 0.65 & 0.98 & 0.79 & 0.50 & 1.12 & 0.71 & 0.54 & 0.49 & 0.54 & 1.34 & 1.21 & 0.98 & 0.48 & 0.20 \\
10 & 1.32 & 2.95 & 1.73 & 1.25 & 2.33 & 0.73 & 1.17 & 1.67 & 7.71 & 1.59 & 2.28 & 0.48 & 0.84 & 0.92 & 0.75 & 1.14 & 1.02 & 1.67 & 0.60 & 0.95 & 1.04 & 1.57 & 0.73 & 0.72 & 0.47 \\
11 & 1.08 & 1.43 & 1.25 & 1.65 & 1.87 & 1.12 & 1.32 & 0.98 & 1.15 & 0.68 & 0.91 & 1.11 & 0.79 & 1.03 & 1.17 & 0.62 & 0.59 & 0.88 & 0.83 & 1.04 & 0.48 & 0.56 & 0.79 & 0.93 & 0.77 \\
12 & 1.45 & 1.11 & 1.78 & 0.89 & 0.65 & 1.15 & 1.25 & 0.78 & 0.49 & 0.83 & 0.63 & 0.72 & 0.94 & 0.85 & 0.67 & 1.04 & 0.77 & 0.51 & 0.54 & 0.61 & 1.10 & 1.24 & 0.93 & 0.82 & 0.46 \\
\hline
\textbf{Mean} & 1.27 & 1.52 & 1.59 & 1.26 & 1.57 & 1.22 & 1.22 & 0.98 & 1.77 & 1.13 & 1.16 & 1.03 & 1.04 & 0.96 & 0.94 & 0.96 & 0.90 & 0.88 & 0.69 & 1.01 & 1.23 & 1.17 & 1.35 & 0.87 & 0.70 \\
\hline
\end{tabular}
\end{adjustbox}
\caption{Standard deviation of model performance on forgetting rate assessment data with ER.}

\scriptsize 
\centering
\label{tab:full-std-traindata-without-er}
\begin{adjustbox}{max width=\textwidth, max height=\textheight, keepaspectratio}
\begin{tabular}{l|cccccccccccccc}
\hline
\multirow{2}{*}{User $i$}  & \multicolumn{13}{c}{$n_{\text{user,}i}$}                                 \\ \cline{2-15} 
& \multicolumn{1}{l}{05} & \multicolumn{1}{l}{10} & \multicolumn{1}{l}{15} & \multicolumn{1}{l}{20} & \multicolumn{1}{l}{30} & \multicolumn{1}{l}{35} & \multicolumn{1}{l}{50} & \multicolumn{1}{l}{55} & \multicolumn{1}{l}{75} & \multicolumn{1}{l}{100} &\multicolumn{1}{l}{105} &\multicolumn{1}{l}{110} & \multicolumn{1}{l}{125} \\ 
\hline
    1 & 0.86 & 1.93 & 1.33 & 1.60 & 2.95 & 4.15 & 3.56 & 2.87 & 1.82 & 3.37 & 2.01 & 0.93 & 2.40 \\
    2 & 1.92 & 1.21 & 2.87 & 1.52 & 1.60 & 2.48 & 2.31 & 1.86 & 3.48 & 4.77 & 4.15 & 3.10 & 1.95 \\
    3 & 4.82 & 9.54 & 8.78 & 5.45 & 4.03 & 2.42 & 2.19 & 6.18 & 6.12 & 4.26 & 7.03 & 6.98 & 8.80 \\
    6 & 0.78 & 1.27 & 2.43 & 1.35 & 3.58 & 3.82 & 2.04 & 4.36 & 5.04 & 5.55 & 5.07 & 3.88 & 2.98 \\
    7 & 2.8 & 2.79 & 3.33 & 2.66 & 3.78 & 3.80 & 4.48 & 2.52 & 2.80 & 3.23 & 1.93 & 2.39 & 1.92 \\
    8 & 1.19 &  2.26 & 2.08 & 2.39 & 3.45 & 4.20 & 3.58 & 3.09 & 5.43 & 5.87 & 6.67 & 5.38 & 2.75 \\
    9 & 1.59 & 2.19 & 4.31 & 4.41 & 5.34 & 5.43 & 2.52 & 2.97 & 1.93 & 3.01 & 1.28 & 2.49 & 0.53 \\
    10 & 2.78 & 4.15 & 2.56 & 3.11 & 4.53 & 3.03 & 4.89 & 3.70 & 4.80 & 5.40 & 4.40 & 3.21 & 3.05 \\
     11 & 1.92 & 1.27 & 2.85 & 1.50 & 1.69 & 2.45 & 2.38 & 1.88 & 3.31 & 4.52 & 4.13 & 3.09 & 1.93 \\
     12 & 1.60 & 2.33 & 4.15 & 4.48 & 5.21 & 5.38 & 2.54 & 3.03 & 1.91 & 3.02 & 1.32 & 2.55 & 0.60 \\
     \hline
     \textbf{Mean} & 2.03 & 2.89 & 3.47 & 2.85 & 3.62 & 3.71 & 3.05 & 3.25 & 3.66 & 4.30 & 3.80 & 3.40 & 2.69 \\
\hline
\end{tabular}
\end{adjustbox}
\caption{Standard deviation of model performance on forgetting rate assessment data without ER.}
\end{sidewaystable}





 
\end{document}